\documentclass[preprint,review,12pt]{article}
\usepackage{moreverb}
\usepackage[titletoc]{appendix}
\usepackage{authblk}
\usepackage{soul}
\usepackage{graphicx}
\usepackage{epstopdf}
\usepackage{subfigure}% subcaptions for subfigures
\usepackage{subfigmat}% matrices of similar subfigures, aka small mulitples
 \usepackage{soul}  % for crossing out words
\usepackage{graphicx} %to crop, rotate, reflect... images
\usepackage[colorlinks,bookmarksopen,bookmarksnumbered,citecolor=red,urlcolor=red]{hyperref}
\usepackage{multirow} %to create tables with multiple rows
\usepackage[table]{xcolor} %to color a cell in the table
\usepackage{amsmath}
\usepackage{amsfonts}
\usepackage{amssymb}
 \usepackage{amstext}
\usepackage{psfrag}

\usepackage[margin=2cm]{geometry}
\usepackage{color,graphicx,shortvrb}
\usepackage{epsfig}
\usepackage{psfrag}

\usepackage[normalem]{ulem}%para tachar texto

\usepackage{nomencl}
\usepackage{longtable}

\title{Propagation of Uncertainty in a Rotating Pipe Mechanism to Generate an Impinging Swirling Jet Flow for Heat Transfer from a Flat Plate}

\author[1]{F.-J. Granados-Ortiz}
\author[2]{J. Ortega-Casanova}
\author[1] {C.-H. Lai}
\affil[1]{School of Computing and Mathematical Sciences. University of Greenwich. Old Royal Naval College, Park Row, London SE10 9LS, UK.}
\affil[2]{Fluid Mechanics Group. E.T.S. Ingenier\'ia Industrial. Universidad de M\'alaga.
  C/ Dr. Ortiz Ramos s/n. 29071 M\'alaga. Spain.}

\begin{document}
\maketitle
\begin{abstract}
In Computational Fluid Dynamics (CFD) studies composed of the coupling of different simulations, the uncertainty in one stage may be propagated to the following stage and affect the accuracy of the prediction.  In this paper, a framework for uncertainty quantification is applied to the two-step simulation of the mechanical design of a swirling jet flow generated by a rotating pipe (\textit{Simulation 1}) impinging on a flat plate to provide convective heat transfer (\textit{Simulation 2}). The first approach is the Stochastic Collocation Method (SCM) with Clenshaw-Curtis sparse grids.  The conclusion drawn from the analysis is that the simulated system does not exhibit a significant sensitivity to stochastic variations of model input parameters, over the tested uncertainty ranges.

Additionally, a set of non-linear regression models for the  stochastic velocity and turbulent profiles for the pipe  nozzle are created and tested, since impinging jets at Reynolds number of $Re=23000$ are very frequent in the literature, but stochastic inlet conditions have never been provided.  Numerical results demonstrate a negligible difference in the predicted convective heat transfer with respect to the use of  the profiles simulated via CFD. These suggested surrogate models can be directly embedded onto other CFD applications  (e.g arrays of jets or jet flows impinging on plates with different shapes) in which a realistic swirling flow under uncertainty can be of interest.
\\


\textbf{\textit{Corresponding author(s)}}: F.-J. Granados-Ortiz (frangranados@live.com).

\end{abstract}

\section{Introduction and Motivation} \label{intro}
\vspace{-2pt}
One of the key aspects of using Computational Fluid Dynamic (CFD) simulations is that these are, often, a cheaper option than experiments for product design and development. In some cases, testing experimentally the performance of a new design by engineers could be both risky and expensive. Such development tests may involve the building of several prototypes and may become dangerous if some conditions are extreme (for instance, in a nuclear reactor). For this reason, CFD simulations are a powerful tool in fields such as optimisation \cite{song2003two, ICIAM2015}, aerospace \& aerodynamics industry \cite{mcdaniel2007comparisons, GranadosAIAA2016wash, granados2018influence}, fire safety modelling \cite{taylor1997smartfire}, heat transfer \cite{5} or nuclear energy \cite{mahaffy2007best}, amongst many others. Much effort has been spent to develop numerical algorithms for CFD, leading to more reliable simulations for decision-making and validation purposes, where uncertainty plays an important role.

In experimental work, uncertainty and error measurements are often given but, when performing CFD simulations, this is not a regular practice. If one needs to provide reliable results, this should be a must to offer the most complete overview by including confidence measures.
Generally speaking, in CFD simulations, boundary conditions and geometries are often imposed, without considering the effect on performance that real-life stochastic variations in geometry and boundary conditions may have. Therefore, an option to take into account the stochasticity of some parameters is to use a stochastic analysis instead of a deterministic approach. For this purpose, uncertain inputs are mathematically modelled by using probabilistic distributions derived from experimental data. This is sometimes unavailable and scientists may model the relevant input uncertainties by means of intervals, as well as the study of this propagation, mostly based on experience.

\begin{figure}[!h]
	\begin{center}
		\includegraphics[width=10cm]{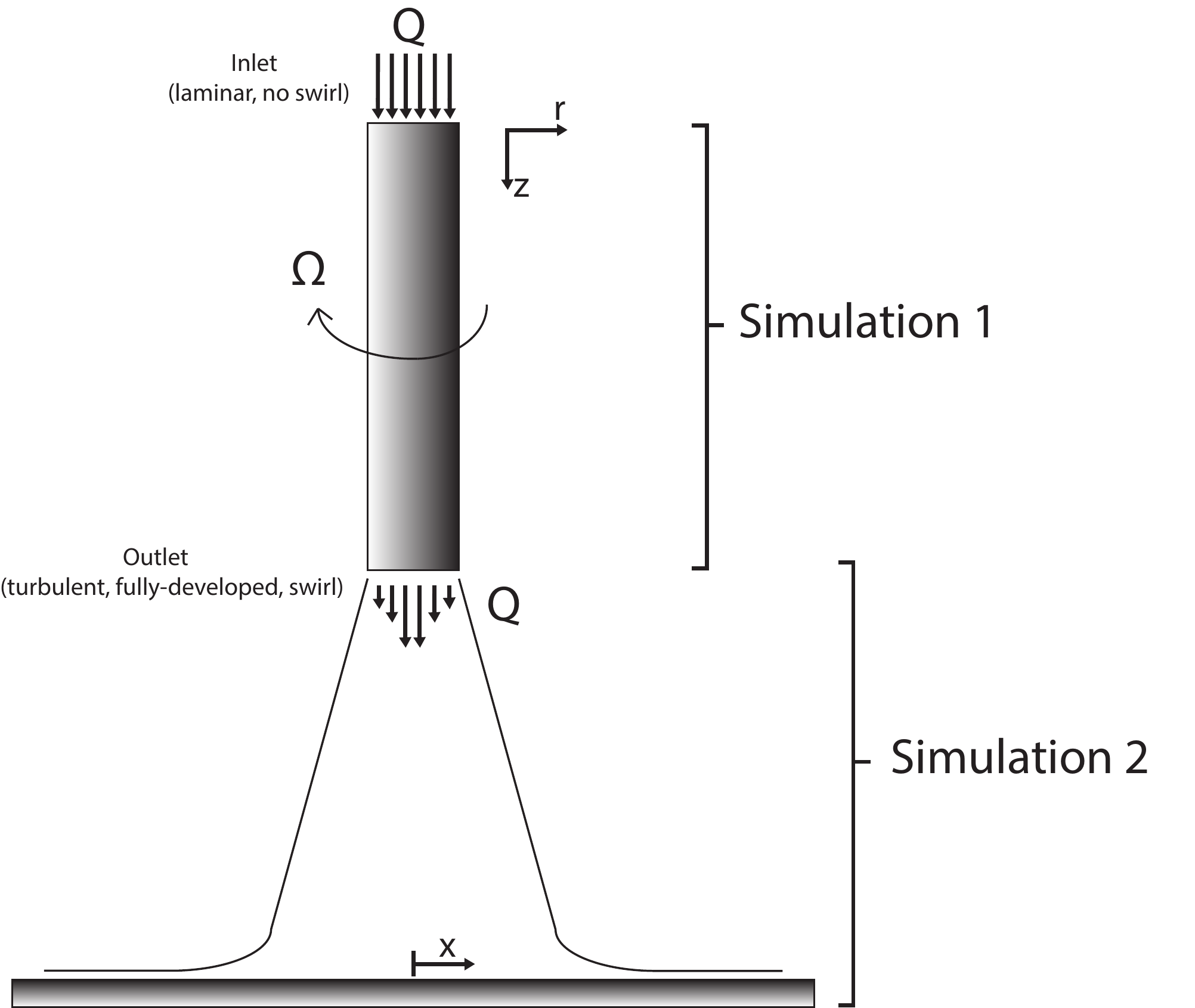}
		\caption{Sketch of \textit{Simulation 1 \& 2}. Note that the $x$ and $r$ axis are the same in practice and have their respective origins in the axisymmetry axis, but for the sake of avoiding confusion in the following plots, $r$ is used in \textit{Simulation 1} and $x$ for \textit{Simulation 2}.} \label{fig: impinging_sketch}
	\end{center} 	
\end{figure}

The work presented in this paper is based on the deterministic CFD simulation studied in \cite{Granados2019I}, whose turbulence model and mesh discretisation errors (Grid Convergence Index, GCI \cite{Celik}) were also discussed and validated in several computational works of heat transfer by swirling impinging jets by the authors \cite{5,ortega2011numerical,ortega2012cfd}. In \cite{Granados2019I}, the main contribution was to propose a new computational simulation to generate the swirl by means of the rotation of a pipe. This simulation is also briefly described in this manuscript. For an efficient computation, the CFD simulation is carried out in two stages: \textit{Simulation 1} \& \textit{2} (see Figure \ref{fig: impinging_sketch}). \textit{Simulation 1} generates the swirling flow to be used as inlet condition in \textit{Simulation 2}, where the heat transfer from the flat plate to the swirling jet is computed. This two-step approach is an efficient alternative to work with different turbulent models (the flow regime in the pipe is different to the flow regime over the plate) as well as to impose periodic boundary conditions onto the pipe to get a fully-developed flow (this avoids a large computational domain) \cite{Granados2019I}.  To properly solve both problems, different turbulent models are tested. The turbulence models with the best performance were the Reynolds Stress Models (RSM) for \textit{Simulation 1}, and the Shear Stress Transport (SST) $k-\omega$ for \textit{Simulation 2}. More details are given in Section \ref{sec:descrip}, by means of a brief description of the set-up, but the reader is referred to \cite{Granados2019I,5,ortega2011numerical,ortega2012cfd} for further information about the successful application of these turbulent models in swirling jet flows.
 
As aforementioned, the main interest is to simulate a swirling flow to enhance the heat transfer on the flat plate. It has been shown in \cite{5, wen2003impingement,ortega2012cfd} that the addition of swirl to impinging jets can increase heat transfer. Note that the swirl can be generated in different ways, such as by using spiral ducts \cite{Lee,bakirci2007visualization,wen2003impingement}, angled blades at the tip \cite{ortega2011experimental}, agitation by stirrer blades, or by a rotating pipe \cite{Imao}. Depending on the generating mechanism, the outflow will have different patterns and, thereby, the jet spread rate will be different. This feature is very influential and the degree of swirl has a dramatic effect on the heat transfer by modifying the shear layer growth and instabilities, entrainment of ambient air and other properties \cite{bakirci2007visualization}. 

For the purpose of simulation, a probabilistic approach to estimate uncertainty provides a more complete overview on the reliability of the numerical computation than a deterministic single-point simulation. This is because the stochastic variance of some parameters is taken into account in the simulation and may have effect on the performance. To our knowledge, there is no previous literature on the effect of simulating impinging jet flows for heat transfer under uncertain conditions, apart from the early stage work presented by the authors in a conference \cite{UNCECOMP2015}. This is a motivation to provide a framework for these complex problems, since the quantification of uncertainties should be an important common practice in CFD. 

Regarding the sources of uncertainty, these are classified as aleatoric or epistemic. Aleatoric uncertainty is considered as implicit to the natural stochasticity in a physical system or quantity. It is also referred to in the literature as irreducible uncertainty, inherent uncertainty, variability and stochastic uncertainty \cite{oberkampf2002error}. On the other hand, epistemic uncertainty is product of the imprecision in the modelling that as a result of a lack of knowledge. This type of uncertainty could, in theory, be reduced if additional information can be added \cite{grossi2005catastrophe}. Epistemic uncertainty associated to the fidelity of the simulation can have a significant effect, and different sources of uncertainty may be actually related. For instance, in \cite{carnevale2013uncertainty} uncertainty is quantified by using both Reynolds Averaged Navier Stokes (RANS) and LES in a heat transfer problem demonstrating that there is an important link between aleatoric and epistemic uncertainties. 

Although the application of Uncertainty Quantification (UQ) to CFD is increasing in popularity exponentially, there are just few documented applications on applying uncertainty quantification to swirling flows \cite{ndiaye2015uncertainty, Congedo}. In \cite{ndiaye2015uncertainty} uncertainty in thermoacoustic instabilities in a swirled stabilized combustor were studied. The motivation of analysing such stochasticity is that the impact of uncertainty could be noticed in the stability modes. This exhibits the importance of a stochastic modelling approach to measure the probability of a mode to be unstable with respect to the input random variables. This is of high relevance in combustor science, since extreme combustion instabilities can highly damage the system, as mentioned in their work. In \cite{Congedo}, a swirling flow with swirl intensity, $ S $, ranging from $ 0 $ to $ 0.6 $ confined in a pipe is simulated by means of both RANS and LES. The pipe is rotating with a Reynolds number of $ Re=30000 $, and it undergoes a sudden expansion. The numerical quantification of uncertainty was found to be very close to the reported experimental one, by means of both RANS with the $ k-\epsilon $ model and LES with the Smagorinsky model. However, due to the lower fidelity of RANS simulations, these provided the least accurate and most sensitive results in the simulations. This work is very close to our problem under study (containing RANS simulations, rotating pipe, swirling flow suddenly expanded, similar $ Re $, similar $ S $ values). Their outcomes encouraged us to undertake the quantification of experimental uncertainties in our simulations. 
The impact of uncertainty in CFD simulations of jets has been also studied by the authors in \cite{granados2018influence}, where the simulation of a compressible jet flow under uncertain conditions is analysed, demonstrating that there is a relationship between the input random variables and the spatial distribution of pressure and velocity arising due to the propagation of uncertainty in the simulation. Other papers that also offered a reference and motivation are \cite{adya2011uncertainty}, where synthetic jets by means of polynomial chaos are studied, \cite{gorle2011epistemic} where underexpanded jets in a crossflow for turbulent mixing are investigated, and \cite{iaccarino2017}, where uncertainty estimation is developed in RANS simulations of high-speed aircraft nozzle jets. In several papers \cite{gorle2011epistemic, iaccarino2017, iaccarino2017eigenspace} a methodology to deal with the well-known epistemic uncertainty in turbulence models is outlined and tested, by means of eigenvalue and eigenvector perturbations. In their work, the perturbation of the eigenvalues of the Reynolds stress anisotropy tensor is modelled by the position in a barycentric triangle map, whose corners stand for the limiting states of turbulence anisotropy. The eigenvector perturbation is made to change the Reynolds stress tensor alignment to find the extremal alignments with the mean strain. Throughout these papers it is highlighted the importance of providing uncertainty bounds in RANS simulations, and the results suggest that their uncertainty estimation method can account most of the model inadequacy. During the literature survey no applications of UQ to heat transfer by impinging jet flows were found. 

The structure of this paper is as follows.  
In Sections \ref{sec:method} and \ref{sec:descrip}, the methodology and a description of the problem are given as an overview. In Section \ref{subsec_SC}, the uncertainty quantification process is described, in order to understand Stochastic Collocation Method and the use of sparse grids. The type of uncertainties considered are also described in this part. In Section \ref{sec: coupling}, the coupling between \textit{Simulation 1} and \textit{Simulation 2} and the suggested models is explained.  This leads to the following comparison between the resulting uncertainties with and without implementing the models for the pipe outflow profiles in Section \ref{sec: UQresults}. In this section, different probabilistic distributions for the inputs on the surrogate models are also tested, for the UQ purposes of this work. Finally, in Section \ref{sec: conclusions}, the relevant conclusions of this work are given. In Appendix \ref{append_A}, polynomial models for the coefficients of the non-linear regression models are shown, and in Appendix \ref{append_B} a piece of code is provided for the implementation of the models by a User Defined Function (UDF) in FLUENT.  \\

\section{Methodology} \label{sec:method}

\begin{figure}[h!]
	\begin{center}
		\includegraphics[width=10cm]{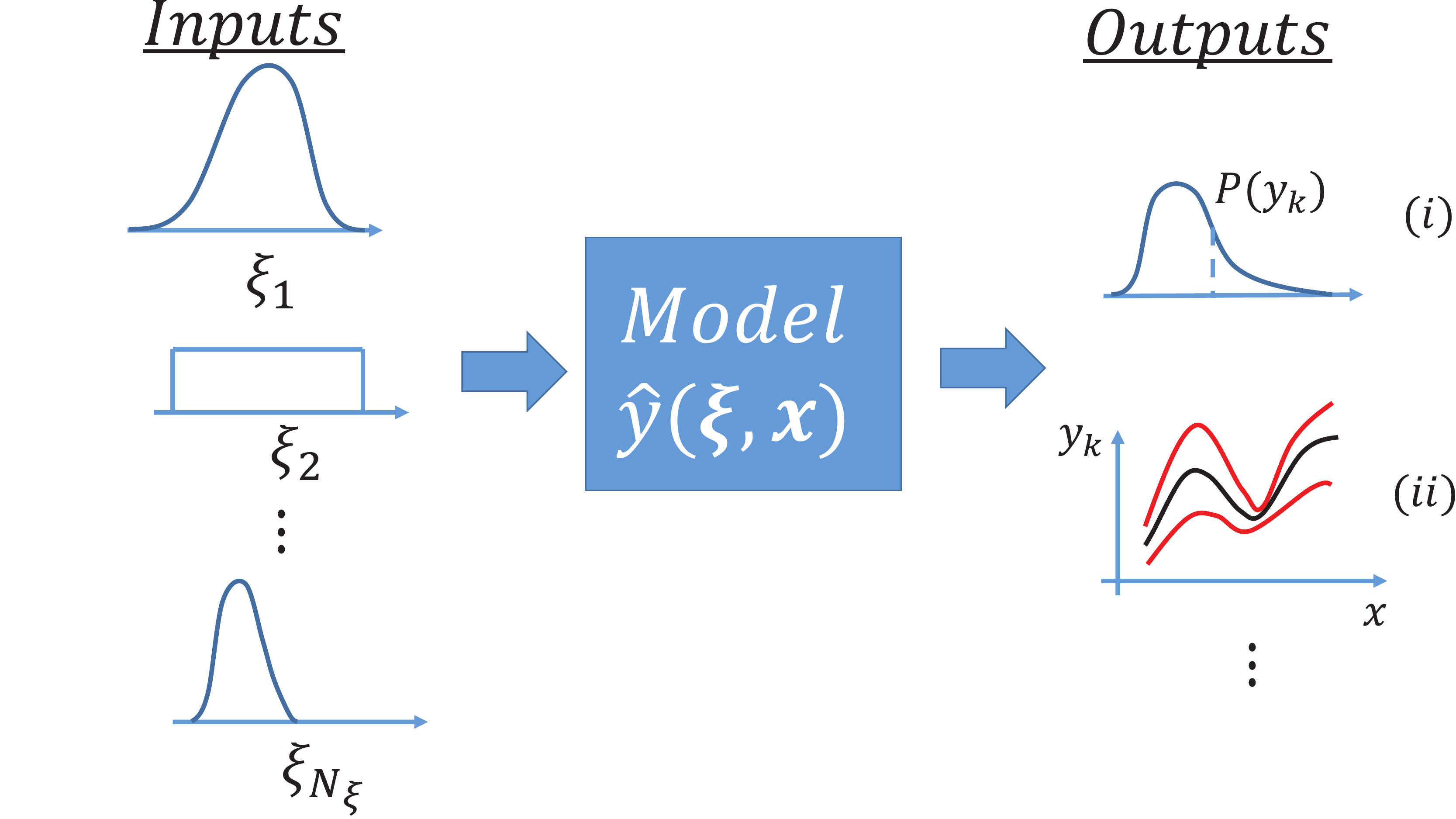}
		\caption{Propagation of uncertainties in a mathematical model.} \label{fig:prop_uncert}
	\end{center}
	
\end{figure}

The aim of the Uncertainty Quantification is to provide confidence measures on how the output of a model, say $\hat{y}$, is varied due to the variability of its inputs, say $\hat{\xi}_{i}$ \cite{Saltelli2004} (see Fig. \ref{fig:prop_uncert}). In the present work, our main investigation is restricted to aleatoric uncertainties arising due to experimental errors or variability. Any epistemic uncertainty from the use and calibration of turbulence models is not considered. However, this study contains modelling work, which consists of the search of mathematical models for the inlet profiles for \textit{Simulation 2}. Since the considered aleatoric uncertainties are actually the same with and without the models, and the UQ method is also the same, the impact of the epistemic uncertainty associated to the non-linear regression models is being quantified. Due to the fact that the concept of epistemic uncertainty may evoke to several sources to the reader, this specific inaccuracy is referred to as modelling uncertainty in this manuscript.\\

Once the input uncertainties are modelled, it is necessary to find an appropriate UQ method. These methods can be either intrusive or non-intrusive. A non-intrusive approach is chosen since this does not require additional code implementation in the solver and can deal with any model as a black-box (FLUENT software in our case). 
On that basis, Monte-Carlo simulations \cite{metropolis1949monte} are a reliable possibility. This non-intrusive method is based on a random sampling on the input uncertainties in order to obtain enough outputs to build the statistical output data. As the convergence of the method is slow, being of order $ O(1/\sqrt{N_{s}})$, with $N_{s}$ the number of samples, in order to reduce $N_{s}$ and increase the efficiency of the method, other sampling methods are available in literature such as the Latin Hypercube \cite{helton2003latin}, the quasi-random Halton \cite{faure2009generalized} or the Sobol sequences \cite{burhenne2013uncertainty}, amongst others. These sampling techniques optimise the sampling by taking also into account previous positions of the samples in the stochastic space, and the error is essentially of order $ O(1/N_{s})$. It is important to point out that this efficiency is mostly noticed for moderate dimensions of the stochastic space, as for high dimensional problems they behave similarly to Monte-Carlo (which is dimension independent).

Sampling based methods are, therefore, a good choice for UQ. However, these usually require a large number of model evaluations. This is the main disadvantage when performing UQ on CFD, as each model evaluation requires a large computational time, being often unaffordable. To overcome this drawback, the Stochastic Collocation Method (SCM) \cite{Mathelin} is implemented. This calculates the statistical moments of the output by dedicated quadrature techniques on tensor grids. SCM has better performance than sampling methods especially for low dimensions in the stochastic parameter space. When the dimension is high, the cost leads to the so-called \textit{Curse of Dimensionality}, as the number of collocation points to evaluate increases exponentially, and it is preferable to consider other methods. An efficient way to mitigate the needs of using many collocation points is to use sparse grids \cite{smolyak63}, for which the accuracy (that is to say the number of collocation points) can be consistently increased until convergence is achieved.

SCM based approaches have been used in several applications in the literature. This non-intrusive method has been successfully applied to problems such as elliptic partial differential equations with random input data \cite{babuvska2007stochastic}, supersonic aircraft jets \cite{GranadosAIAA2016wash} or cardiovascular research \cite{sankaran2011stochastic}. In some UQ methods, including SCM, the exact response can be approximated by creating a surrogate model. Then, sampling techniques can be applied on the surrogate to recover more statistical information, such as probabilistic distributions of the outputs. For this reason, two types of input probabilistic distributions were tested in the present work to generate the random inputs on the surrogates: a uniform and a normal distribution. As experimental uncertainty data is not available to build the empirical probabilistic functions, to try two different distributions provide some information about the impact of non-linearity and high-order effects in the propagation of uncertainty through the CFD simulation. The random variables are sampled to evaluate the SCM response surface and get converged probabilistic distributions of the outputs. \\

Since the swirling flow generation in \textit{Simulation 1} is decoupled from \textit{Simulation 2}, it is interesting for reliability reasons to find a way to characterise the outflow velocity and turbulence profiles from \textit{Simulation 1} under uncertain conditions, to impose that data onto \textit{Simulation 2}. This is the objective in Section \ref{sec_models}, where non-linear regression models are sought. A numerical computation of the uncertainty of the outlet profiles from \textit{Simulation 1} is done by means of SCM in the present paper, at different values of the normalised radial coordinate $r/R$, being $r$ the radial coordinate and $R$ the radius of the pipe. However, to quantify the uncertainty in the output of \textit{Simulation 2}, several profiles should be systematically generated from \textit{Simulation 1} and then input as boundary condition to \textit{Simulation 2}. The alternative to avoid \textit{Simulation 1} and provide functions to generate the profiles as input in \textit{Simulation 2} is investigated in this work. These functions can be coded into FLUENT by means of a User Defined Function (UDF) to any CFD problem, in contrast to coding a SCM surrogate for every $r/R$ location, which is a very cumbersome option, as well as potential source of human errors.  

Impinging jets at Reynolds number of $Re=23000$ have been one of the most studied in the literature for years on end, e.g. in \cite{Lee, behnia1998prediction, yan1998heat, ortega2017using, uddin2013simulations, cooper1993impinging, Baughn, uddin2008turbulence, Granados2019I, wen2003impingement, shum2014large}. Hence, it is useful for future research to have stochastic/non-stochastic profiles ready to input as boundary condition to CFD simulations. The use of algebraic functions for modelling jet flow profiles is an established practice \cite{crow1971orderly, yen1998application,balakumar1998prediction,tam1984sound}, for instance to perform stability analysis of the jet plume \cite{piot2006investigation, airiau2015non}; or in topics closer to the present paper as the jet in \cite{ortega2011experimental}, whose swirling jet empirical functions were successfully used as inlet profiles to simulations in \cite{5,ortega2011numerical,ortega2012cfd}. \\
 
\section {A brief description of the set-up simulated by CFD} \label{sec:descrip}
In this paper, few details on the CFD configuration are given, since a detailed numerical investigation on this swirling flow was developed in \cite{Granados2019I}. In this study, the impinging swirling jet is created by using a rotating pipe with a fully-developed flow at its exit. The inlet is a uniform flow with Reynolds number $ Re =\frac{4 \rho Q}{\pi D\mu} = 23000 $, whilst the outflow is a fully developed turbulent flow. This flow spreads from the exit of the pipe (nozzle) and impinges on the heated flat plate below, located orthogonally at a dimensionless distance $ H/D = 5$, where $H$ is the distance between the nozzle and the plate, and $D$ is the diameter of the pipe. The Prandtl number is $Pr=\frac{\nu}{\alpha}=0.71$, where $\nu$ is the kinematic viscosity of the fluid and $\alpha$ is the thermal diffusivity. The Swirl number is set to $S= \frac{\pi D^3\Omega}{8Q} = 1 $, with $Q$ the volume-flow rate and $ \Omega $ the angular velocity of the pipe.

\begin{figure} [ht!]
\begin{center}
	\begin{subfigmatrix}{2}
		\subfigure[]{{
				\includegraphics[width=0.48 \textwidth]{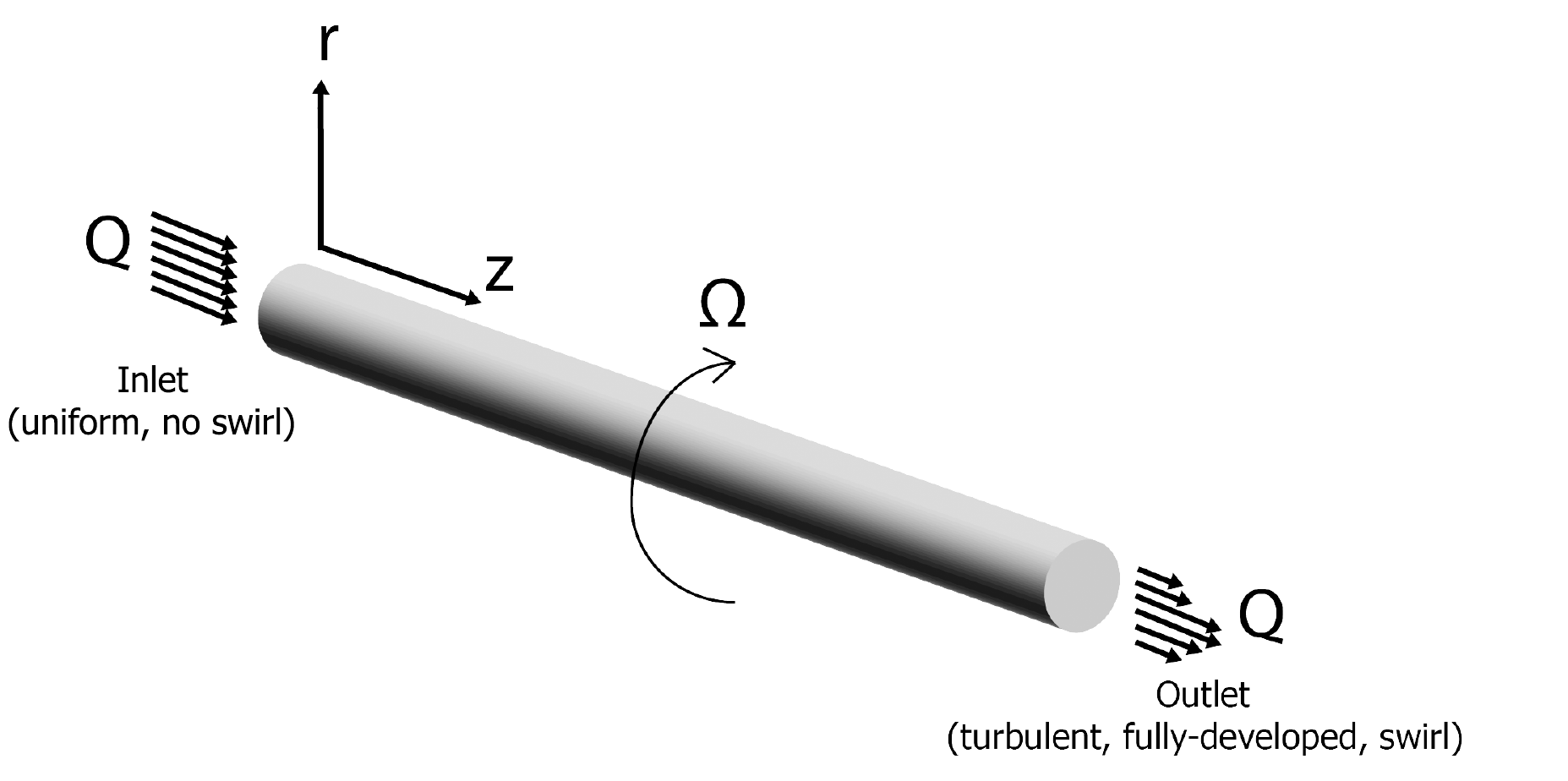}}}
		\subfigure[]{{
				\includegraphics[width=0.48\textwidth]{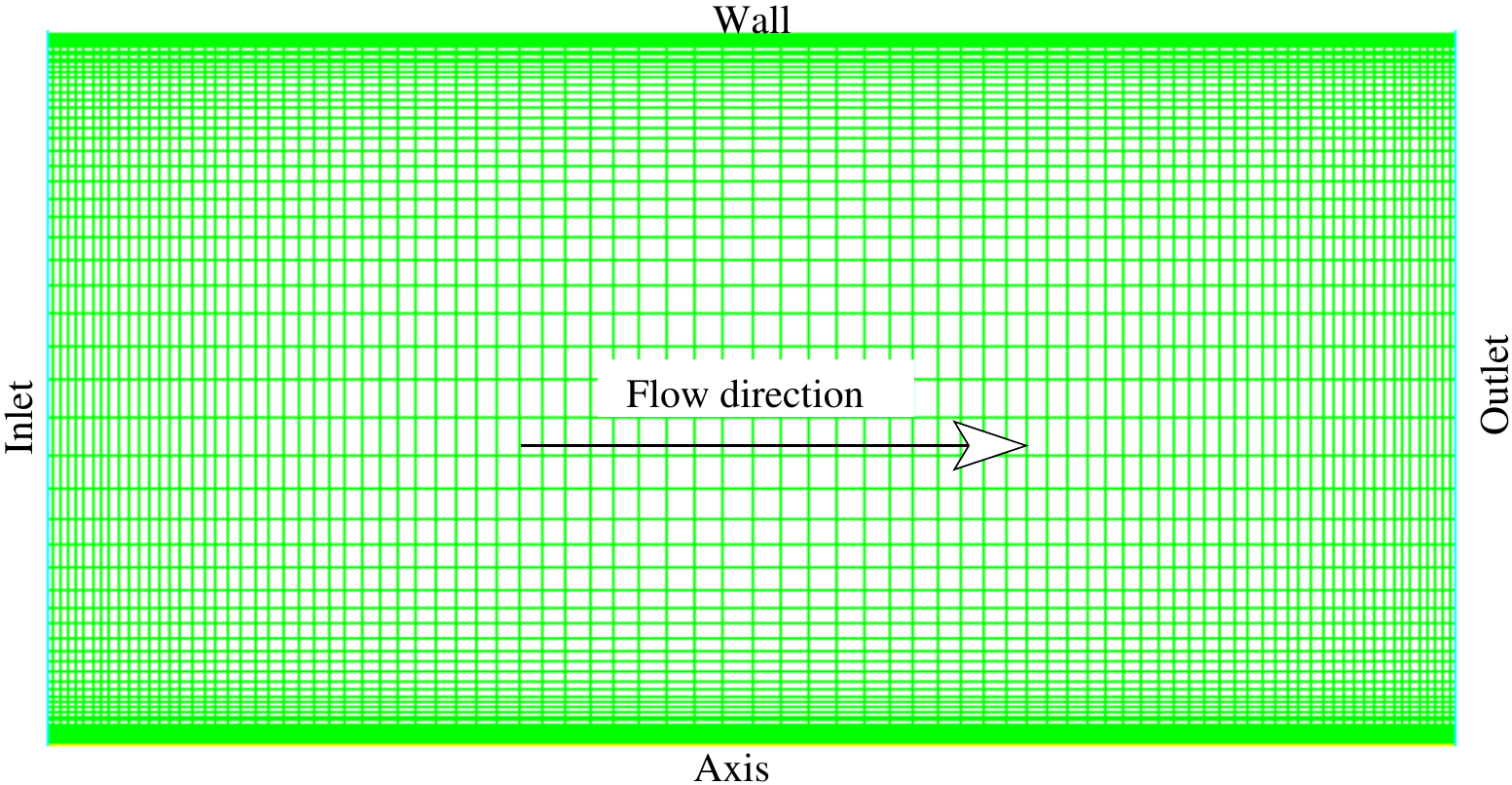}}}			
	\end{subfigmatrix}

	\caption{(a) Sketch of the swirling flow generator by a rotating pipe. The axial periodic flow is initialised with a uniform profile and without swirl, and develops over time as turbulent fully-developed pipe flow with swirl. (b) Axisymmetric 2D mesh of the rotating pipe. In the simulation, the flow goes from left (inlet boundary condition) to right (outlet boundary condition) to impose the periodic condition. The bottom side is the axis (axis boundary condition) and the top side is the pipe wall (wall boundary condition with rotation imposed).} \label{rot_pipe}
\end{center}
\end{figure}

Prior to the simulation of the heat transfer, the velocity and turbulent profiles of the impinging swirling jet are to be produced in a separate simulation of a rotating pipe. This is \textit{Simulation 1}, shown in Fig. \ref{fig: impinging_sketch}. In Fig. \ref{rot_pipe}(a), the pipe problem is depicted, for which a 2D RANS simulation was developed in FLUENT. The flow under study is axisymmetric, steady, incompressible and becomes fully-developed. To obtain the fully-developed pipe flow from a uniform inflow requires a pipe length greater than a specific characteristic one. In order to reduce the cost of this computation, a piece of pipe has been simulated with periodic boundary conditions, with a mass-flow rate corresponding to $Re=23000$ and $S=1$. The turbulence model used in this simulation was the Reynolds Stress Model (RSM). The finite-volume discretisation used the axisymmetric  mesh of the pipe shown in Fig. \ref{rot_pipe}(b). The mesh size is $[n_{r} \times n_{z}]=68\times450$ cells, having a dimensionless wall distance of $y^{+}$ $<$ $1$ along the wall of the pipe. \\

\begin{figure}[t]
	\begin{center}
		\includegraphics[width=12cm]{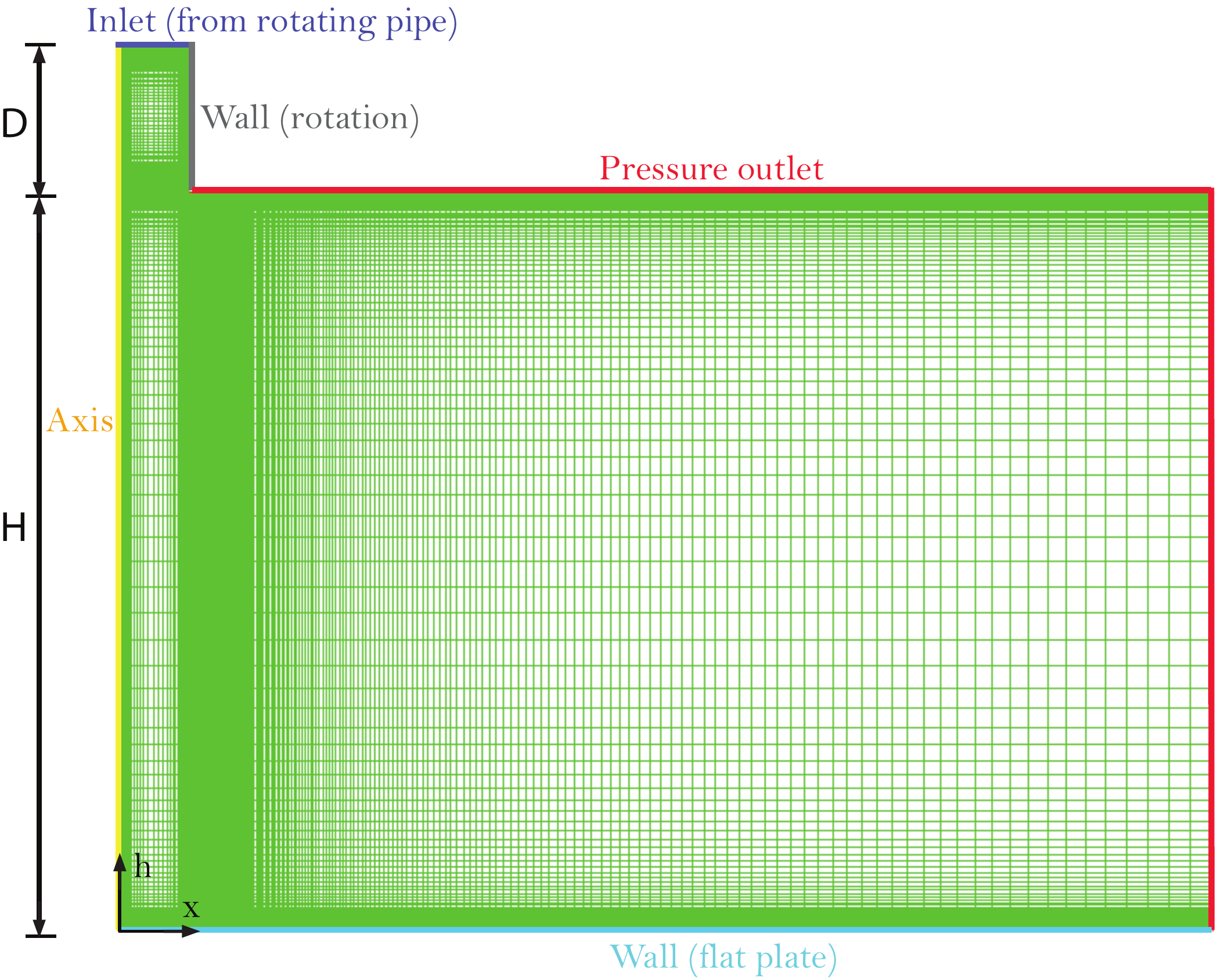}
		\caption{Sketch of the impinging jet flow and the boundary conditions chosen in FLUENT. A portion of the pipe of length $D$ is simulated for a less abrupt adaptation of the inlet from \textit{Simulation 1} in \textit{Simulation 2}.}  \label{ht_domain}
	\end{center}
	
\end{figure}

Once \textit{Simulation 1} is completed, its output is used as inlet boundary condition for \textit{Simulation 2}. In Fig. \ref{ht_domain}, the detail of its discretisation and boundary conditions of the 2D axysimmetric simulation are depicted. The computational mesh is in this case $[n_{x} \: \times \: n_{h}]$ = $140$ $\times$ $250$ cells, ensuring an $y^{+}$ $<$ $1$ all along the plate. The dimensionless nozzle-to-plate distance is $H/D=5$ and a piece of the swirling pipe of length $D$ has been included in the domain in order to enable the pressure field to properly develop inside the pipe prior the expansion \textit{Simulation 1}. For the sake of clarity, the $x$ axis is the same as the $r$ one, but different notation is used to avoid confusion between the parameters at the exit of the pipe and those on the flat plate. Regarding the turbulence model, the SST $k-\omega$ has been used as shown in \cite{Granados2019I}. 

Both the turbulent models RSM for the rotating pipe and the SST $k-\omega$, for the impinging problem, have been exhaustively analysed and validated in \cite{Granados2019I}, and the impinging jet simulation also in \cite{5,ortega2011numerical,ortega2012cfd}. The authors suggest to see these publications for further information about the computational features in these simulations, including the discretisation errors, comparisons against experimental results and other tested turbulence models.

Once the deterministic CFD simulations are ready, these can be run a number of times with different values of the input parameters to perform the uncertainty analysis. In this work, the chosen method is the Stochastic Collocation, which is formally described next. 

%\vspace{-6pt}

\color{black}
\section{Uncertainty Quantification} \label{subsec_SC}

\subsection {Stochastic Collocation Method} 

The method implemented here for UQ is the Non-Intrusive Stochastic Collocation Method (SCM). It was originally developed by Mathelin and Hussaini \cite{Mathelin} at NASA as alternative to other UQ methods, which demanded higher costs.  SCM represents a very efficient option for lower dimension problems in comparison with sampling techniques such as \textit{e.g.} Monte-Carlo. For higher dimension problems, sampling techniques tend to be more suitable.\\

Consider the differential operator on an output of interest of a stationary problem, $y$(\textbf{x}, $\boldsymbol\xi$($ \eta $)) as
\begin{eqnarray}
\mathcal{L}( \textbf{x}, \boldsymbol\xi(\eta); y(\textbf{x}, \boldsymbol\xi(\eta)) ) = \mathcal{Q}( \textbf{x}, \boldsymbol\xi(\eta)),
\label{eq:diff_operator}
\end{eqnarray}
with $ \mathcal{L}$ and $\mathcal{Q}$ differential operators on $ \mathcal{D} \times \Xi  $, where $ \textbf{x} \in \mathcal{D} \subset \mathbb{R}^d $, $ d \in \{1, 2, 3\} $. $ \eta $ denotes events in the complete probabilistic space $ (\hat{\Omega}, \mathcal{\hat{F}}, \hat{P}) $, with $ \mathcal{\hat{F}} \subset 2^{\hat{\Omega}} $ the $ \sigma $-algebra of subsets of $ \hat{\Omega} $ and $ \hat{P} $ a probability measure. $ \Xi \subset \mathbb{R}^{N_\xi} $, is the stochastic space on which the random variables $ \boldsymbol\xi(\eta) $ are defined and $N_{\xi}$ stands for the number of random variables (two in our case under study). The objective is to find the $i$-th statistical moments, $ \mu (\textbf{x}, \boldsymbol\xi)^y _{i} $, of the output of the model by
\begin{eqnarray}
\mu (\textbf{x}, \boldsymbol\xi)^y _{i} = \int_{\Xi} y(\textbf{x}, \boldsymbol\xi)^i f_\xi (\boldsymbol\xi) d\boldsymbol\xi,
\label{eq:stat_moments}
\end{eqnarray}
%\begin{eqnarray}
%\mathbb{E}^{SC} (y) = \int_{\Xi} y(\textbf{x}, \boldsymbol\xi) f_\xi (\boldsymbol\xi) d\boldsymbol\xi,
%\label{eq:stat_moments}
%\end{eqnarray}
%
%\begin{eqnarray}
%\mathbb{V}^{SC} (y) = \int_{\Xi} y(\textbf{x}, \boldsymbol\xi)^2 f_\xi (\boldsymbol\xi) d\boldsymbol\xi-\mathbb{E}^{SC} (y),
%\label{eq:stat_moments}
%\end{eqnarray}
with $ f_\xi (\boldsymbol\xi) $ standing for the density function $ f_\xi : \xi \mapsto \mathbb{R}_{+} $. In many applications this is not sufficient and incomplete information of the variable of interest is given. The reason is that two very different probabilistic distributions can have, for instance, the same mean and standard deviation. Therefore, this can be misleading when further information is required, such as finding out whether the solution is multimodal. Consequently, the probabilistic distribution of $y$(\textbf{x}, $\boldsymbol\xi$) is often sought by sampling surrogates as in this manuscript.

When implementing SCM, special attention must be paid to the probabilistic density function of the random variables $\xi$ $\in$ $\Xi$, as we have to perform a mathematical transformation from the physical random variable space to an artificial stochastic space, known as $\alpha$-domain or $\alpha$-space:
\begin{equation} 
\alpha= \mathcal{S(\xi)}.
\label{eq:alpha_space}
\end{equation}
This transformation is a difference with respect to other UQ methods, as the stochastic space $ \alpha $ is defined in the domain of Lagrange interpolating polynomials in $ [-1, 1]$.\footnote{It is interesting to point out that in several papers \cite{Mathelin,loeven2007probabilistic}, $\alpha$ is referred to as an artificial stochastic space, but the space is in fact $\Gamma$, as $\alpha$ $\in$ $\Gamma$. In this work, $\Gamma$ will be referred to as the $\alpha$-space in order to be consistent with the existing notation in the literature.} It is hence useful that, for each collocation point, the CFD problem is solved deterministically and the solution can be reconstructed by  
\begin{eqnarray}
\hat{y}(\textbf{x},\alpha) \simeq \sum\limits_{j=1}^{N_{q}} y_j(\textbf{x})\:l_j(\mathcal{S}^{-1}(\boldsymbol\alpha)) = \sum\limits_{j=1}^{N_{q}} y_j(\textbf{x})\:l_j(\alpha),
\label{eq:interp}
\end{eqnarray}
where $ y_j(\textbf{x}) $ are the deterministic solutions from $y(\textbf{x}, \xi_j)$ now transformed into $y(\textbf{x}, \alpha_j)$, $N_{q}$ the number of collocation points and $ l_j $ the Lagrange interpolation polynomials defined in the new stochastic space. Note that $\mathcal{S}^{-1}(\boldsymbol\alpha)$ denotes the inverse mathematical transformation to $\alpha$ $\in$ $\Gamma$. The integral of Eq. (\ref{eq:stat_moments}) now can be approximated as
%\begin{eqnarray}
%\mu (\textbf{x}, \boldsymbol\xi)^{\hat{u}} _{i} = \sum\limits_{j=1}^{N_{q}} u_j(\textbf{x})^i \int_{\Gamma} l_j(\mathcal{S}^{-1}(\boldsymbol\alpha)) \: f_\xi (\mathcal{S}^{-1}(\boldsymbol\alpha)) \: \frac{1} {\mathcal{S}'(\boldsymbol\alpha)} d\boldsymbol\alpha,
%\label{eq:quad}
%\end{eqnarray}
\begin{eqnarray}
\mu (\textbf{x}, \boldsymbol\alpha)^{\hat{y}} _{i} = \sum\limits_{j=1}^{N_{q}} y_j(\textbf{x})^i \int_{\Gamma} l_j(\boldsymbol\alpha) \: f_\xi (\boldsymbol\alpha) \: J(\boldsymbol\alpha) \: d\boldsymbol\alpha,
\label{eq:quad}
\end{eqnarray}
where $J(\boldsymbol\alpha)$ the Jacobian of the differential transformation. Finally, it can be numerically computed by quadrature as
%\begin{eqnarray}
%\mu (\textbf{x}, \boldsymbol\xi)^{\hat{u}} _{i} = 
%\sum\limits_{j=1}^{N_{q}} u_j(\textbf{x})^i 
%\sum\limits_{k=1}^{N_{q}} l_j(\mathcal{S}^{-1}(\boldsymbol\alpha_k)) \: f_\xi (\mathcal{S}^{-1}(\boldsymbol\alpha_k)) \: \frac{1} {\mathcal{S}'(\boldsymbol\alpha_k)} z_k,
%\label{eq:quad2}
%\end{eqnarray}
\begin{eqnarray}
\mu (\textbf{x}, \boldsymbol\alpha)^{\hat{y}} _{i} \simeq
\sum\limits_{j=1}^{N_{q}} y_j(\textbf{x})^i 
\sum\limits_{k=1}^{N_{q}} l_j(\alpha_k) \: f_\xi (\alpha_k) \: J(\alpha_k) \: z_k,
\label{eq:quad2}
\end{eqnarray}
with $ z_k $ and $\alpha_k$ standing for the quadrature weights and points respectively. Both the quadrature and collocation points are the same in this work. For the Lagrange interpolating polynomials, must be taken into account that $l_i (\alpha_j) = \delta_{ij}$ for $i,j =[1, 2, ..., N_q]$. Therefore, with these assumptions, the integral from Eq. (\ref{eq:quad2}) above can be finally rewritten for an uniform probabilistic input defined in $[-1, 1]$ as
\begin{eqnarray}
\mu (\textbf{x}, \boldsymbol\alpha)^{\hat{y}} _{i} \simeq \frac{1} {2} \sum\limits_{j=1}^{N_{q}} y_j(\textbf{x})^i z_j.
\label{eq:stat_momentsquad}
\end{eqnarray}

\begin{figure}[h!]
	\begin{center}
		\includegraphics[width=12cm]{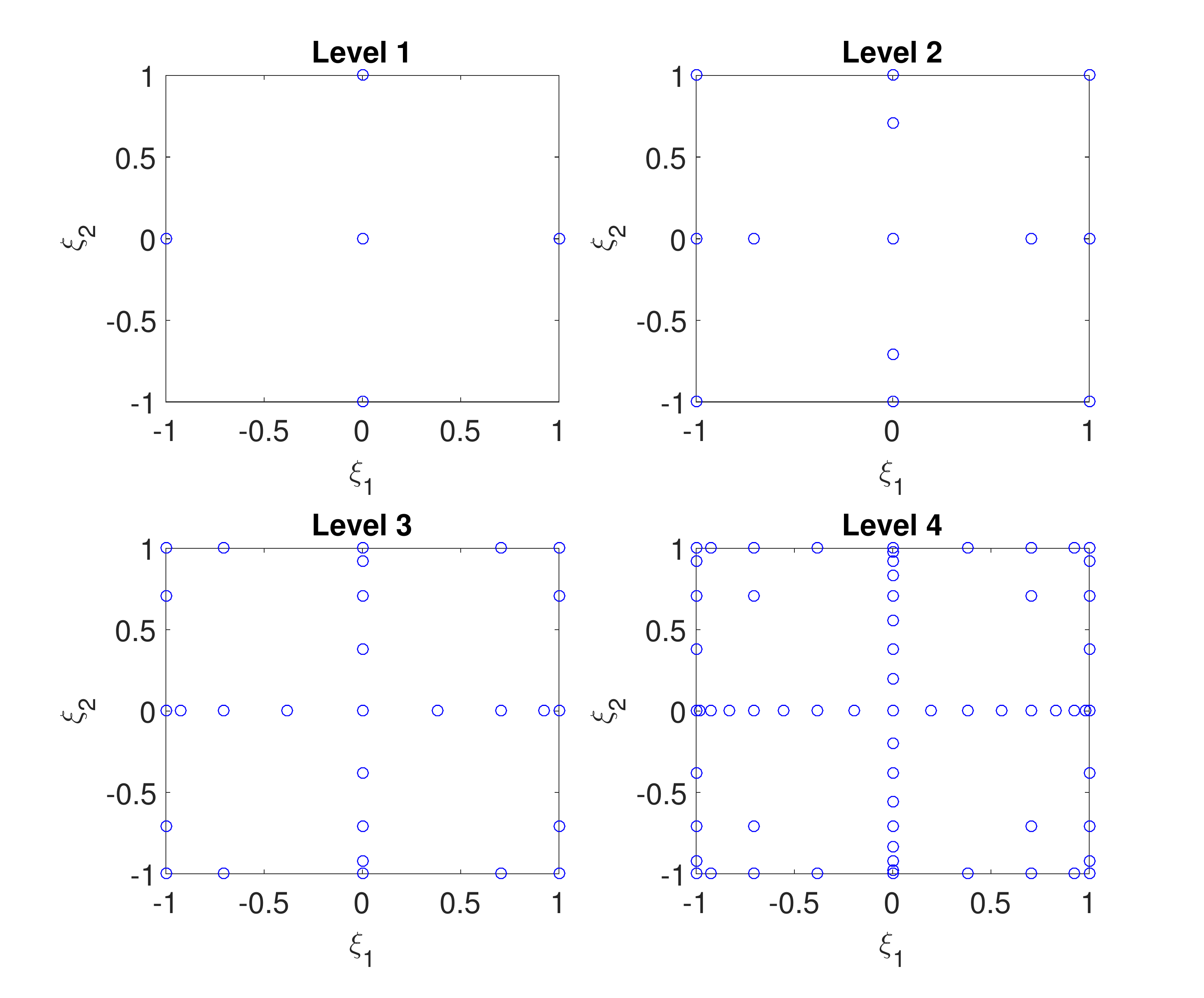}
		\caption{Clenshaw-Curtis Smolyak's nested Sparse Grids for different levels of accuracy.} \label{fig: SG}
	\end{center} 	
\end{figure}

The choice of collocation points is an important matter. If tensor grids are chosen, the computational cost would be very high, especially when $N_\xi$ is not low. A very efficient alternative is the use of sparse grids. In the present paper, the collocation points of the sparse grids have been determined according to the Clenshaw-Curtis (C-C) quadrature nested rule \cite{Clenshaw-Curtis} and Smolyak construction \cite{8}. With sparse grids, several levels of accuracy were tried in the problem, in order to test the convergence. In Fig. \ref{fig: SG} the evolution of the sparse grid collocation points is shown up to the fourth level for $ N_\xi=2 $. \\ 

\subsection{Sources of Uncertainty} \label{sec:sources}

The sources of uncertainty in the present work are based on a mixture of outcomes of literature review and information extracted from experimentalists through critical analysis and discussions on the mechanical characteristics of a similar rotating pipe experimental facility at University of M\'alaga. The idea of this research is to conduct the propagation of uncertainties in the computational problem, with the view to improve the reliability of deterministic simulations. 
To define our sources of uncertainty, we referred to a set of relevant literature. \cite{Congedo} is a very related paper and motivation for this work, where a $ 2.5\% $ of variance in uniform distribution in the swirl number and in the inlet velocity was applied, that means a $2.5\% $ of variance in $Q$. In \cite{turbine} a $ \pm3\% $ of variation as uniform inlet velocity was applied. Also, in \cite{Pitot} velocity was measure within a $ \pm 1\% $ with the described measurement techniques. As a consequence of the mentioned reviews, a conservative value of uncertainty for $Q$ has been chosen as a $5\% $ of variance. In addition, in the Fluid Mechanics laboratory at University of M\'alaga, the mechanical tolerance of the pipe angular velocity measurement was found to be $0.5\% $. We have set this uncertainty for the angular velocity of the simulated pipe.

Let $\bar{Q}$ and $\bar{\Omega}$ denote the deterministic mean values of the two uncertain parameters $Q$ and $\Omega$. Within this framework, and being conservative with respect to the literature results aforementioned, the source of uncertainties have been determined as the uniform distributions $Q \sim$ $Unif$$(0.95\bar{Q},\: 1.05\bar{Q}\:)$ and $ \Omega \sim$ $Unif$$(0.995\bar{\Omega}, 1.005\bar{\Omega})$. The main motivation in modelling the inputs by this type of distribution is conservativeness. The same probability is assigned to every value in the ranges of the variables, so it is ensured that the worst performance scenario is analysed. This approach has been used in the literature when insufficient information about experimental uncertainty is available, as e.g. in \cite{Congedo, witteveen2009comparison}. In \cite{witteveen2009comparison} a study of the impact of both Gaussian and uniform distributions was developed in CFD simulations of the flow over an airfoil. In our work, in order to observe the response to other distributions, truncated Gaussian distributions were also input in Section \ref{sec:UQ_impinging}, besides the uniform distributions.   

The advantage of dealing with $Q$ as stochastic input is that several uncertainties such as those related to measurement tools tolerance, pipe diameter, loss of pressure or density variations can be accounted in only one parameter. This also supports the idea of being conservative. If one wants to study each parameter separately, the dimension of the stochastic space would be unnecessarily increased and correlation-based sampling techniques must be used. That is, both the cost and difficulty would increase. In this work, there is no correlation amongst the chosen random variables, and this is important because the SCM applied in this work assumes independence amongst variables.  

\section{Coupling the Two-Step CFD Simulations} \label{sec: coupling}

Since the turbulence closure used to model the pipe flow and the one used for the impinging heat transfer problem are both different, a single CFD simulation of the whole system cannot be developed. Moreover, the computational cost of the problem would be unnecessarily increased if the computed duct is long enough to ensure a fully-developed flow, resulting to an unnecessarily expensive domain to simulate. 
 
To overcome this problem, two-step CFD simulations are coupled in the following way: Firstly, the CFD simulation of the swirling flow confined in a rotating pipe solved with the RSM turbulent model are completed (\textit{Simulation 1}). Secondly, both the velocity and some turbulent dimensionless profiles at the exit of the pipe are input as inlet boundary conditions for the heat transfer simulation (\textit{Simulation 2}). In particular, the turbulent parameters are defined by 
\begin{eqnarray} \label{eq:k}
k = \frac{2}{3} (U I)^{2}, 
\end{eqnarray}
\begin{eqnarray} 
\omega= \rho \frac{k}{\mu} \left(\frac{\mu_t}{\mu}\right)^{-1}=\rho \frac{k}{\mu} \beta^{-1},   \label{eq:omega}
\end{eqnarray} \\
where $k$ is the turbulent kinetic energy, $I$ is the turbulence intensity, $U$ is the time-average velocity of the flow, $\rho$ is the density of the fluid, $\omega$ is the specific turbulent dissipation rate and ${\mu_t}/{\mu}\equiv\beta$ is defined as the turbulent viscosity ratio, represented by $\beta$ throughout this paper. The turbulent kinetic energy is available from the RSM simulations, and the turbulent dissipation rate can be evaluated by Eq. (\ref{eq:omega}) prior to running \textit{Simulation 2}.

\begin{figure}[h!]
	\begin{center}
		\includegraphics[width=12cm]{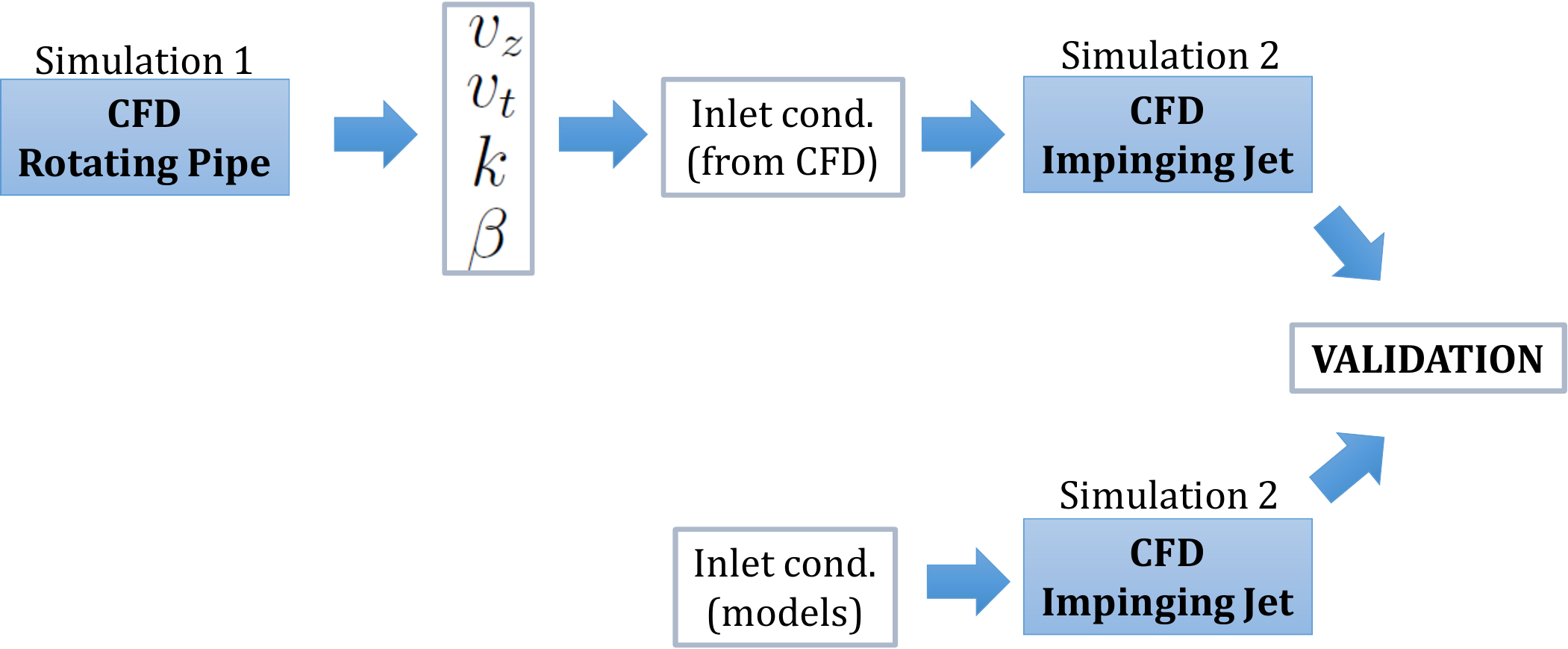}
		\caption{Two alternatives for conducting \textit{Simulation 2}: using the CFD profiles at the exit of the nozzle or to impose mathematical models.} \label{fig: two_paths}
	\end{center} 	
\end{figure}

In order to couple the two simulations for the modelling, two options have been tested in this work for \textit{Simulation 2}, as shown in Fig. \ref{fig: two_paths}. The first one is to impose the simulated profiles at the exit of the rotating pipe in \textit{Simulation 1} as inlet conditions onto \textit{Simulation 2}, for each required deterministic simulation for uncertainty quantification. The second option is to develop non-linear regression models for the fully-developed swirling turbulent flow profiles emerging from the pipe, avoiding to compute \textit{Simulation 1}.  The differences between these approaches will be analysed in this work to test whether these provide equivalent results in the heat transfer by impinging jets. If swirling jet flow profiles with/without uncertainty estimates are required for other applications (for instance, for a plate with variations on its surface \cite{5}), the use of these empirical models would avoid to simulate the rotating pipe flow. 

%%%%%%%%%%%%%%%%%%%%%%%%%%%%%%%%%%
%\section{Proposed mathematical Models for the dimensionless profiles from the rotating pipe} \label{sec: models}

\section{Modelling of Dimensionless Profiles From the Rotating Pipe Flow of Simulation 1} \label{sec: models4}

To quantify the uncertainty of the output of \textit{Simulation 2}, several profiles should be systematically generated from \textit{Simulation 1} and then be input as boundary condition into \textit{Simulation 2}. In Section \ref{sec:UQ_pipe}, SCM is applied $r/R$ point by $r/R$ point to the output profiles from \textit{Simulation 1}. Then, in Section \ref{sec:UQ_impinging}, the uncertainty in the Nusselt number is quantified. Despite the fact that SCM provides a $Q-\Omega$ dependent response surface for the profiles, the dependence of these profiles on the spatial coordinate $r/R$  makes inefficient to code hundreds of surrogates in the UDF, one for each $r/R$ point. 

Another option to compute uncertainty could be to include $r/R$ as an additional parameter of the stochastic space. This would be even more cumbersome, because the number of collocation points (CFD simulations) would be dramatically increased to deal with this new parameter of high non-linearity in the profiles. The suggested approach by means of non-linear regression models overcomes both problems, providing easy functions to code in any CFD software. The differences between the direct SCM approach and the use of the models is investigated in this work.
 
This type of surrogate model has been studied in \cite{ortega2011experimental} and successfully implemented in \cite{5,ortega2011numerical,ortega2012cfd}. The use of such modelling is in fact useful in industrial and academic applications, and one can find this, e.g. in stability analysis of jets flows as in \cite{piot2006investigation, airiau2015non}, where researchers used empirical functions of jet flows from \cite{crow1971orderly, yen1998application,balakumar1998prediction,tam1984sound} to test their approaches.

In the investigation of the non-linear regression models, a Reynolds number of $Re=23000$ is considered. Despite the modelling may seem focused on a particular application, this Reynolds number is one of the most studied test cases in the literature with/without swirl, e.g. in \cite{Lee, behnia1998prediction, yan1998heat, ortega2017using, uddin2013simulations, cooper1993impinging, Baughn, uddin2008turbulence, Granados2019I, wen2003impingement, shum2014large}. Thus, the proposed models for the swirling flow under uncertainty can be directly input as boundary condition to other investigations with similar set-up such as systems with arrays of impinging jets \cite{chang2006heat, geers2004experimental}, swirling jets impinging on a plate with bumps/dimples \cite{5}, or swirling turbulent flows in an axisymmetric sudden expansion \cite{guo2001simulation,Congedo}, amongst others. The computational efforts to perform UQ/CFD on these new applications would be only restricted to a single simulation. It is also relevant to mention that an analytical solution based on physics ground was not considered and this is the main limitation of this approach, as the flow is turbulent and this makes a pure mathematical analysis not feasible.\\
\color{black}

In the present study, four dimensionless models are given: axial velocity ($ v_{z} $), azimuthal velocity ($ v_{t} $), turbulent kinetic energy ($ k $) and turbulent viscosity ratio ($ \beta $) profiles. These models approximate the response of the CFD RANS profiles for the different $Q$ and $\Omega$, although initially introduced in this section for the deterministic base case ($Re=23000$ and $S=1$). In the literature can find often that inlet boundary conditions for turbulence quantities are modelled as a percentage (turbulence intensity), as in \cite{Congedo,turbine,Plat_MSc_2008}. Therefore, to provide specific profiles for both the turbulent kinetic energy and viscosity ratio instead of broad percentages, intends a more accurate insight.

These models are composed of several functions chosen according to the features of the profiles and trained against data from \textit{Simulation 1}. Interpolation/splines fit was conseidered, however, a new model should be built for each CFD simulation profile, not being a good alternative to model a link between the profiles and $Q$ and $\Omega$. For this reason, a non-linear parametric regression approach for the profiles plus a polynomial regression for the coefficients is preferred, as this approach makes possible to build profiles dependent on both the physical parameters $Q$ and $\Omega$ under uncertainty. These coefficients are found by a MATLAB custom code which calls Curve Fitting Toolbox functions that use a Non-Linear Square method with Trust-Region algorithm \cite{TrustRegion}. The introduced piecewise-like process to fit these models can be reproduced by other researchers in CFD/experimental profile modelling purposes. 

\begin{figure}[htb!]
	\begin{center}
		\includegraphics[width=16cm]{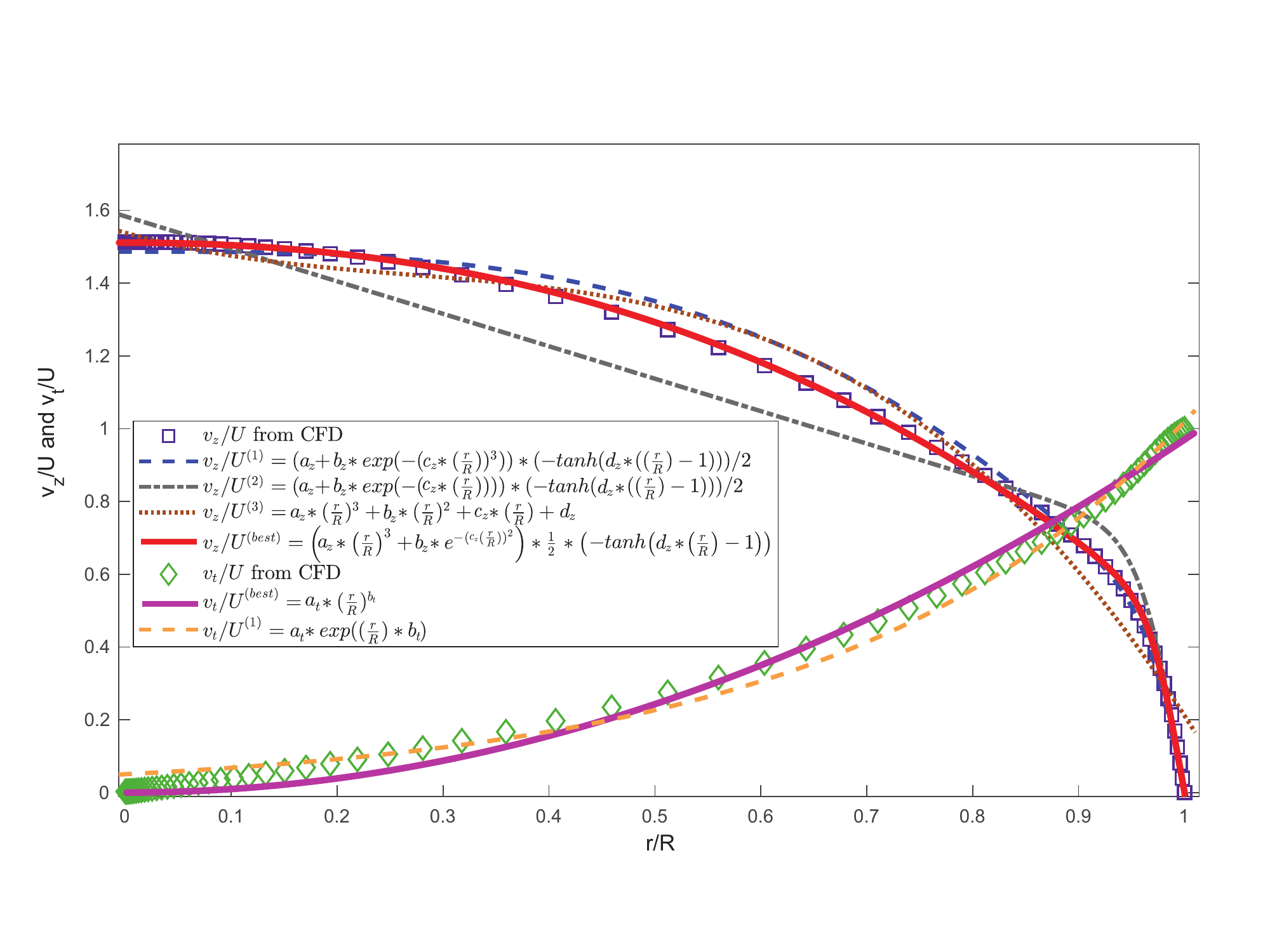}
		\caption{Axial and azimuthal velocity profiles from CFD and the non-linear regression models for $Re=23000$ and $S=1$.} 
		\label{fig: vz_vt_tests}
	\end{center} 	
\end{figure}

Several parametric models were tested for the axial and azimuthal radial velocities as shown in Fig. \ref{fig: vz_vt_tests} \footnote{In this manuscript the superscript $(i)$ denotes the i-\textit{th} fitted profile, and $(best)$ stands for the best fitting function amongst the tested ones.}. For the axial velocity radial profile, it can be observed that the profile shape is non-linear with different piecewise curvatures. It is specially relevant for the proper modelling of the boundary layer, since this plays an important role in the further development of the flow. This led to select a parametric function with a hyperbolic tangent for $v_z/U^{(1)}, v_z/U^{(2)}, v_z/U^{(best)}$. This is consistent with the past work of \cite{manneville2016transition, ortega2011experimental} that reports boundary layer of flows under rotation having a hyperbolic tangent shape. In jet flows this feature has also been observed since a hyperbolic tangent profile takes place when two flows are in contact with different but uniform axial velocities \cite{michalke1964inviscid}. As the flow is turbulent and with swirl, the axial velocity radial profile differs from the classic parabolic profile of a Hagen-Poiseuille flow. Nevertheless, this modelling was also tried with a cubic polynomial in $v_z/U^{(3)}$, which requires only three constants to estimate, but had a poor performance.
Amongst $v_z/U^{(1)}$ and $v_z/U^{(2)}$ there are some clear differences. A cubic power factor in $v_z/U^{(1)}$ is introduced to deal with the non linearity, which is not introduced in $v_z/U^{(2)}$, undergoing a bad fit. The combination of the power function featured in $v_z/U^{(1)}$ with the flexibility of the third order polynomial $v_z/U^{(3)}$ led to the selected model in Eq. (\ref{eq:vzmodel}), the $v_z/U^{(best)}$, which performs the best fit among the tested ones. 

The procedure to model the azimuthal velocity profile was simpler. It was already shown in \cite{Imao} that this radial profile is nearly parabolic. However, an exponential constant was also tried. In Fig. \ref{fig: vz_vt_tests}, it is shown that the power function fitted the best, and the predicted values of the constants are $a_t=0.9709$ and $b_t=2.0052$, thus $v_t/U \sim (r/R)^2$. 

% \begin{table}[h]
% 		\caption{Fitting coefficients of the discarded mathematical models for the axial and azimuthal dimensionless velocity profiles.}
% 	\begin{center}
% 
% 		\begin{tabular}{c | c c c c}
% 			\hline
% 			Model  & a & b & c & d \\
% 			\hline
% 			%format long values			
% 			$ v_z/U^{(1)} $ & 1051  & -1048   & -0.1276  & 32.230  \\
% 			$ v_z/U^{(2)} $ & -596.3 & 599.5 & 2.980E-03 &  24.710 \\
% 			$ v_z/U^{(3)} $ & -2.682 &  2.174 & -0.819 &  1.539 \\
% 			\hline
% 			$ v_t/U^{(1)} $ & 5.032E-02 &  3.009 & - &  - \\
% 			\hline		
% 		\end{tabular}
% 	\end{center}
% 
% 	\label{table: vz_vt_coeff}
% \end{table}

\begin{figure}[htb!]
	\begin{center}
		\includegraphics[width=16cm]{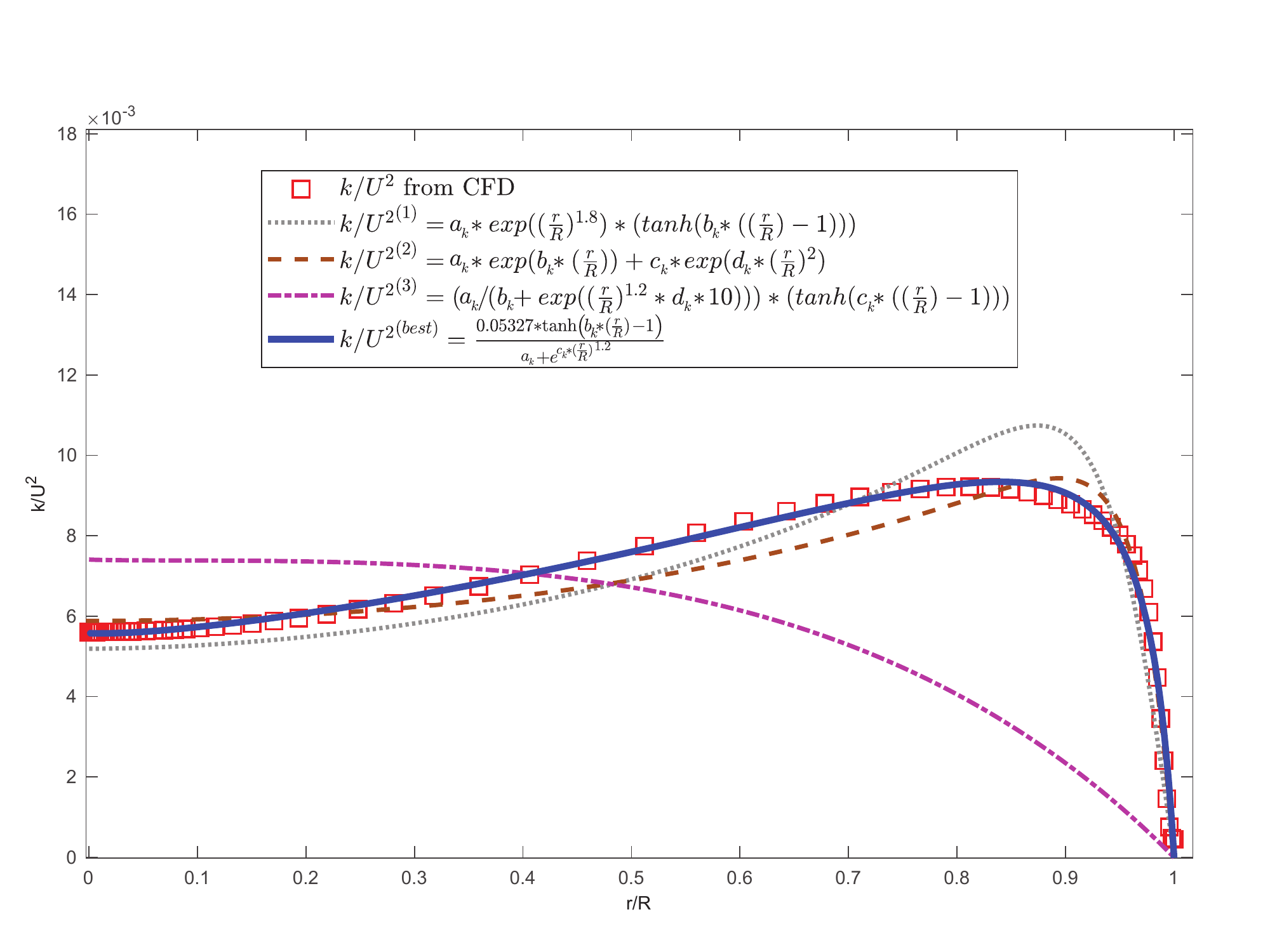}
		\caption{Dimensionless turbulent kinetic energy profile from CFD and the non-linear regression models for $Re=23000$ and $S=1$.} 
		\label{fig: k_tests}
	\end{center} 	
\end{figure}

The modelling of $k/U^2$ was the most challenging one. Four different regression models were tested. shown in Fig. \ref{fig: k_tests}. The best function for the region near to the wall was the hyperbolic tangent. $k/U^{2\:(2)}$ did not have a bad fit in that region without that term, however, its fit was not good for the rest of the profile. Thus, efforts were focused on improving the exponential term of $k/U^{2\: (1)}$, consistently to its shape. As one can see in  Fig. \ref{fig: k_tests}, $k/U^{2\: (3)}$ is far from the training points.

The coefficients were found to be difficult to regress using the MATLAB curve fitting algorithm. The reason may be that the algorithm struggles to fit the $k/U^2$ data with four coefficients, being required to reduce the dimension in the search space. An accurate initial guess may solve this issue, but even with the $k/U^{2\: (best)}$ coefficients as initial guess, the fitting search was not possible. In $k/U^{2\: (3)}$, it was observed that the $a$ coefficient tended to a value close to $ 0.05327 $. Despite the fact that the fitting is very accurate, to fix the  $ 0.05327 $ and  $ 1.2 $ coefficients in the equation is reducing the flexibility of the model to adapt to changes in $Q$ and $\Omega$. %The constants of the discarded models can be found in Table \ref{table: k_coeff}.\\
% \begin{table}[h]
% 		\caption{Fitting coefficients of the discarded mathematical models for the kinetic energy dimensionless profile. $k/U^{2\: (3)}$ does not appear because the search was not possible.}
% 	\begin{center}
% 
% 		\begin{tabular}{c | c c c c}
% 			\hline
% 			Model  & a & b & c & d \\
% 			\hline
% 			%format long values			
% 			$ k/U^{2\:(1)}$ & 5.190E-03  & -14.090   & -  & -  \\
% 			$ k/U^{2\:(2)} $ & -4.552E-17 & 33.140 & 5.883E-03 &  0.6341 \\
% 			\hline		
% 		\end{tabular}
% 	\end{center}
% 
% 	\label{table: k_coeff}
% \end{table}

\begin{figure}[htb!]
	\begin{center}
		\includegraphics[width=16cm]{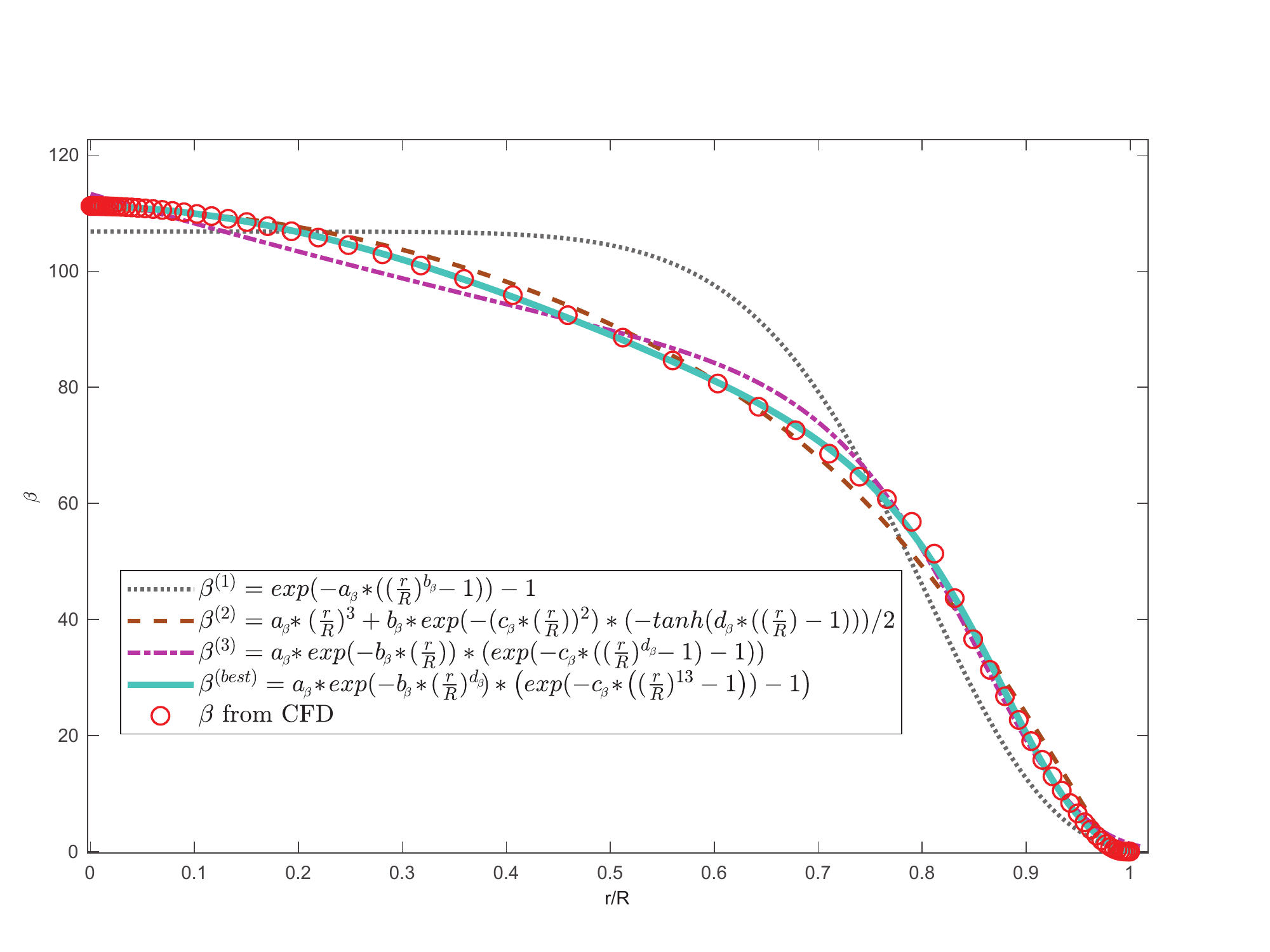}
		\caption{Viscosity ratio profile and the non-linear regression models for $Re=23000$ and $S=1$.} 
		\label{fig: w_tests}
	\end{center} 	
\end{figure}

The turbulent viscosity ratio, $\beta$, was also modelled with four regression models and their performance can be seen in Fig. \ref{fig: w_tests}. An exponential decay was tested, $ \beta^{(1)} $, which reproduces well the shape of the profile. The power function was suitable to model the curvatures. Since the shape in the intermediate region is similar to the shape observed in the axial velocity profile, the same power functions were tried with a hyperbolic tangent in $ \beta^{(2)} $. But unfortunately the fit close to the wall was fully linear, with no curvature approaching $\beta=0$. This led to testing the product of exponential functions in $ \beta^{(3)} $. Some difficulties in fitting the $r/R < 0.8$ region were noticed and the addition of a coefficient $d$ in the power of the left hand exponential solved that successfully. In order to reduce the dimension in the search space, the $d$ coefficient in $ \beta^{(3)} $ was fixed to $ 13 $, as it mainly controls the curvature approaching $\beta=0$. %The fitting coefficients of the discarded models are shown in Table \ref{table: w_coeff}.

%\begin{table}[h]
%		\caption{Fitting coefficients of the discarded mathematical models for the turbulent viscosity profile.}
%	\begin{center}
%
%		\begin{tabular}{c | c c c c}
%			\hline
%			Model  & $a$ & $b$ & $c$ & $d$ \\
%			\hline
%			%format long values			
%			$ \beta^{(1)}$ & 4.681  & 7.743   & -  & -  \\
%			$ \beta^{(2)} $ & -4.561 & 221.5 & 0.8234 &  4.513 \\
%			$ \beta^{(3)} $ & 5.785 & 0.458 & 3.975 &  10.170 \\
%			\hline		
%		\end{tabular}
%	\end{center}
%
%	\label{table: w_coeff}
%\end{table}

%The models of Eqs.  (\ref{eq:vzmodel})-(\ref{eq:tvrmodel}) are valid in the intervals $Q\in[0.95\bar{Q}, 1.05\bar{Q}]$ and $ \Omega \in[0.995\bar{\Omega}, 1.005\bar{\Omega}]$, which contain the prescribed uncertainties. 

After fitting the non-linear regression models to the CFD data profiles, the proposed ones are those labelled as best in Figs. \ref{fig: vz_vt_tests}-\ref{fig: w_tests}:
\begin{equation}
\frac {v_{z}} {U}=\left( a_z \: \left( \frac {r} {R}\right) ^{3} + b_z \: e^{-(c_z \: (\frac {r} {R}))^{2}}\right)  \: \frac {1} {2} \: \left( -\tanh\left( d_z \: \left( \frac {r} {R}\right) -1\right) \right) , %FITTING FUNCTION FOR AXIAL VELOCITY
\label{eq:vzmodel}
\end{equation}
\begin{equation}
\frac {v_{t}} {U}= a_t \: \left( \frac {r} {R}\right) ^{b_t}, %FITTING FUNCTION FOR AZIMUTAL VELOCITY
\label{eq:vtmodel}
\end{equation}
\begin{equation}
\frac {k} {U^2} = \frac {0.05327 \: \tanh\left( b_k \:(\frac {r} {R})-1\right) } {a_k+e^{c_k \:(\frac {r} {R})^{1.2}}}  ,
\label{eq:kmodel}
\end{equation}
\begin{equation}
\beta = a_\beta \, e^{-b_\beta \, (\frac {r} {R})^{d_\beta}} \,\left( e^{-c_\beta \, \left( (\frac {r} {R})^{13}-1\right) }-1\right) .
\label{eq:tvrmodel}
\end{equation} 

\begin{figure}[htb!]
	\begin{center}
		\includegraphics[width=10cm]{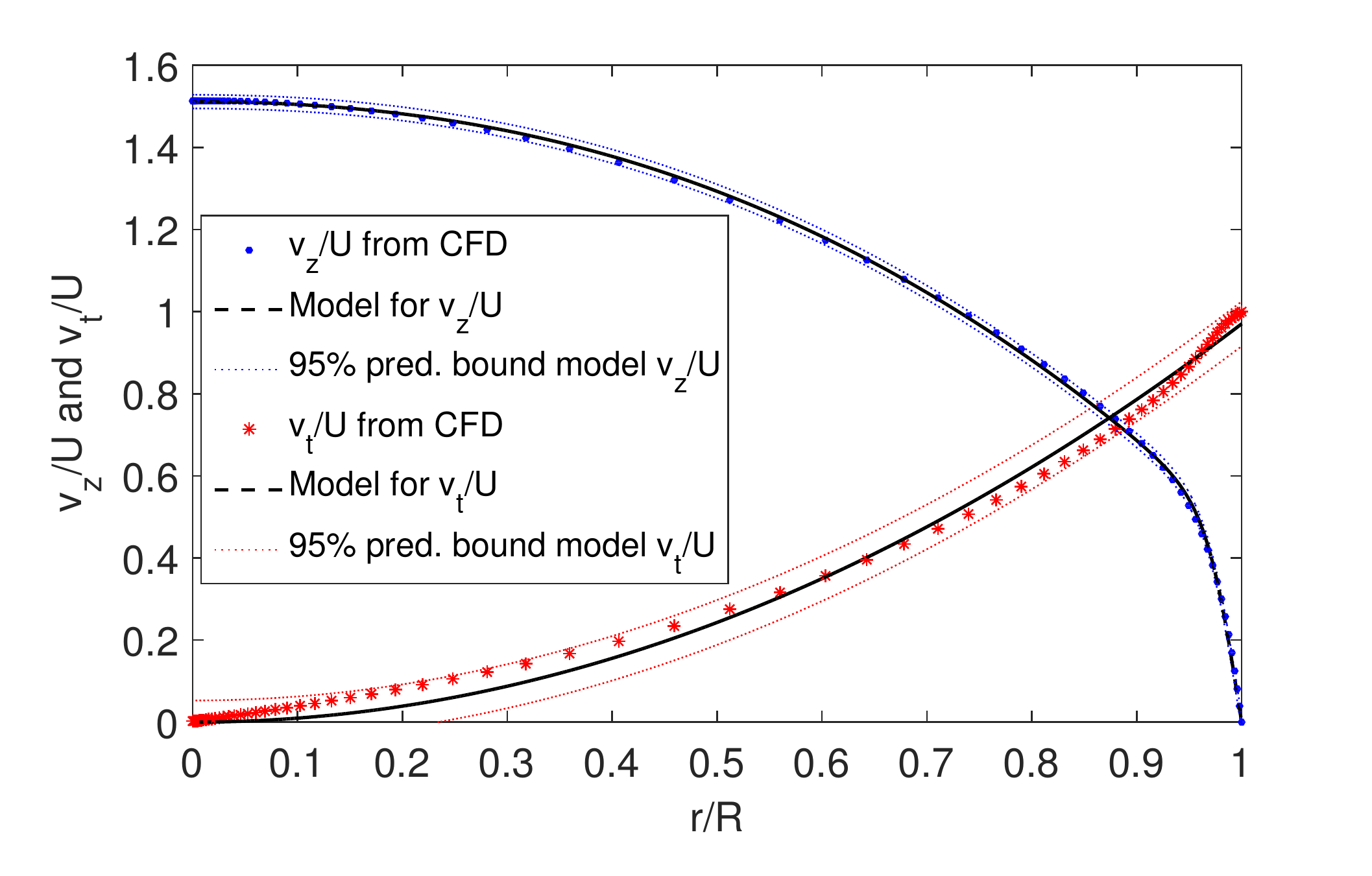}
		\caption{Axial and azimuthal velocity profiles from CFD and non-linear regression model fits with prediction bounds for $Re=23000$ and $S=1$.} \label{fig: model_SG_velocity}
	\end{center} 	
\end{figure}

\begin{figure}[htb!]
	\begin{center}
		\includegraphics[width=10cm]{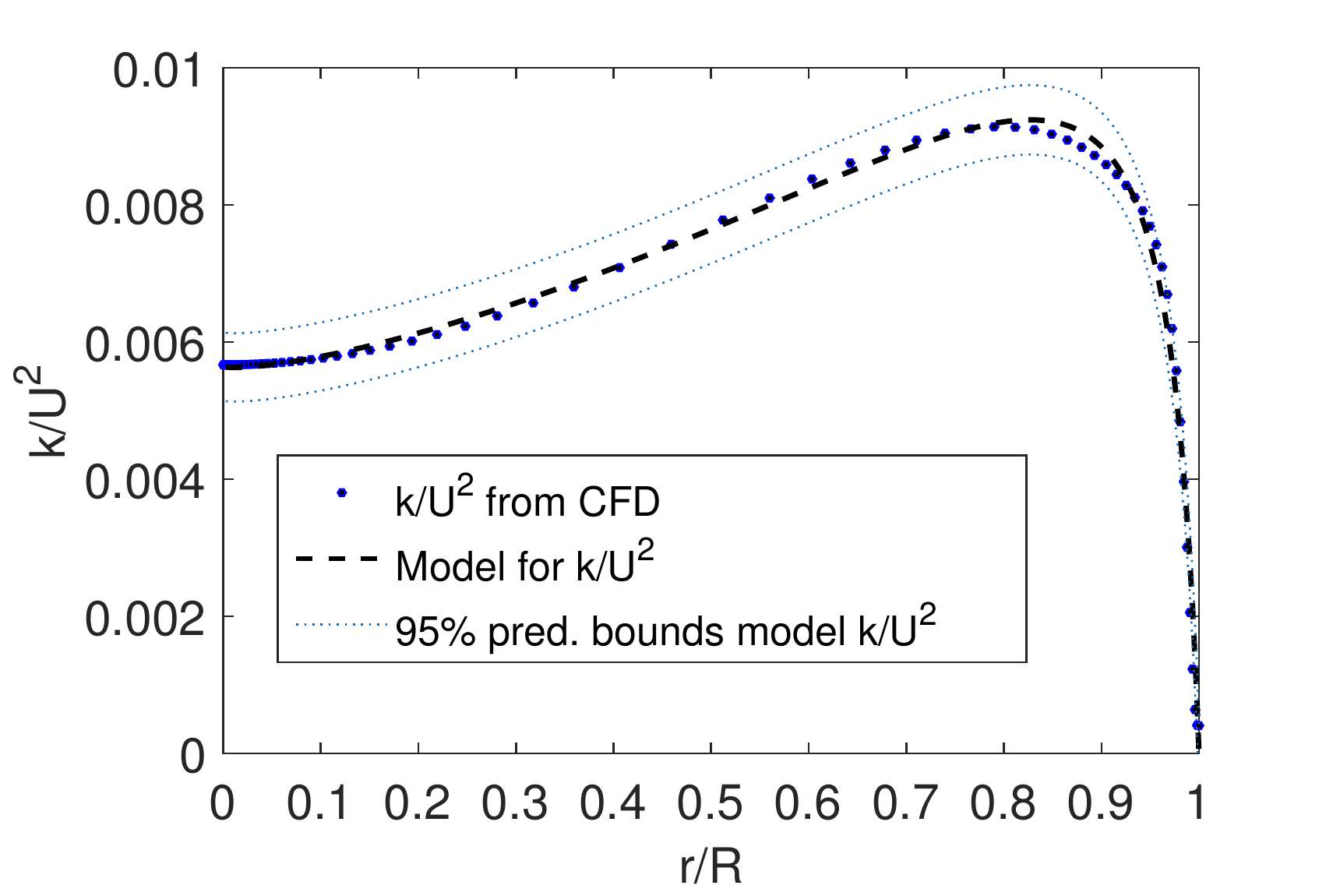}
		\caption{Dimensionless kinetic energy profile from CFD simulations and non-linear regression model fits with prediction bounds for $Re=23000$ and $S=1$.} 
		\label{fig: model_SG_k}
	\end{center} 	
\end{figure}

\begin{figure}[htb!]
	\begin{center}
		\includegraphics[width=11cm]{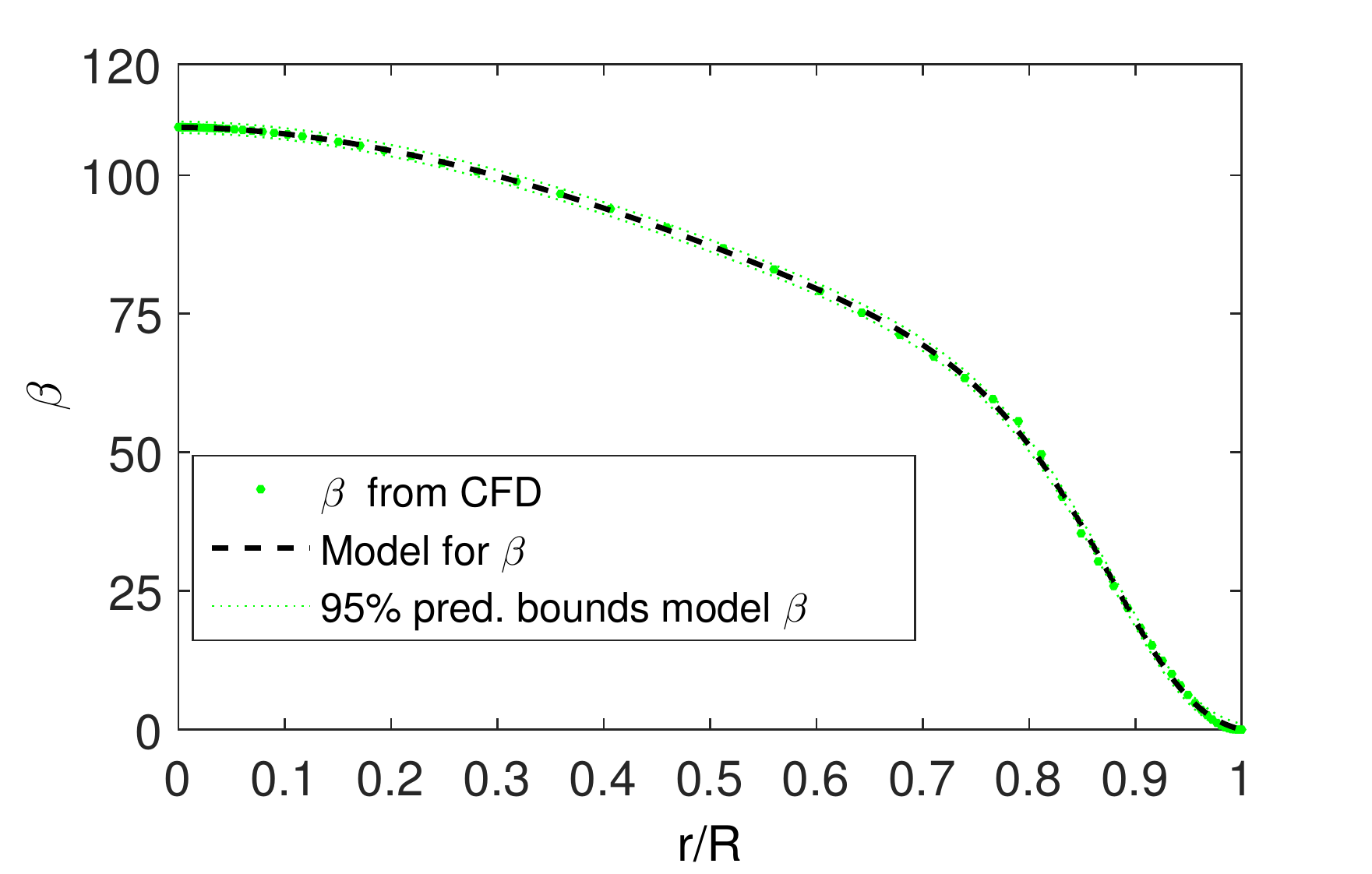}
		\caption{Turbulent viscosity ratio profile from CFD simulations and non-linear regression model fits with prediction bounds for $Re=23000$ and $S=1$.}  \label{fig: model_SG_tvr}
	\end{center} 	
\end{figure}

The coefficients of the mathematical models, addressed for sake of notation as $\gamma_i$, where $\gamma= a, b, c, d$ and $i=z, t, k, \beta$,  are shown in Table \ref{table: goodness_coeff_bench}. In Figs. \ref{fig: model_SG_velocity}-\ref{fig: model_SG_tvr}, the goodness of fit is plotted by the 95\% confidence prediction bound and goodness measures. The model with the widest confidence interval is $k/U^2$. The model for $v_t/U$ has a notable prediction bound as well, which likely results from the relatively simple formulation (only two unknown coefficients). A more complex parametric function would be required to capture the negative curvature shown by the CFD data over the range $0.95 < r/R < 1$.

It is important to ensure that the regression coefficients are enclosed within a short confidence bound, which denotes a good fit. The confidence bounds for the regression coefficients are calculated as 
\begin{equation}
c_b= \gamma_i \pm t_{0.95,v} SE_{\gamma_i},
\label{eq: coeff_conf_bound}
\end{equation}
%The critical value t that should be used depends on the number of degrees of freedom for error (the number data points minus number of parameters estimated, which is n-1 for this model) and the desired level of confidence. See also references:
%http://people.duke.edu/~rnau/mathreg.htm
%http://www.stat.tamu.edu/~hart/652/poly.pdf
%http://www.stat.tamu.edu/~hart/652/model.pdf
%MATLAB: $ c_b= \gamma_i \pm t_{0.95,\:v} \sqrt{S}= \gamma_i \pm t_{0.95,\:v} \sqrt{diag((X^TX^{-1})\: MSE^2)}=\gamma_i \pm t_{0.95,\:v} \sqrt{diag((X^T X^{-1})\: (\frac{SSE}{v})^2)}$ 
%http://uk.mathworks.com/help/curvefit/confidence-and-prediction-bounds.html
where $t_{0.95,v}$ refers to the two-tailed value of the inverse of t-Student distribution for $v$ degrees of freedom and a 95\% level of confidence, and $ SE_{\gamma_i} $ refers to the estimated standard error of $ \gamma_i $.  The results can be seen in Table \ref{table: goodness_coeff_bench} for the deterministic base case ($ Re=23000 $ and $ S=1 $). These values, as well as the goodness of fit metrics in Table \ref{table: goodness_bench}, were monitored for all the fittings of the collocation points when studying the propagation of uncertainty, showing almost identical performance. In Table \ref{table: goodness_bench} it can also be found the goodness indicators of the fit for the deterministic base case. As the coefficient of determination (defined in this manuscript as $\hat{R}^2$, to avoid confusion with $R$, the radius of the pipe) is not the best measure for the goodness of a fit in non-linear responses or with overfitting, the Sum of Squares due to Error (SSE), Adjusted-$\hat{R}^2$ and Root Mean Squared Error (RMSE) are also given. When increasing the number of terms in the models, $\hat{R}^2$ increases, despite of the model can be overfitted. It is hence important to consider the Adjusted-$\hat{R}^2$, whose value in case of overfitting starts to decrease. 
 
\begin{table}[h]
	\caption{Fitting coefficients of the dimensionless profiles for the deterministic base case ($Re=23000$ and $S=1$).}
	\begin{center}
		\begin{tabular}{c | c c }
			\hline
			Coefficient  & Value, $ \gamma_i $ &  95\% confidence bound, $c_b$ \\
			\hline
			%format long values
			
			$ a_z $ & -1.0474 & (-1.1231, -0.9718)\\		   
			$ b_z $ & 3.0230 & (3.0169, 3.0291) \\
			$ c_z $ & -0.6538 & (-0.6765, -0.6310)\\
			$ d_z $ & 34.1414  & (32.6899, 35.5929) \\
			\hline			
			$ a_t $ & 0.9709  & (0.9563, 0.9854) \\
			$ b_t $ & 2.0052  & (1.9048, 2.1056) \\ 
			\hline
			$ a_k $ & -0.1708  & (-0.1815, -0.1602) \\ 
			$ b_k $ & -0.0880  & (-0.0890, -0.0870)\\ 
			$ c_k $ & -1.6985  & (-1.7538, -1.6432) \\
			\hline
			$ a_\beta $ & 1.9314  & (1.7141, 2.1486) \\
			$ b_\beta $ & 0.7933  & (0.7712, 0.8154) \\
			$ c_\beta $ & 4.0472  & (3.9363, 4.1581) \\
			$ d_\beta $ & 1.8621  & (1.8048, 1.9194) \\			
			
			\hline
		\end{tabular}
	\end{center}

	\label{table: goodness_coeff_bench}
\end{table}

\begin{table}[ht!]
\caption{Goodness of fit for the non-linear regression model profiles for the deterministic base case ($Re=23000$ and $S=1$).}
	\begin{center}
		\begin{tabular}{c | c c c c c c}
			\hline
			Profile & SSE & $ \hat{R}^2 $ & Adjusted$-\hat{R}^2 $ & RMSE & $ m $ & $ v $ \\
			\hline
			%format long values
			
			$ v_z/U $ & 0.0044 & 0.9997	& 0.9997 & 0.0082 & 4 & 65\\
			$ v_t/U $ & 0.0536 & 0.9947	& 0.9946 & 0.0283 & 2 & 67\\
			$ k/U^2 $ & 3.9766e-06 & 0.9865 & 0.9861 & 2.4546e-04 & 3 & 66\\
			$ \beta $ & 16.6794 & 0.9999 & 0.9999 & 0.5066 & 4 & 65 \\
			\hline
		\end{tabular}
	\end{center}
	
	\label{table: goodness_bench}
\end{table}

%\begin{figure}[h!]
%	\begin{center}
%		\includegraphics[width=11cm]{bench_fit_vz_vt.png}
%		\caption{(\color{blue} $\bullet$ \color{black}) Axial velocity profile from CFD. (\color{red} $\ast$ \color{black}) Azimutal velocity profile from CFD. (- -) Mathematical models. } \label{fig: bench_fit_k_tvr}
%	\end{center}
%\end{figure}

%\begin{figure}[h!]
%	\begin{center}
%		\includegraphics[width=14cm]{bench_fit_k_tvr.png}
%		\caption{(\color{blue} $\ast$ \color{black}) Dimensionless $k$ profile from CFD. (\color{red} $\bullet$ \color{black}) Dimensionless $\beta$ from CFD. (- -) Mathematical models. } \label{fig: bench_fit_vz_vt}
%	\end{center}
%	
%\end{figure}

In the coefficient estimation process described above, the regression models do not depend on $Q$ and $\Omega$. Therefore, Eqs. (\ref{eq:vzmodel})-(\ref{eq:tvrmodel}) are written as
\begin{eqnarray}
(v_z/U, \: \: \: v_t/U, \: \: \: k/U^2, \: \: \: \beta)^t = \mathbf{F}(\gamma_i).
\label{eq:model1}
\end{eqnarray} 

To turn Eqs. (\ref{eq:vzmodel})-(\ref{eq:tvrmodel}) into $Q$ and $\Omega$ dependent functions, polynomial regression models can be built as
\begin{eqnarray}
(v_z/U, \: \: \: v_t/U, \: \: \: k/U^2, \: \: \: \beta)^t = \mathbf{F}(\gamma_i)=\mathbf{G}(Q,\Omega).
\label{eq:model2}
\end{eqnarray} 
As a consequence of this approach, the suggested models for the profiles are linked to the input random variables. These polynomial fits are shown in Appendix \ref{append_A}, with the corresponding equations, goodness of fit and plots. In addition to this, in Appendix \ref{append_B} the User Defined Function (UDF) for its implementation in FLUENT is given.

\begin{figure} [ht!]
	\begin{subfigmatrix}{3}
		\subfigure[]{{
				\includegraphics[width=0.50 \textwidth]{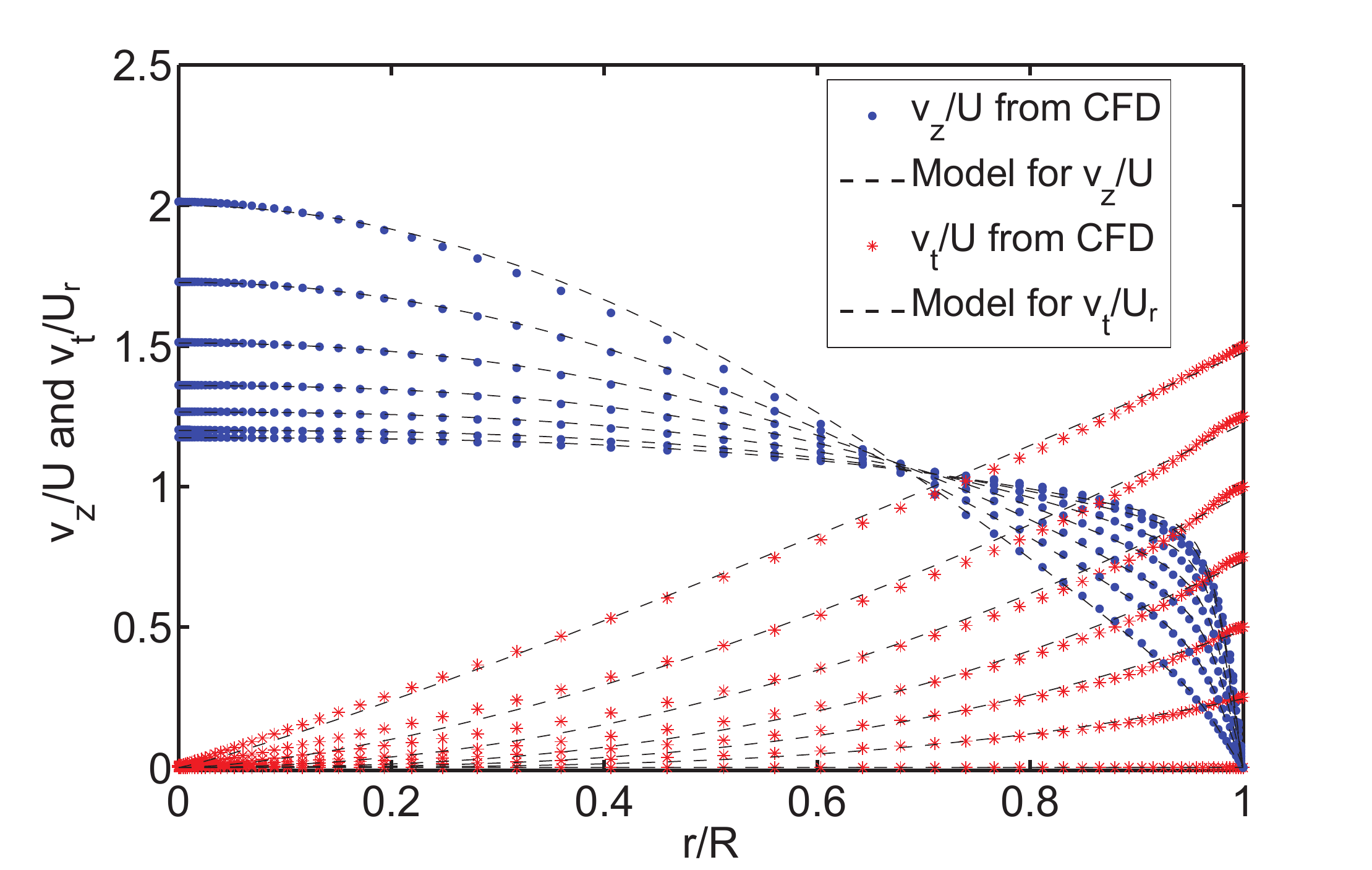}}}
				\put (-160,90) {S} 
				\put (-170,135) {\vector(0,-1){60}}
				\put (-80,40) {S} 
				\put (-70,20) {\vector(0,1){60}}				
		\subfigure[]{{
				\includegraphics[width=0.50\textwidth]{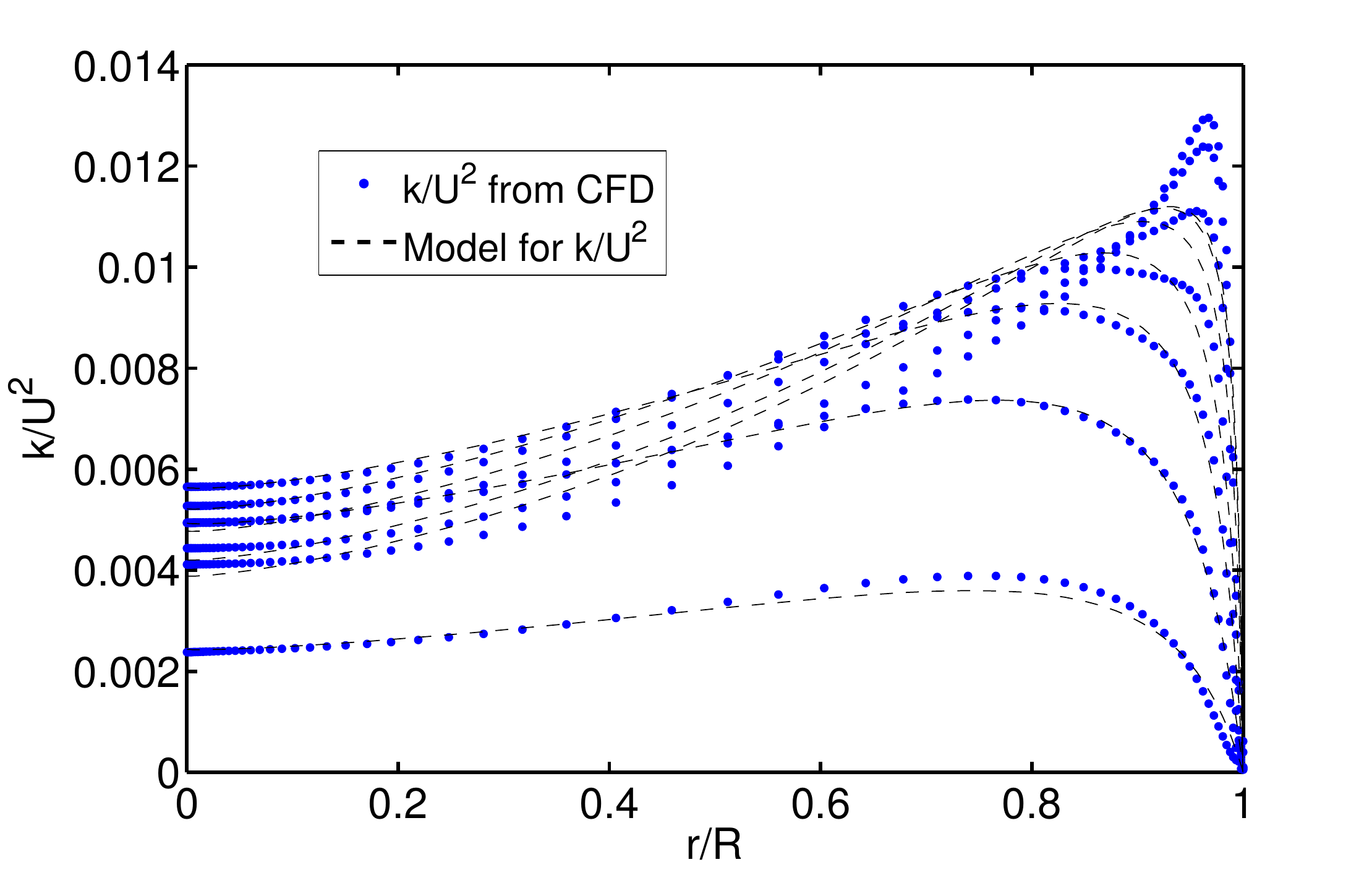}}}
				\put (-110,70) {S} 
				\put (-100,40) {\vector(0,1){60}}				
		\subfigure[]{{
				\includegraphics[width=0.5\textwidth]{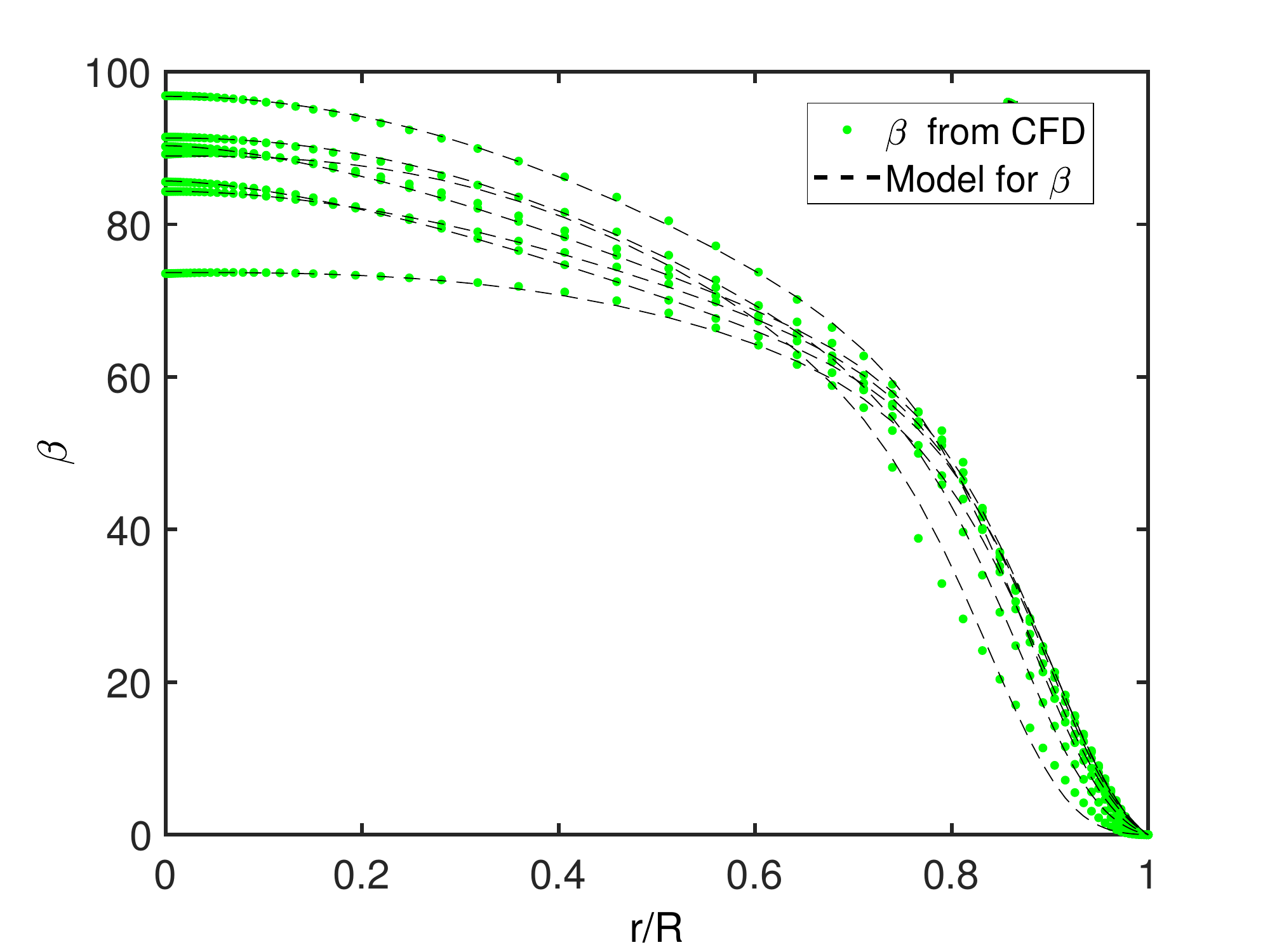}}}
				\put (-170,150) {S} 
				\put (-180,172) {\vector(0,-1){40}}								
	\end{subfigmatrix}
	\caption{Dimensionless profiles from CFD and their suggested models for (a) axial and azimuthal velocity, (b) turbulent kinetic energy and (c) turbulent viscosity ratio, for $S$ varied from 0 up to 1.5.} \label{fig:OPTmodels}
\end{figure}

\begin{figure}[h!]
	\begin{center}
		\includegraphics[width=12cm]{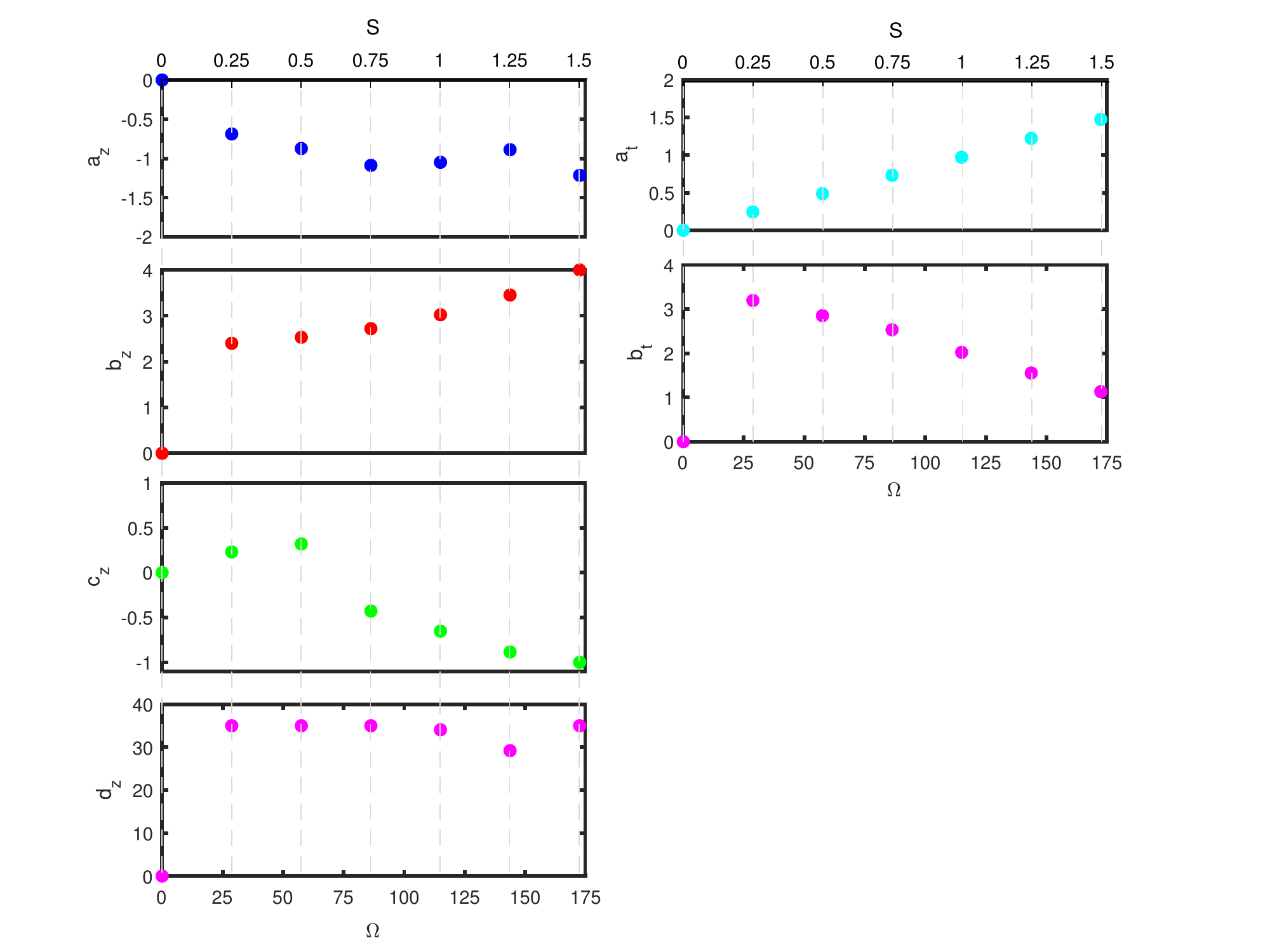}
		\caption{Coefficients for the dimensionless velocity profiles in Figure \ref{fig:OPTmodels}.} \label{fig:VelCoeffOPT}
	\end{center}
\end{figure}

\begin{figure}[h!]
	\begin{center}
		\includegraphics[width=12cm]{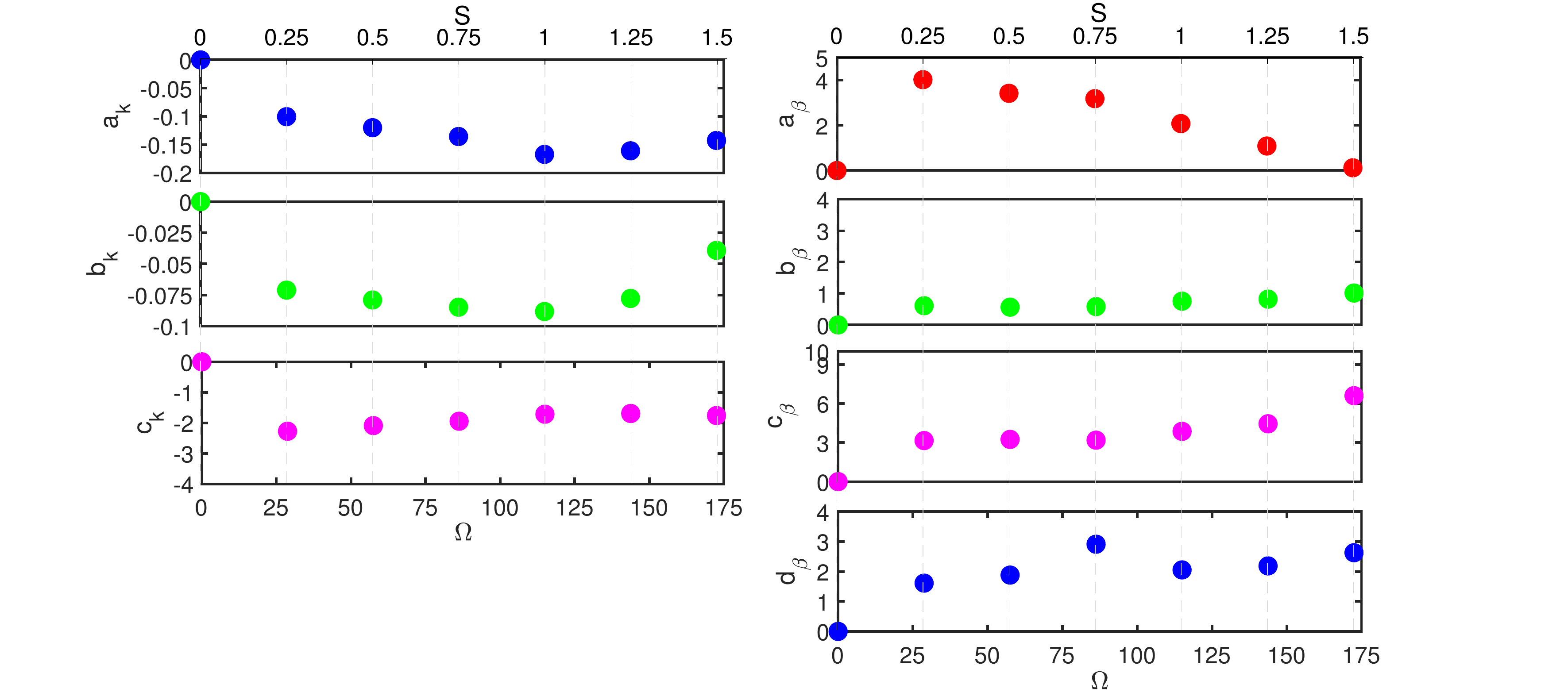}
		\caption{Coefficients for the dimensionless turbulent profiles in Figure \ref{fig:OPTmodels}.} \label{fig:TurbCoeffOPT}
	\end{center}
\end{figure}

The domain of Eq. (\ref{eq:model2}) for the uncertainty analysis is constrained to $Q\in[0.95\bar{Q}, 1.05\bar{Q}]$ and $ \Omega \in[0.995\bar{\Omega}, 1.005\bar{\Omega}]$, which is set by the domain of the uniform input uncertainties introduced in Section \ref{sec:sources}. The performance of the proposed models for larger values of $S$ (and therefore, $\Omega$), can be seen in Fig. \ref{fig:OPTmodels}(a)-(c). In these plots, it is shown a comparison between the CFD and modelled profiles with a very good match. In these figures it is observed that $S$ can be varied at least from $0$ to $1$, with $Q $ fixed at its deterministic base value, and these fitting results provide a promising fit \cite{ICIAM2015}. Amongst the studied profiles, the turbulent kinetic energy develops a near-wall peak that is not rendered by the current fit. Further numerical models can be investigated to suitably model this feature, which is out of scope in this work. The coefficients of these fittings are provided in Figs. \ref{fig:VelCoeffOPT} and \ref{fig:TurbCoeffOPT}.


\section{Uncertainty Quantification Results of the Two-Step CFD Simulation.} \label{sec: UQresults}

\subsection {Simulation 1. Uncertainty Quantification of the Fully-developed Turbulent Swirling Flow Generated by the Rotation of a Pipe} \label{sec:UQ_pipe}

Regarding the uncertainty quantification process, in Section \ref{subsec_SC} the Stochastic Collocation method has been introduced and the results of analysing \textit{Simulation 1} are shown in Table \ref{pipe_means} \& \ref{pipe_variances}. It can be observed that even using few points for the C-C Sparse Grid, the statistical moments are not changing remarkably.
This is because despite of the non-linear behaviour behind the equations that govern the problem, the outputs of interest have a linear response with respect to the ranges of the random inputs in the simulation of the pipe. A budget (Design of Experiment, DoE) of few collocation points was enough, as increasing the number of collocation points there is no change, as expected (see Table \ref{pipe_means} \& \ref{pipe_variances}). However, the budget of 65 collocation points was necessary to monitor the convergence of the stochastic variance of the Nusselt number in \textit{Simulation 2}, and therefore, 65 profiles for each velocity and turbulent parameter are required from \textit{Simulation 1} when the mathematical models coded in the UDF are not used as substitute.

\begin{table}[h!]
	\begin{center}
		\begin{tabular}{*{6}{c}}
			\hline
			Level & Points & $\lambda_{\: r/R=1}$ & $I (\%)_{\: r/R=0.5}$ & $(v_{z}/U)_{\: r/R=0.5}$  & $(v_{t}/U)_{\: r/R=0.5}$ \\
			\hline
			%format long values
			% 1 & 5  & 0.013734087758336 &  0.084578336801094 & 1.283503880028882 & 0.266772950806308 \\
			% 2 & 13 & 0.013733990493924 &  0.084577413746483 & 1.283505931237867 & 0.266769989809882 \\
			% 3 & 29 & 0.013734040422995 &  0.084577342870457 & 1.283505668666422 & 0.266769570250826 \\
			% 4 & 65 & 0.013734017038625 &  0.084577395694921 & 1.283504358180147 & 0.266769611248173 \\
			1 & 5  & 0.01373401 &  8.45783368 & 1.28350388 & 0.26677295 \\
			2 & 13 & 0.01373399 &  8.45774137 & 1.28350593 & 0.26676999 \\
			3 & 29 & 0.01373404 &  8.45773429 & 1.28350567 & 0.26676957 \\
			4 & 65 & 0.01373401 &  8.45773957 & 1.28350436 & 0.26676961 \\
			
			\hline
		\end{tabular}
	\end{center}
	\caption{Stochastic means of the friction factor ($ \lambda $), turbulent intensity ($I$) , dimensionless axial ($v_z/U$) and azimutal ($v_t/U$) velocity at $r/R=0.5$, at the exit of the rotating pipe.}
	\label{pipe_means}
\end{table}

\begin{table}[h!]
	\begin{center}
		\begin{tabular}{*{6}{c}}
			\hline
			Level & Points & $\lambda_{\: r/R=1}$ & $I (\%)_{\: r/R=0.5}$ & $(v_{z}/U)_{\: r/R=0.5}$  & $(v_{t}/U)_{\: r/R=0.5}$ \\
			\hline
			% 1 & 5  & 0.221520100479947e-06 &  0.726302890971878e-05 & 0.757052364697142e-04 & 0.229386667435286e-03 \\
			% 2 & 13 & 0.220594689902758e-06 &  0.725208289475976e-05 & 0.752566724313564e-04 & 0.226398090432886e-03 \\
			% 3 & 29 & 0.220666321606690e-06 &  0.725533821812219e-05 & 0.752458767339093e-04 & 0.226365409345536e-03 \\
			% 4 & 65 & 0.220675501157229e-06 &  0.725546593365913e-05 & 0.752613651886502e-04 & 0.226366003307499e-03 \\
			
			1 & 5  & 0.22152010e-06 &  0.72630289e-03 & 0.75705236e-04 & 0.22938667e-03 \\
			2 & 13 & 0.22059469e-06 &  0.72520829e-03 & 0.75256672e-04 & 0.22639809e-03 \\
			3 & 29 & 0.22066632e-06 &  0.72553382e-03 & 0.75245877e-04 & 0.22636540e-03 \\
			4 & 65 & 0.22067550e-06 &  0.72554659e-03 & 0.75261365e-04 & 0.22636600e-03 \\
			
			\hline
		\end{tabular}
	\end{center}
	\caption{Stochastic variances of the friction factor ($ \lambda $), turbulent intensity ($I$) , dimensionless axial ($v_z/U$) and azimutal ($v_t/U$) velocity at $r/R=0.5$, at the exit of the rotating pipe.}
	\label{pipe_variances}
\end{table}

\begin{figure}[h!]
	\begin{center}
		\includegraphics[width=14cm]{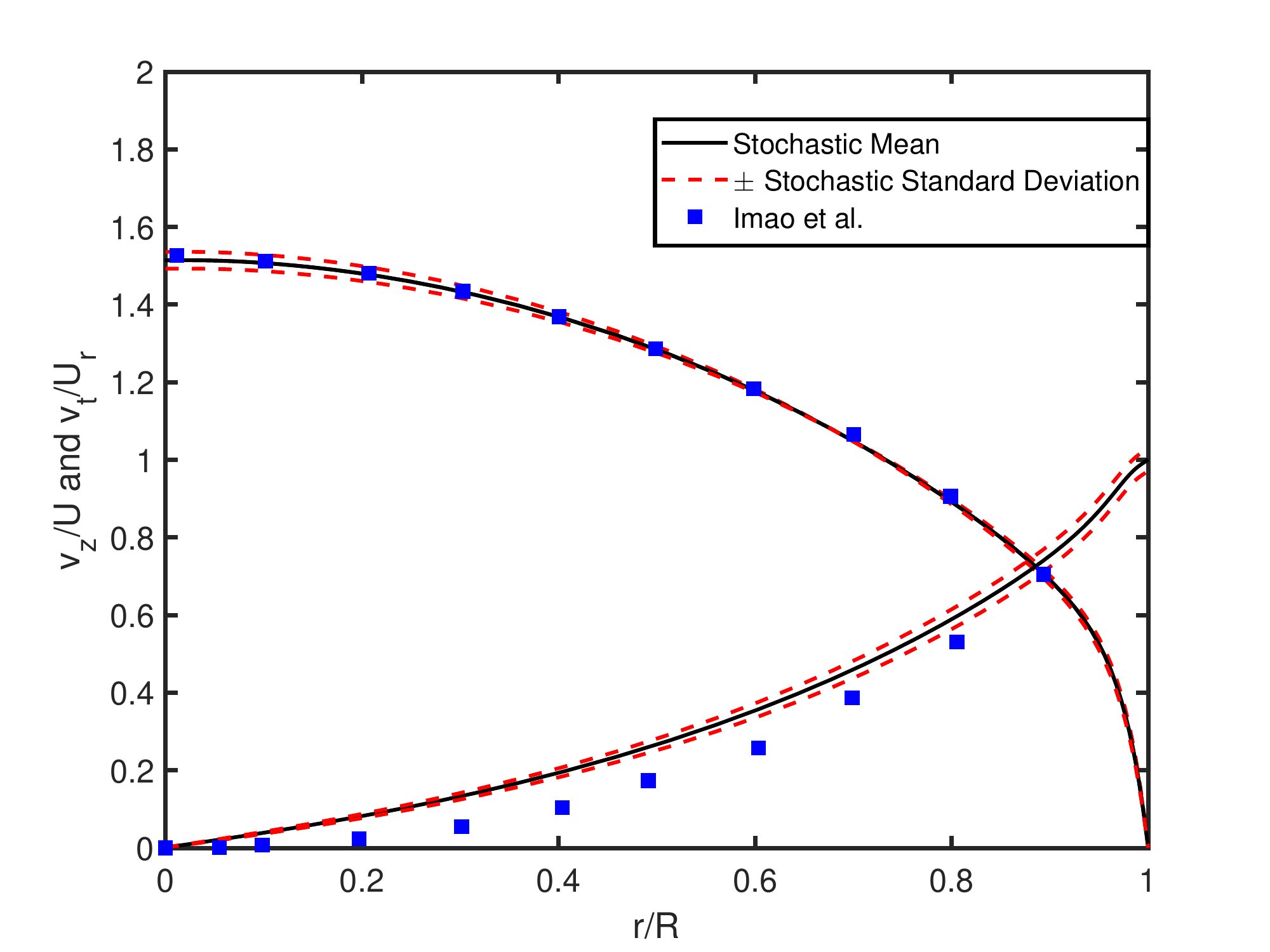}
		\caption{Radial distribution of the dimensionless axial and azimuthal velocity profiles at the exit of the pipe for the level 4 of the C-C Sparse Grid. Experimental data from \cite{Imao}.} \label{pipe_vel_prof}
	\end{center}
	
\end{figure}

\begin{figure}[h!]
	\begin{center}
		\includegraphics[width=14cm]{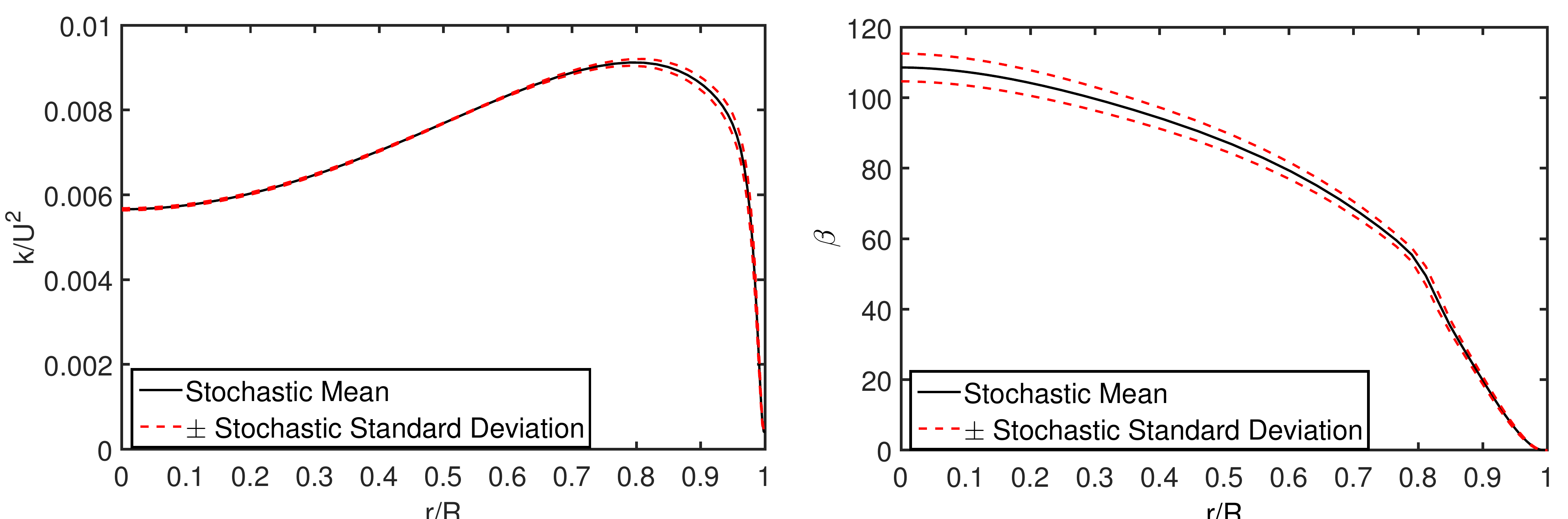}
		\caption{Radial distribution of the dimensionless turbulent kinetic energy and turbulent viscosity ratio profiles at the exit of the pipe for the level 4 of the C-C Sparse Grid.} \label{pipe_turb_prof}
	\end{center}
	
\end{figure}

The outputs of interest from \textit{Simulation 1} are the dimensionless velocity and turbulent profiles at the exit of the pipe, as the pipe becomes the nozzle for the impinging problem and these profiles will be used as inlet boundary conditions. The uncertainty analysis of these profiles can be seen in Fig. \ref{pipe_vel_prof} and \ref{pipe_turb_prof}, where the mean and standard deviation envelopes of those profiles are shown. Experimental data was only available for the velocity profiles, used to validate the CFD simulations in \cite{Granados2019I} and also included in Fig. \ref{pipe_vel_prof}. The plots show that for the over the selected $Q$ and $\Omega$ ranges, the variance in the profiles are relatively small, as evidenced by the modest width between the dashed lines. Fig. \ref{pipe_vel_prof} shows that the CFD predictions of axial velocity are in good agreement with the experimental data. Regarding the azimuthal velocity, the deterministic CFD simulation in \cite{Granados2019I} had a good match with the experimental data, but systematically over-predicted $v_t/U$. The standard deviation was not large, meaning that the simulation of the pipe under uncertainty is not sensitive to the modelled random inputs. 

As these output uncertainties will be propagated to the next simulation, these may affect to the further behaviour of the impinging jet. For this reason, the uncertainty analysis in \textit{Simulation 2} is undertaken.

%  \begin{figure}[h!]
%    \begin{center}
%      \includegraphics[width=9cm]{tvr_std.png}
%      \caption{Radial distribution of the Turbulent Viscosity Ratio, $\beta$, at the exit of the pipe. }
%    \end{center}
%    \label{pipe_tvr_prof}
%  \end{figure}

\subsection {Simulation 2. Uncertainty Quantification on the Impinging Swirling Jet for Heat Transfer: Boundary Conditions from first-step CFD simulations.} \label{sec:UQ_impinging}

In \textit{Simulation 2}, a heat transfer process from a heated flat plate to an impinging swirling jet takes place. For uncertainty quantification purposes, the same Stochastic Collocation method with the Clenshaw-Curtis nested rule with sparse grid quadrature points is used as in \textit{Simulation 1}. In order to be consistent with the deterministic simulations from the rotating pipe for different values of $Q$ and $\Omega$, the same collocation points simulated in that previous step are systematically used.

\begin{figure}[h!]
	\begin{center}
		\includegraphics[width=12cm]{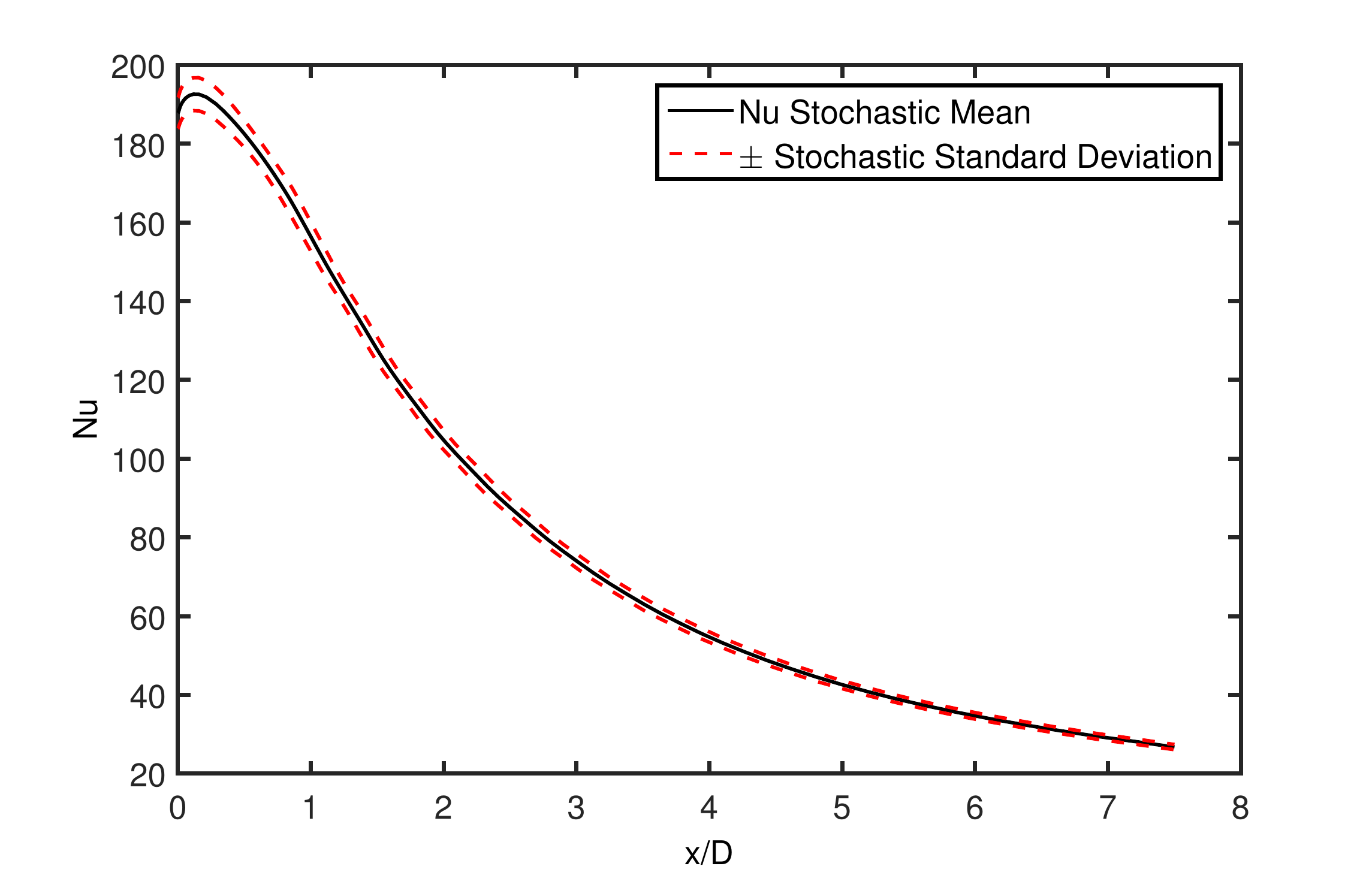}
		\caption{Radial distribution of the stochastic mean of the Nusselt number over the plate $\pm$ its stochastic standard deviation for level 4 of the C-C sparse grid.} \label{Nusselt_std}
	\end{center}
	
\end{figure}

In this framework, uncertainties are propagated from the input of \textit{Simulation 1} to the output of \textit{Simulation 2}. The convective heat transfer is quantified by the Nusselt number $Nu$ both at the stagnation point and its surface averaged value on the plate, denoted respectively by $Nu_0$ and $Nu_{avg}$, with the latter defined by
\begin{eqnarray} \label{eq:Nua}
Nu_{avg} = \frac {1} {\pi R_{int}^2} \int_{0}^{R_{int}} Nu(s) \: 2 \pi s \: ds,
\end{eqnarray}
% \begin{eqnarray} \label{eq:RSM}
% Local \: Time \: Derivate + C_{ij} = D_{T,ij} + D_{L,ij} + P_{ij} + G_{ij} + \phi_{ij} - \epsilon_{ij} + F_{ij} + S_{user},
% \end{eqnarray}
where $R_{int}$ is  the radius of the impinged flat plate and its value is set to $7.5D$. The stochastic mean and variance, for the $Nu$ are presented in Tables \ref{ht_means} \& \ref{ht_variances}. The radial distribution of the Nusselt number along the plate is shown in Fig. \ref{Nusselt_std}, where it is plotted with its standard deviation as uncertainty measure. It can be observed that the most sensitive part to the input uncertainties is the stagnation area, whereas for $x/D$ values far from the stagnation, input uncertainties are irrelevant. 
  
In the same manner as in the UQ study in \textit{Simulation 1}, it can be noted that even by using fewer points for the C-C Sparse Grid, statistical convergence in $Nu_0$ and $Nu_{avg}$ is generally achieved. Only the stochastic variance needed 29 collocation points to achieve convergence. In addition, since the results are integrated quantities, it is frequent to exhibit some numerical variability due to precision error that do not necessarily mean any oscillation in the convergence rate of the UQ technique.
\begin{table}[h!]
	\begin{center}
		\begin{tabular}{*{4}{c}}
			\hline
			Level & Points &$Nu_{0}$ & $Nu_{avg}$  \\
			\hline
			%format long values
			%1 & 5  & 187.8183333333346 &  51.454739515241798 \\
			%2 & 13 & 187.7914166666679 &  51.454138253348077 \\
			%3 & 29 & 187.8031984546113 &  51.454574875198190 \\
			%4 & 65 & 187.7993255381768 &  51.454970832858578 \\   
			1 & 5  & 187.8183333 &  51.45473951 \\
			2 & 13 & 187.7914166 &  51.45413825 \\
			3 & 29 & 187.8031984 &  51.45457487 \\
			4 & 65 & 187.7993255 &  51.45497083 \\        
			\hline
		\end{tabular}
	\end{center}
	\caption{Stochastic means of the Nusselt number at the stagnation point and of the surface average over the flat plate.}
	\label{ht_means}
\end{table}

\begin{table}[h!]
	\begin{center}
		\begin{tabular}{*{4}{c}}
			\hline
			Level & Points &  $Nu_{0}$ & $Nu_{avg}$  \\
			\hline
			%1 & 5  & 15.243481888661336 &  1.509455409450311 \\
			%2 & 13 & 15.568816376158793 &  1.508783017682617 \\
			%3 & 29 & 15.472847192140762 &  1.509090357563764 \\
			%4 & 65 & 15.476214312162483 &  1.508224731997416 \\
			
			1 & 5  & 15.24348189 &  1.50945541 \\
			2 & 13 & 15.56881637 &  1.50878302 \\
			3 & 29 & 15.47284719 &  1.50909036 \\
			4 & 65 & 15.47621431 &  1.50822473 \\                 
			\hline
		\end{tabular}
	\end{center}
	\caption{Stochastic variance of the Nusselt number at the stagnation point and of the surface average over the flat plate.} \label{ht_variances}
	
\end{table}

\begin{figure}[h!]
	\begin{center}
		\includegraphics[width=10cm]{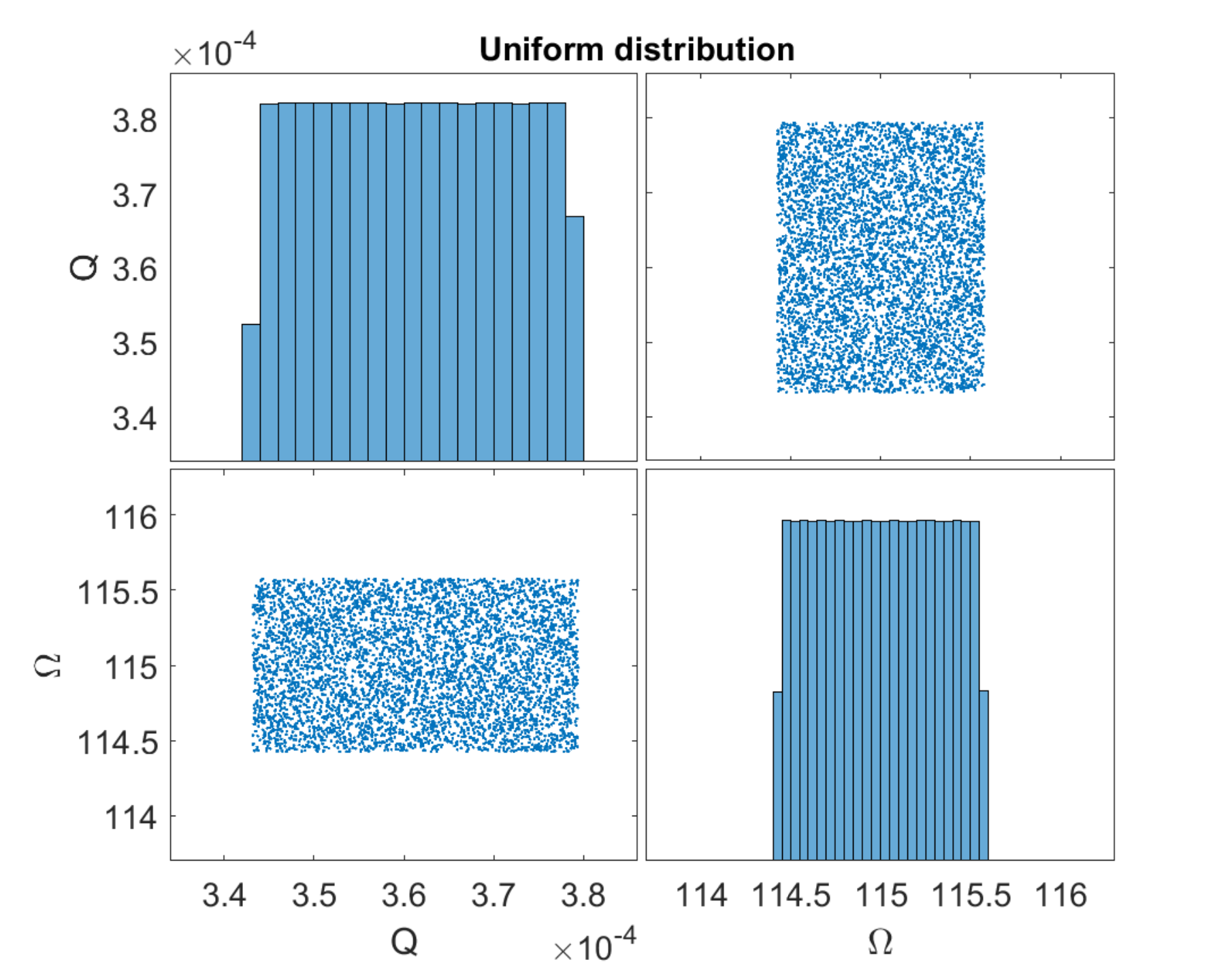}
		\caption{PDFs for $ Nu_0 $ (left) and $ Nu_{avg} $ (right) when sampling the uniform probabilistic distributions. } \label{fig: LHSunif}
	\end{center}
\end{figure}

%\begin{figure}[h!]
%	\begin{center}
%		\includegraphics[width=9.5cm]{modulo_velocidad_adim_Re23000S1_coord.jpg}
%		\caption{Detail of the contour plot of the dimensionless velocity nearby the coordinate of the peak in the evolution of the Nusselt number.} \label{Nusselt_peak}
%	\end{center}
%\end{figure}
In order to give a wider insight of the propagation of uncertainties, it is recommended to plot the Probabilistic Distribution Functions (PDF) of the outputs. To obtain such information, Latin Hypercube Sampling (LHS) (see Fig. \ref{fig: LHSunif}), has been applied on the construction from Eq. (\ref{eq:interp}). With this technique, since the response surface has a negligible cost of evaluation, it has been sampled with 5 million samples to get converged PDFs. 

\begin{figure}[h!]
	\begin{center}
		\includegraphics[width=17cm]{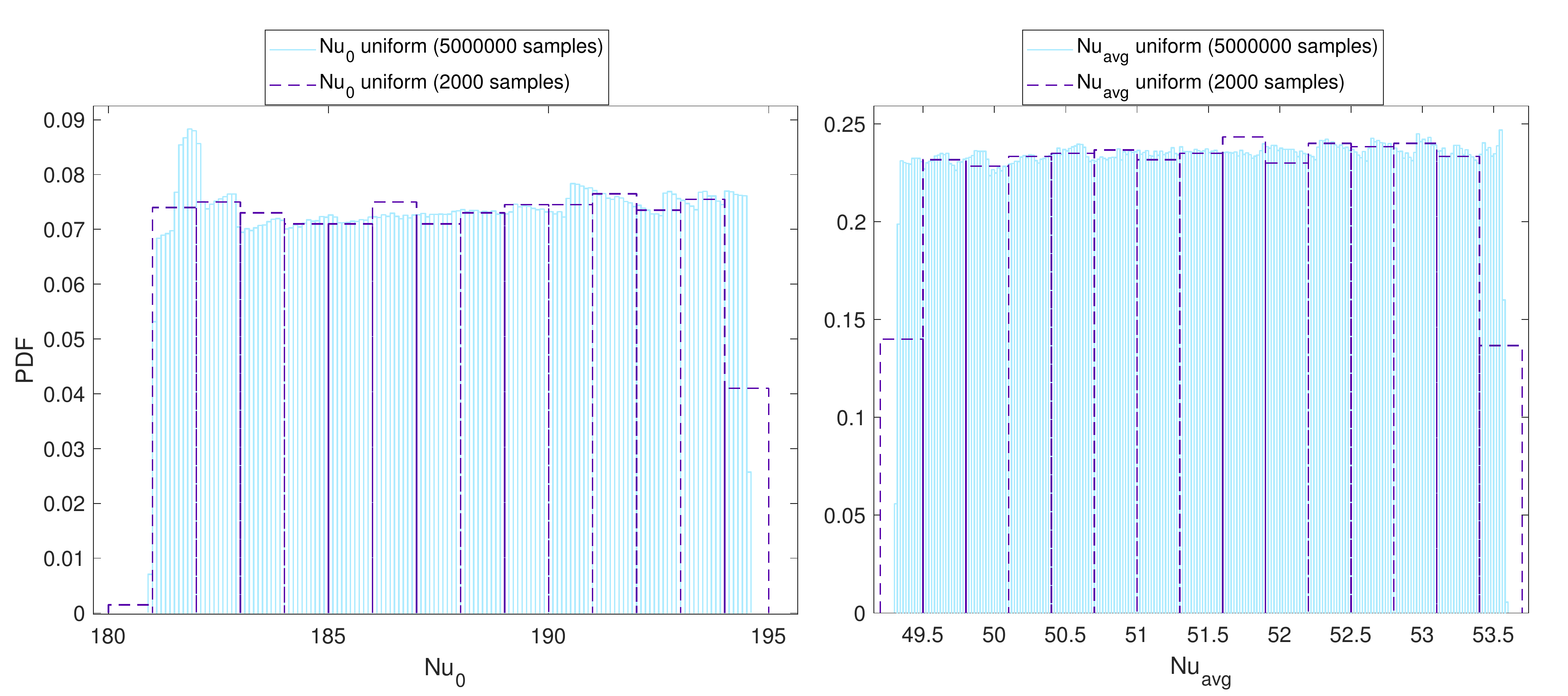}
		\caption{PDFs for $ Nu_0 $ (left) and $ Nu_{avg} $ (right). } \label{fig: PDFunif}
	\end{center}
\end{figure}

The PDFs for the Nusselt average and at the stagnation point are plotted in Fig. \ref{fig: PDFunif}. These outputs accurately approximate a $ Nu_0 \sim Unif(181, 194.5)$ and $ Nu_{avg} \sim Unif(49.32, 53.6)$ distributions. The reasons of having  uniform distributions also for the output variables seems to be related to the already discussed predominant linear input-output relationship. The input uncertain parameters have been also modelled as truncated Gaussian distributions to observe their contribution to the probabilistic functions of the output. For such constructions, the definition of the input from a standard Gaussian distribution is defined as
\begin{eqnarray}
Q = \bar{Q} + \sigma_Q \theta, \\
\Omega = \bar{\Omega} + \sigma_\Omega \theta, 
\label{eq: gaussian_PDF}
\end{eqnarray}
with 
\begin{eqnarray}
\theta \sim N(0, 1), \\
\sigma_Q = CoV_Q \: \bar{Q}, \\
\sigma_\Omega = CoV_\Omega \: \bar{\Omega}, 
\label{eq: moregaussian_PDF}
\end{eqnarray}
where $N$ refers to the Gaussian distribution, $CoV$ the coefficient of variation ($2.5\%$ for $Q$ and $0.25\%$ for $\Omega$), $\theta$ the normal distribution of zero value mean and unit variance, and $\sigma$ the standard deviation. 

In this paper, the truncated Gaussian distribution is denoted as $ \mathcal{N}(\mu_o, \sigma_o^2; \Delta_l, \Delta_u) $, where $\mu_o $ and $\sigma_o^2 $  are the mean and variance of the original Gaussian distribution respectively, and $\Delta_l$ and $\Delta_u$ the lower and upper limits that define the truncation interval.

\begin{figure}[h!]
	\begin{center}
		\includegraphics[width=10cm]{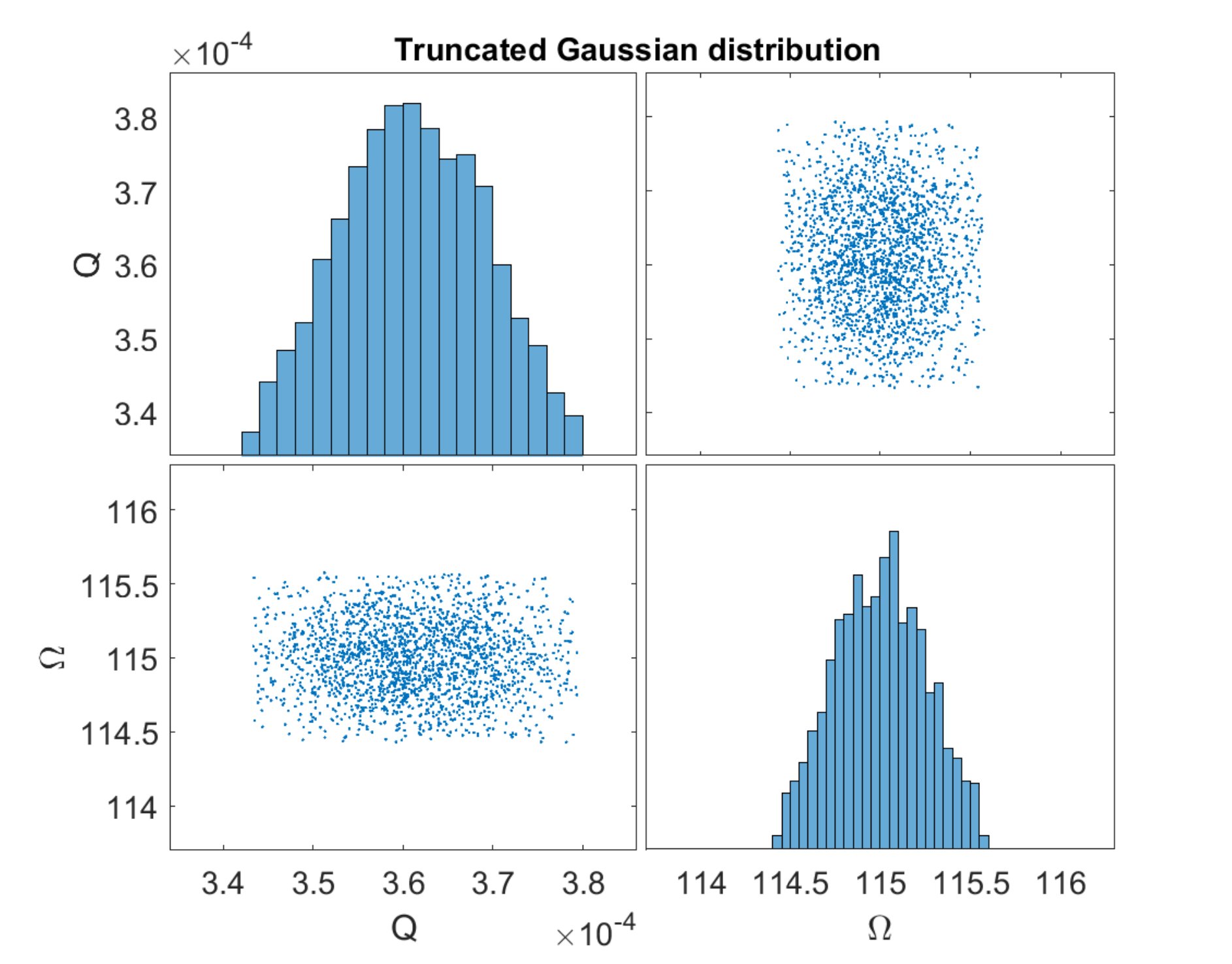}
		\caption{Sampling with $N_s=2000$ samples. } \label{fig: rand_gauss}
	\end{center}
\end{figure}

The coefficient of variation of the original Gaussian distributions have been chosen as half of the corresponding ranges in the considered uniform probabilistic distributions. Since the objective is just to observe the propagation of the uncertainty by different distributions, the coefficients of variation can be set at a lower value if necessary. Also, the shorter the variance, the less the effect of truncation. In our truncated Gaussian functions, the distributions lie within an interval, which is the one defined by the uniform distributions. In other words, these are restricted to the intervals $Q\in(0.95\bar{Q}, 1.05\bar{Q})$ and $ \Omega \in(0.995\bar{\Omega}, 1.005\bar{\Omega})$. An example of their sampling is shown in Fig. \ref{fig: rand_gauss} for $N_s =2000$ samples. 

\begin{figure}[h!]
	\begin{center}
		\includegraphics[width=17cm]{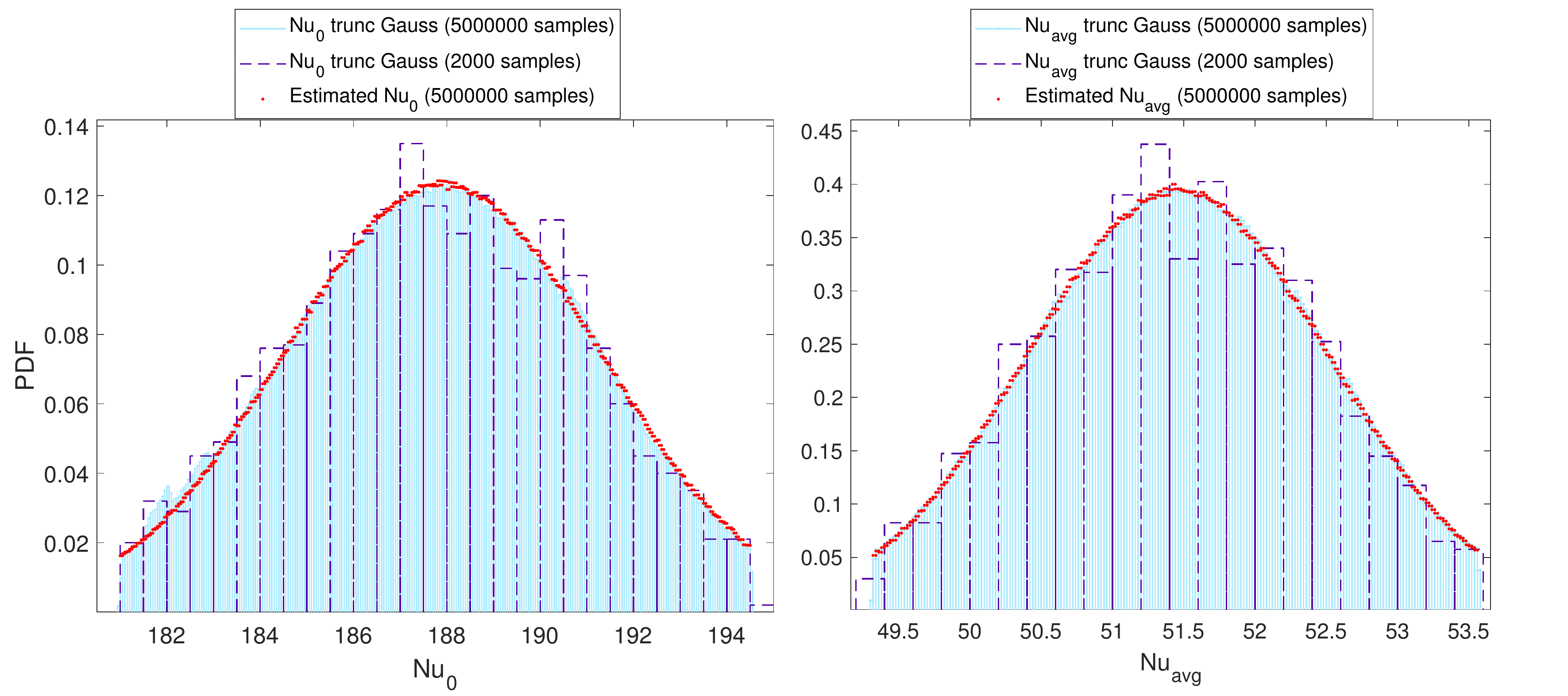}
		\caption{PDFs for $ Nu_0 $ (left) and $ Nu_{avg} $ (right) when sampling the truncated normal probabilistic distributions. } \label{fig: PDFgauss}
	\end{center}
\end{figure}

The distributions of the output variables are plotted in Fig. \ref{fig: PDFgauss} for two different number of samples by means of the LHS method. Again, $N_s = 5000000$ samples have been used to evaluate the response surface and get a converged probabilistic distribution. The output variables follow the truncated Gaussian distributions  $ Nu_0 \sim \mathcal{N}(187.9, \: 3.4; 181, \: 194.5)$ and $ Nu_{avg} \sim \mathcal{N}(51.4, \:1.06; 49.32, \: 53.6)$. The sampling of these estimated distributions is plotted in Fig. \ref{fig: PDFgauss} in red, with an accurate match. The type of distribution of the inputs has been again preserved, as expected.
 In addition, one can see in Table \ref{gauss_vs_unif_means} and \ref{gauss_vs_unif_variances} a comparison
 of the mean and variance of the Nusselt number predicted with the uniform and with the Gaussian inputs. It can be noticed in Table \ref{gauss_vs_unif_variances} that the variances from the Gaussian input are almost half of the corresponding values using a uniform one. This is because the coefficient of variance was halved for the original Gaussian ones and the predominant linearity preserves this proportion. 

\begin{table}[h!]
	\begin{center}
		\begin{tabular}{*{3}{c}}
			\hline
			Input uncertainties & $Nu_{0}$ & $Nu_{avg}$  \\
			\hline
			Uniform & 187.80547444 &  51.45682557 \\ 
			Gaussian & 187.82027016 &  51.45935403 \\ 
			\hline       

		\end{tabular}
	\end{center}
	\caption{Stochastic means of the Nusselt number at the stagnation point and its average value along the flat plate for the two different input uncertainties by sampling the response surface from the SCM.}
	\label{gauss_vs_unif_means}
\end{table}

\begin{table}[h!]
	\begin{center}
		\begin{tabular}{*{3}{c}}
			\hline
			Input uncertainties & $Nu_{0}$ & $Nu_{avg}$  \\
			\hline
			Uniform & 15.44029670 &  1.50541134 \\  
			Gaussian & 9.01274706 &  0.87362700 \\ 
			\hline       
			
		\end{tabular}
	\end{center}
	\caption{Stochastic variance of the Nusselt number at the stagnation point and its average value along the flat plate for the two different input uncertainties by sampling the response surface from the SCM.}
	\label{gauss_vs_unif_variances}
\end{table}

\subsection{Simulation 2. Uncertainty Quantification of the Impinging Swirling Jet for Heat Transfer: Use of Non-linear Regression Models for Simulation 1 input.} \label{sec_models}

In Section \ref{sec:UQ_impinging}, the computed CFD profiles have been used as new inlet conditions for \textit{Simulation 2}. However, the use of the modelled regression profiles is another option to run \textit{Simulation 2}. The outputs by this approach are compared to those by running \textit{Simulation 1}. Since the collocation points are the same, the difference between these outputs is pointing out the impact of the modelling errors. 

\begin{figure}[h!]
	\begin{center}
		\includegraphics[width=13cm]{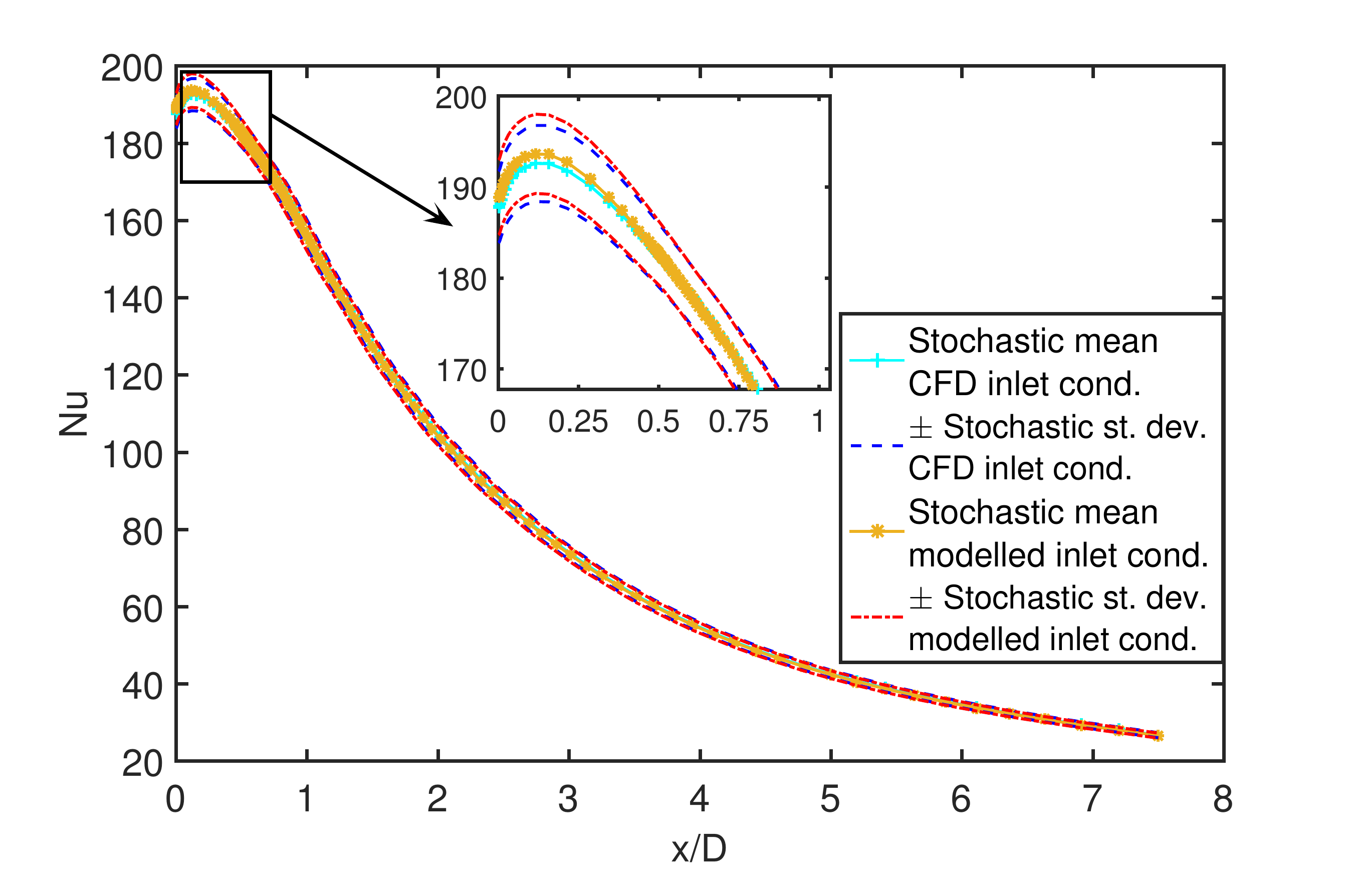}
		\caption{Comparison between radial distribution of the Nusselt number when using the CFD profiles and regression models for the inlet boundary conditions.} \label{fig: CFDvsModel}
	\end{center}
\end{figure}

To quantify the propagation of uncertainties, the Stochastic Collocation method has been applied for levels 1 and 2 of the Clenshaw-Curtis sparse grid, in order to compare the use of the inlet regression models with the direct CFD inlet condition case (coupling between \textit{Simulation 1} and \textit{2}). 
The numerical results are shown in Tables \ref{table_model_means} and \ref{table_model_variances}. Note that negative values in the relative error show an increase of the Nusselt number when using the regression models. It can be seen that, despite the models had a very accurate goodness of fit with CFD data profiles in the modelling stage, the small fitting errors are propagated through the simulation leading to a relative error of almost a $8.7\%$ in the Nusselt number at the stagnation point with respect to the direct CFD inlet condition case. Such percentage of error may seem large in principle, however, by plotting the radial distribution of $Nu$ for both the modelled and direct CFD approach (Fig. \ref{fig: CFDvsModel}), it can be noticed that the results are not remarkably different. Therefore, the suggested regression models for the inlet of \textit{Simulation 2} are introducing very low uncertainty. 

\begin{table}[h!]
	\begin{center}
		\begin{tabular}{*{8}{c}}
			\hline
			Level & Points & $Nu_0$ & $Nu_{avg}$ & $Nu_{0,model}$ & $Nu_{avg, model}$ & $\epsilon_{r,Nu_0} (\%)$ & $\epsilon_{r,Nu_{avg}} (\%)$ \\
			\hline
			%format long values
			%1 & 5  & 187.8183333333346 &  51.454739515241798 & 188.8865000000013 & 51.245621440328193 \\
			%2 & 13 & 187.7914166666679 &  51.454138253348077 & 188.8598277777790 & 51.244494261924352 \\\  
			1 & 5  & 187.81833 &  51.454739 & 188.886500 & 51.245621 & -0.5687 & 0.4064 \\
			2 & 13 & 187.79141 &  51.454138 & 188.859827 & 51.244494 & -0.5689 & 0.4074 \\
			\hline
		\end{tabular}
	\end{center}
	\caption{Stochastic means of the Nusselt number case at the stagnation point and its average value along the flat plate. Relative error (in \%) between the CFD and model input results.}
	\label{table_model_means}
\end{table}

\begin{table}[h!]
	\begin{center}
		\begin{tabular}{*{8}{c}}
			\hline
			Level & Points & $Nu_0$ & $Nu_{avg}$ & $Nu_{0,model}$ & $Nu_{avg, model}$ & $\epsilon_{r,Nu_0} (\%)$ & $\epsilon_{r,Nu_{avg}} (\%)$ \\
			\hline
			%format long values
			%1 & 5  & 15.243481888661336 &  1.509455409450311 & 16.467074583102658 & 1.521298611826751  & -8.026989524955422 & -0.784600972528969  \\         
			%2 & 13 & 15.568816376158793 &  1.508783017682617 & 16.930460364550527 & 1.521590372016362 & -8.745969863687765 & -0.848853138363252 \\
			1 & 5  & 15.243481 &  1.509455 & 16.467074 & 1.521298 & -8.026989 & -0.784601 \\
			2 & 13 & 15.568816 &  1.508783 & 16.930460 & 1.521590 & -8.745970 & -0.848853 \\              
			\hline
		\end{tabular}
	\end{center}
	\caption{Stochastic variances of the Nusselt number case at the stagnation point and its average value along the flat plate. Relative error (in \%) between the CFD and model input results. }
	\label{table_model_variances}
\end{table}

\begin{figure}[h!]
	\begin{center}
		\includegraphics[width=15cm]{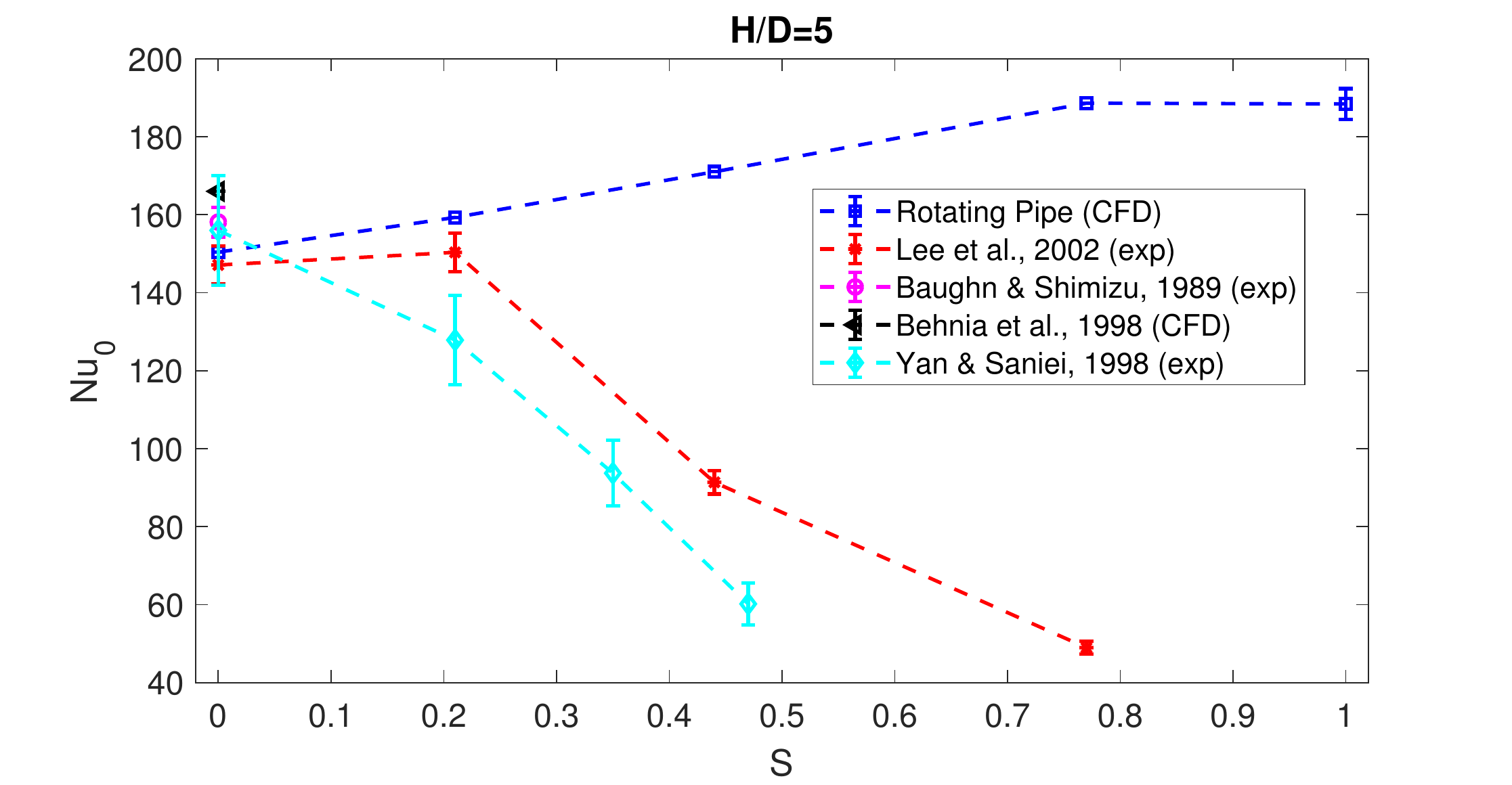}
		\caption{Comparison of Nusselt at stagnation point and various swirl numbers with other existing data from the literature, with different swirl generation mechanisms for $ H/D =5$ and $Re=23000$. In the legends, \textit{exp} and \textit{CFD} stands for experimental and computational, respectively. For the rotating pipe, only $S=1$ uncertainty bars are reported in this paper.} \label{result_comp}
	\end{center}
\end{figure}

A limitation to these computations is that there is no experimental work in the literature for this heat transfer set-up to validate the Nusselt output results. Therefore, the computational uncertainty cannot be plotted on experimental data as in Fig. \ref{pipe_vel_prof}. On the other hand, as done in \cite{Granados2019I}, one can compare this set-up with other mechanisms to generate swirl from the literature. In Fig. \ref{result_comp}, the swirl number $S$ is varied and the output of \textit{Simulation 2} is compared with results from the literature using pipes with spirals/vane-type swirl generators \cite{Lee, bakirci2007visualization}, swirless impinging jets \cite{Baughn, behnia1998prediction}, and using tangential jets to impart the swirl \cite{yan1998heat}. Fig. \ref{result_comp} also includes, for the rotating pipe system analysed in this work, the uncertainty for $S=1$, already computed in Table \ref{table_model_variances}. To report uncertainty for other values of $S$ would be very costly (at least 5-13 collocation points per $S$ value) and not very useful (the uncertainty in $S$ is less influential than uncertainty in $Q$, and the lower the value of $S$ the smaller the angular velocity error). In addition, the modelled inputs are uniform probabilistic distributions, which are conservative. For these reasons, the uncertainty bar plotted is expected to be the largest within the considered ranges. It can be seen in the experimental results reported in Fig. \ref{result_comp} that the greater the value of $Nu_0$, the larger the uncertainty is, since it was typically reported as a percentage of the Nusselt number. Thus, the simulations suggest that the simulated rotating pipe is very insensitive to the studied aleatoric uncertainty.

\section{Conclusions} \label{sec: conclusions}
%Explain the conclusions obtained: sensitivity, parameters to control, etc. \\

A framework for CFD simulations of a two-step heat transfer process has been carried out to model the propagation of uncertainty. The outflow from \textit{Simulation 1} (generation of a fully-developed swirling turbulent flow due to the rotation of a round pipe) has been mathematically modelled, in order to provide stochastic radial profiles for turbulent and velocity variables. The Stochastic Collocation Method with Clenshaw-Curtis nested rule has been applied considering the angular velocity $\Omega$ and the volume flow rate $Q$ as input uncertainties. \\
 
In the UQ study on \textit{Simulation 1}, results have shown that the considered experimental uncertainties in $Q$ and $\Omega$ slightly vary the friction factor, $\lambda$, and the turbulent intensity, $I$. In the dimensionless velocity profiles, these uncertainties have only a modest impact. The most sensitive part of the dimensionless axial velocity profile is near $r/R = 0$, and near $r/R = 1$ for the azimuthal one. Only experimental results for these velocities have been reported in the literature, and all the axial velocity data is within the standard deviation envelopes. The azimuthal velocity 
profiles matched the trend of the experimental data well, but the magnitude is systematically over-predicted by the base case CFD simulation. In the uncertainty analysis of the dimensionless turbulent kinetic energy, $k/{U^2}$, the most sensitive region is the part located at the beginning of the decay, due to the strong effect of the wall. Finally, for the laminar to turbulent viscosity ratio, $\beta$, the sensitivity trend is very similar to the dimensionless axial velocity one, exhibiting around the nozzle axis the largest variances. This is the most sensitive profile to the input uncertainties, but not remarkable.

The UQ study on the heat transfer process in \textit{Simulation 2} was developed in two different ways: by imposing the dimensionless outflow profiles from \textit{Simulation 1}, and by imposing the non-linear regression models by User Defined Functions as inlet boundary conditions to \textit{Simulation 2}. With both approaches, it was observed that the random inputs have a modest impact on the predicted Nusselt number along the plate. It was also noticed that $Q$ was the most influential parameter, since $\Omega$ had a very low contribution to uncertainty. As a result, in an experimental facility with similar sources of uncertainty, engineers should put more effort in reducing the stochastic variance in the volume-flow rate by, e.g., reducing pressure losses. 

Within the tested ranges of $Q$ and $\Omega$, it has been observed a linear relationship of most quantities of interest analysed, with respect to the uncertain inputs. As a consequence, in most uncertainty results presented in this paper can be noticed that, even by using few collocation points with the Clenshaw-Curtis sparse grid, the statistical moments do not change dramatically. Only the convergence in the stochastic variance of the Nusselt number required a greater number of collocation points (\textit{i.e.} CFD simulations of \textit{Simulation 1} and \textit{2}) with important computational efforts.

Four non-linear parametric regression models have been suggested for the fully-developed state of the swirling flow confined in the rotating pipe and used as jet nozzle (dimensionless profiles $v_{z}/{U}, \: v_{t}/{U}, \: {k}/{U^2}$ and $\beta$). The objective was to find surrogate models for one of the most studied impinging jet regime in the literature with/without swirl: the $Re=23000$. These models fitted well with the computational data from \textit{Simulation 1}. It was shown that the sensitivity of the distribution of the Nusselt number to the modelling errors is negligible, when the regression models are imposed onto \textit{Simulation 2}. Thus, these models are recommended to replace the deterministic or stochastic computation of \textit{Simulation 1} and can be an interesting option to save computational costs in related applications such as systems with arrays of impinging jets, swirling jets impinging on a plate with variations on the surface, or swirling turbulent flows with axisymmetric sudden expansions.

\section*{Acknowledgments}
The first author acknowledges his support from the European Community's Seventh Framework Programme (FP7) Marie Curie AeroTraNet 2 Action, under the grant number PITN-GA-2012-317142. The second author thanks the III Plan Propio at University of M\'alaga, which supported his visit to the University of Greenwich in Summer 2014. The first author is also thankful to Dr Paul Constantine about the fruitful discussions in the beginning of this work and a warm special acknowledgement to Dr Jeroen A. S. Witteveen for his useful suggestions and who sadly passed away.

\bibliography{../../Bibliografia_general/bibliografia_Granados}
\bibliographystyle{wileyj}

%\bibliography{../../Bibliografia_general/bibliografia_Granados}
%\bibliographystyle{elsarticle-num}%unsrtnat}

\newpage
\appendix

\begin{appendices}
  \section{Appendix: Models for the fitting coefficients.} \label{append_A}

\begin{figure}[h!]
    \begin{center}
      \includegraphics[width=13cm]{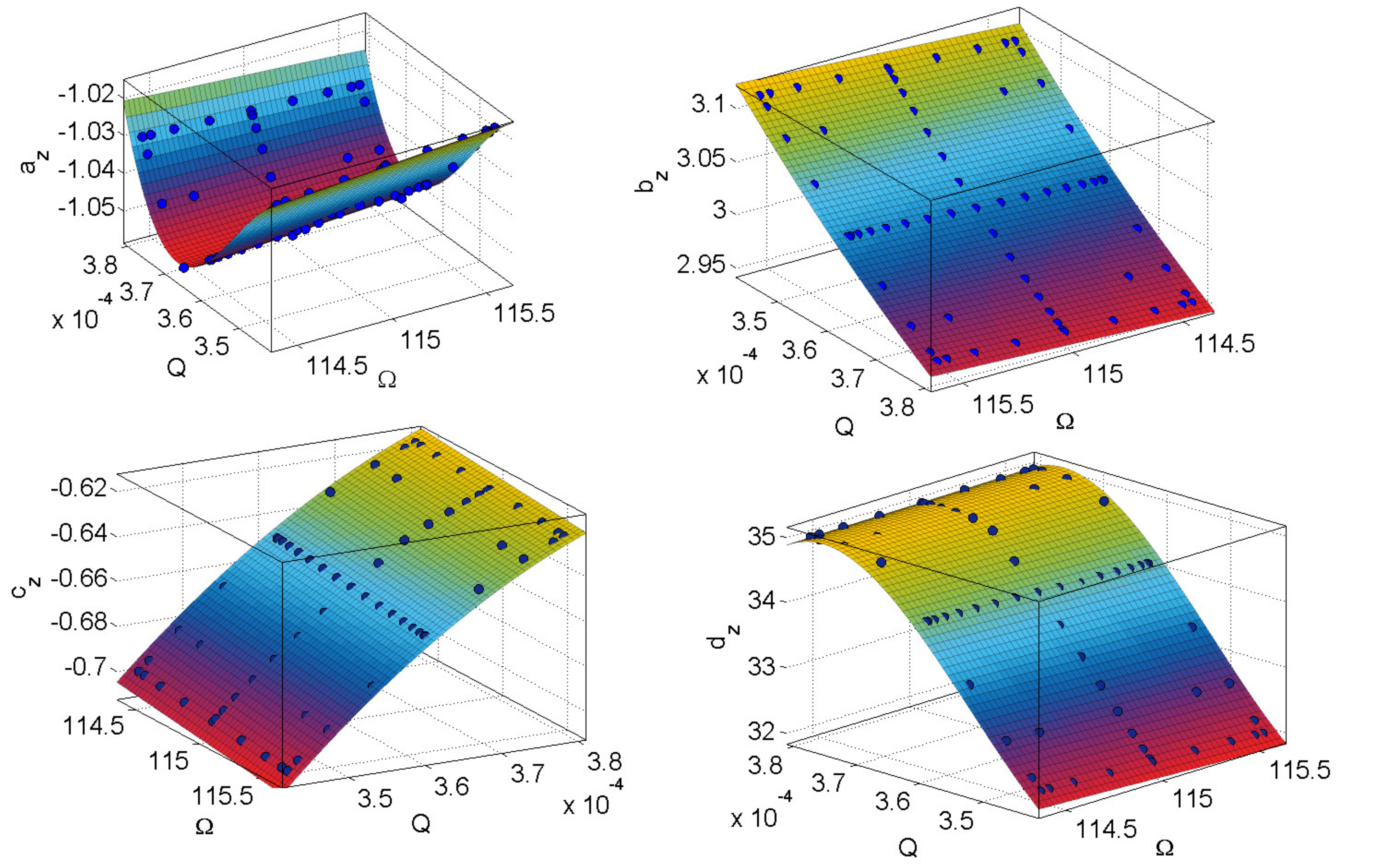}
      \caption{Models for the $\frac {v_{z}} {U}$ coefficients.}  \label{all_vz_coeff}
    \end{center}
   
  \end{figure}

\begin{figure}[h!]
      \begin{center}
        \includegraphics[width=11cm]{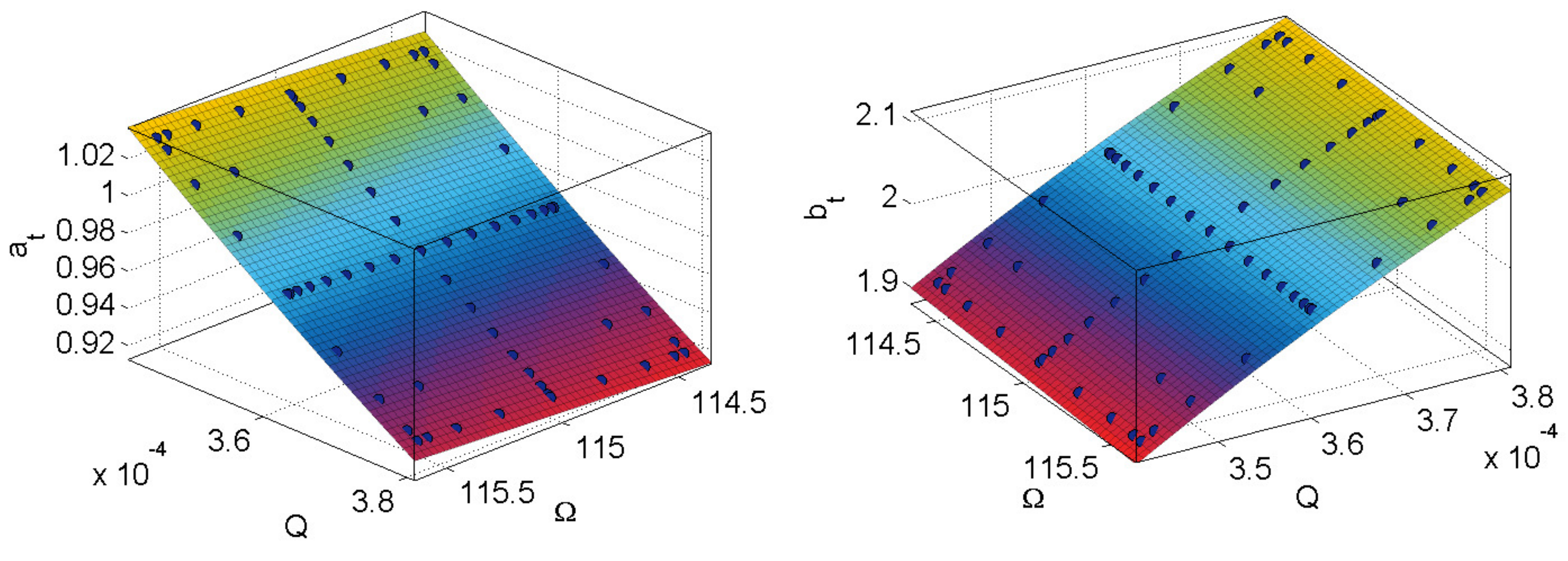}
        \caption{Models for the $\frac {v_{t}} {U}$ coefficients.} \label{all_vt_coeff}
      \end{center}
      
    \end{figure}

\begin{figure}[h!]
        \begin{center}
          \includegraphics[width=13cm]{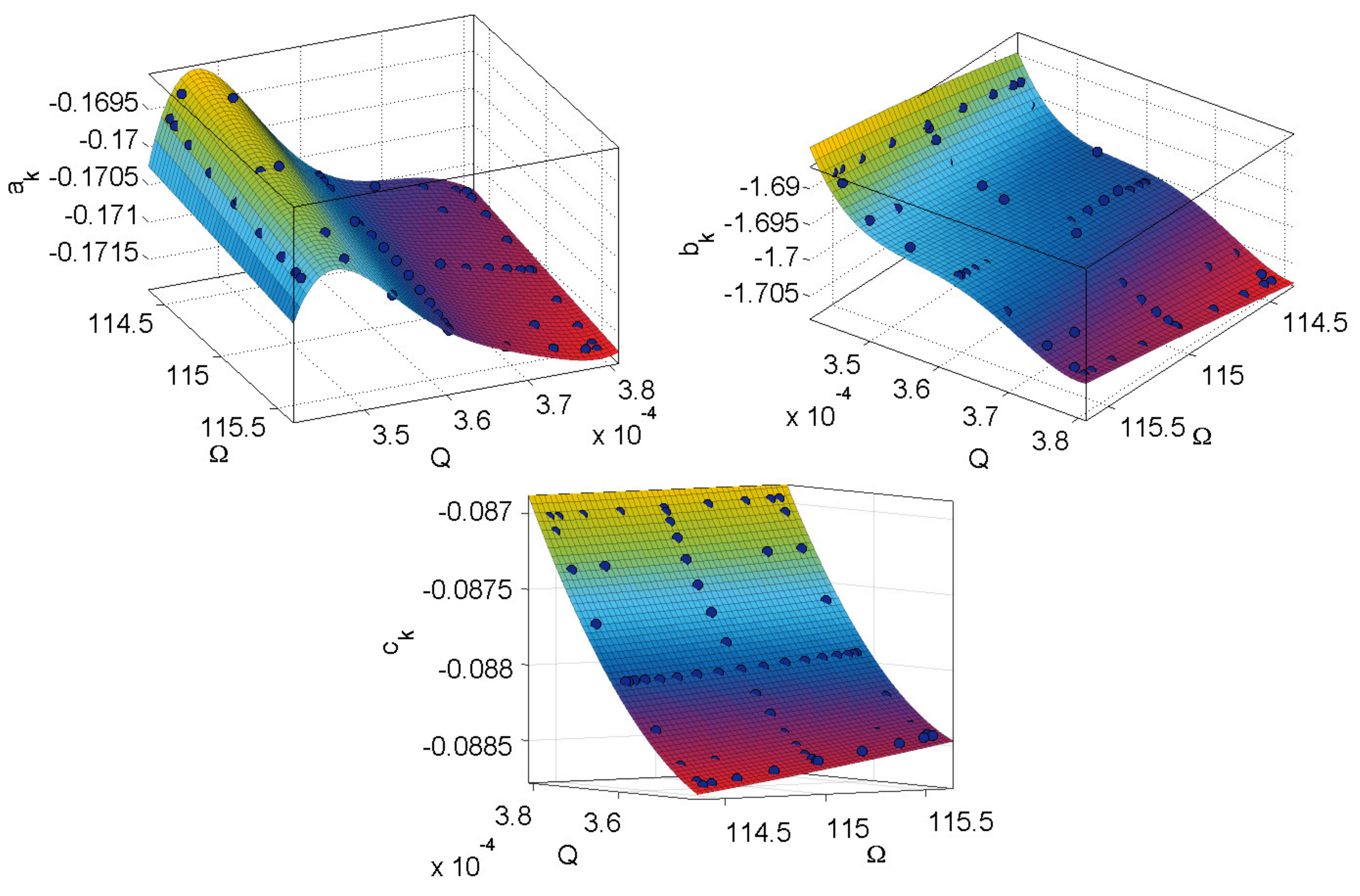}
          \caption{Models for the $\frac {k} {U^2}$ coefficients.} \label{all_k_coeff}
        \end{center}
        
      \end{figure}

\begin{figure}[h!]
          \begin{center}
            \includegraphics[width=13cm]{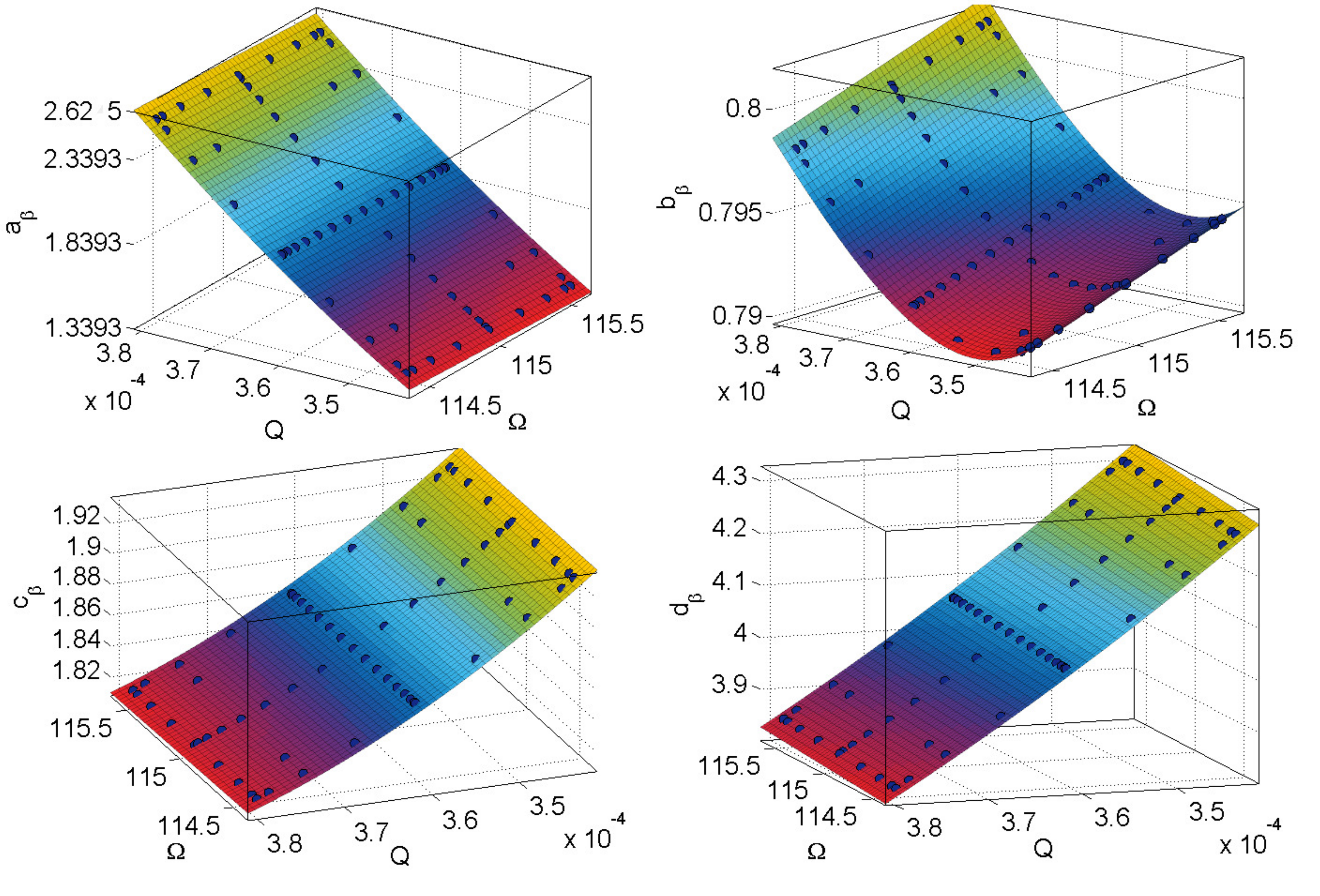}
            \caption{Models for the $\beta$ coefficients.} \label{all_tvr_coeff}
          \end{center} 
          
        \end{figure}

As aforementioned in Section \ref{sec_models}, the $\gamma_i$ parameters are modelled by polynomial regressions. These are introduced in Eqs (\ref{eq:az}) - (\ref{eq:db}). The goodness of the fits can be seen in Figs. \ref{all_vz_coeff}-\ref{all_tvr_coeff}. \\ 
From these results, the equations to model $\frac {v_{z}} {U}, \frac {v_{t}} {U}, \frac {k} {U^2}$ and $\beta$ are only dependent on $Q$ and $\Omega$, and ready to use for any application. Note the equations are normalised via $Q_n= \frac{Q-\bar{Q}}{\sigma'_Q}$ and $\Omega_n= \frac{\Omega-\bar{\Omega}}{\sigma'_\Omega}$, where overbar denotes mean values and $\sigma'$ stands for the standard deviation.

In Table \ref{table: goodness_coeff} can also be found the goodness indicators of the fit. As the coefficient of determination ($\hat{R}^2$) is not the best measure for the goodness of a fit, specially in non-linear cases, the Sum of Squares due to Error (SSE), Adjusted-$\hat{R}^2$ and Root Mean Squared Error (RMSE) are also given. 
Over-fitting has been avoided for all the fits by using the minimum order polynomial that reasonably satisfied the goodness.

\begin{eqnarray}  \label{eq:az}
a_{z} = -1.048 + 0.001532  \Omega_n -0.01444  Q_n - (4.147e-05)  \Omega^2_n-0.0009388  \Omega_n Q_n \\  \nonumber
+ 0.01183  Q^2_n  -(5.997e-05)  \Omega^3_n + (3.648e-05)  \Omega^2_n Q_n    -0.0005627  \Omega_n Q^2_n \\ \nonumber
 + 0.005394  Q^3_n, \nonumber
\end{eqnarray}
\begin{eqnarray}
b_{z} = 3.023   + 0.004779  \Omega_n  -0.05368  Q_n + (5.742e-06)  \Omega^2_n  -0.0005031  \Omega_n Q_n \\ \nonumber
+ 0.003906  Q^2_n, \nonumber
  \label{eq:bz}
\end{eqnarray}
\begin{eqnarray}
c_{z} = -0.6536  -0.002857  \Omega_n + 0.03038  Q_n + (1.346e-05)  \Omega^2_n + 0.0003585  \Omega_n Q_n \\  -0.004234  Q^2_n, \nonumber
  \label{eq:cz}
\end{eqnarray}
\begin{eqnarray}
d_{z} = 34.16   + -0.05319  \Omega_n + 1.362  Q_n + 0.0008092  \Omega^2_n + 0.02154  \Omega_n Q_n  -0.3225  Q^2_n \\ +  0.002316  \Omega^3_n + 0.0001007  \Omega^2_n Q_n  + 0.01276  \Omega_n Q^2_n  -0.1631  Q^3_n, \nonumber
  \label{eq:dz}
\end{eqnarray}
\begin{eqnarray}
a_{t} = 0.9708   + 0.003415  \Omega_n  -0.03677  Q_n  -0.0001444  \Omega_n Q_n + 0.001342  Q^2_n,\\ \nonumber \\
b_{t} = 2.005    -0.006083  \Omega_n + 0.06979  Q_n + 0.0004104  \Omega_n Q_n -0.004283  Q_n^2,
  \label{eq:at_bt}
\end{eqnarray}
\begin{eqnarray}
a_{k} = -0.1709 -0.0001532 \Omega_n  -0.000986  Q_n  + (7.86e-05)  \Omega_n Q_n +  0.0007671  Q^2_n \\ \nonumber -0.0001781  \Omega_n Q^2_n  -(6.527e-05)  Q^3_n -(3.653e-05) \Omega_n Q^3_n -0.0003428  Q^4_n \\ \nonumber + (8.297e-05) \Omega_n Q^4_n + 0.0001397  Q^5_n, \nonumber
  \label{eq:ak}
\end{eqnarray}
\begin{eqnarray}
b_{k} = -1.698   + 0.001364  \Omega_n  -0.002698  Q_n + (3.023e-05)  \Omega^2_n  -0.0002644  \Omega_n Q_n \\ \nonumber -0.002849  Q^2_n + (3.223e-05)  \Omega^2_n Q_n +  0.0001315  \Omega_n Q^2_n  -0.001438  Q^3_n \\ \nonumber -(6.665e-05)  \Omega^2_n Q^2_n + (7.724e-05)  \Omega_n Q^3_n + 0.001602  Q^4_n, \nonumber
  \label{eq:bk}
\end{eqnarray}
\begin{eqnarray}
c_{k} = -0.08801   + (5.075e-05)  \Omega_n + 0.0005716  Q_n + (3.703e-06)  \Omega^2_n \\ \nonumber -(2.574e-05)  \Omega_n Q_n + 0.0001155  Q^2_n, \nonumber
\label{eq:ck}
\end{eqnarray}
\begin{eqnarray}
a_{\beta} =1.932 -0.01402  \Omega_n + 0.4053  Q_n  -0.0006922  \Omega_n Q_n + 0.02551  Q^2_n,
  \label{eq:ab}
\end{eqnarray}
\begin{eqnarray}
b_{\beta}  = 0.7933   + 0.001219  \Omega_n + 0.003093  Q_n  -(7.063e-06)  \Omega_n Q_n + 0.001437  Q^2_n \\ + (1.968e-06)  \Omega_n Q^2_n  -0.0002586  Q^3_n, \nonumber
  \label{eq:bb}
\end{eqnarray}
\begin{eqnarray}
c_{\beta}  = 1.862    -0.000533  \Omega_n  -0.04129  Q_n  -0.0002055  \Omega_n Q_n + 0.005018  Q^2_n,
  \label{eq:cb}
\end{eqnarray}
\begin{eqnarray}
d_{\beta}  = 4.047   + 0.008784  \Omega_n  -0.1622  Q_n  -0.0009475  \Omega_n Q_n + 0.00729  Q^2_n.
  \label{eq:db}
\end{eqnarray}
\\


\begin{table}[h!]
	\begin{center}
		\begin{tabular}{c | c c c c}
			\hline
			Coefficient & SSE & $ \hat{R}^2 $ & Adjusted $ \hat{R}^2 $ & RMSE  \\
			\hline
			%format long values
			
			$ a_z $ & 8.206e-05  & 0.9903  & 0.9887  & 0.001221 \\
			$ b_z $ & 4.892e-07   & 1  & 1  & 9.105e-05 \\
			$ c_z $ & 3.278e-05  & 0.9995  & 0.9994  & 0.0007453 \\
			$ d_z $ & 0.07246  & 0.9991  & 0.999  & 0.0363 \\
			\hline
			$ a_t $ & 6.173e-08  & 1  & 1  & 3.208e-05 \\
			$ b_t $ & 4.583e-07  & 1  & 1  & 8.74e-05 \\ 
			\hline
			$ a_k $ & 3.357e-07  & 0.9904  & 0.9887  & 7.884e-05 \\ 
			$ b_k $ & 1.371e-05  & 0.9925  & 0.991   & 0.0005086 \\ 
			$ c_k $ & 1.238e-08  & 0.9994  & 0.9994  & 1.448e-05 \\
			\hline
			$ a_\beta $ & 4.878e-06  & 1  & 1  & 0.0002851 \\
			$ b_\beta $ & 2.248e-08  & 1  & 1  & 1.969e-05 \\
			$ c_\beta $ & 1.954e-06  & 1  & 1  & 0.0001805 \\
			$ d_\beta $ & 7.532e-06  & 1  & 1  & 0.0003543 \\			
			
			\hline
		\end{tabular}
	\end{center}
	\caption{Goodness of the fitting for the coefficients of the models. }
	\label{table: goodness_coeff}
\end{table}

%\newpage
\clearpage
  \section{Appendix: User-Defined Functions to implement the models.} \label{append_B}

The models have been implemented in FLUENT by means of a User-Defined Function (UDF) coded in $ C $. In this appendix, a simple example about how to implement the UDF is given. Please note that the mathematical expressions of the coefficients, introduced in Appendix \ref{append_A} should be implemented in the UDF as well. For sake of simplicity, these models are not implemented and the coefficients are supposed to be calculated and used as defined variables in the UDF. \\

%\begin{figure} [h!]
%	\setlength{\fboxsep}{0pt}%
%	\setlength{\fboxrule}{0pt}%
	\begin{center}
		\begin{verbatim}
		/****** UDF FOR THE PROFILES ******/
#include "udf.h"

#define az  <input_value>  
#define bz  <input_value>  
#define cz  <input_value>  /* These are the constants for v_z/U */
#define dz  <input_value>

#define at  <input_value>   
#define bt  <input_value> /* These are the constants for v_t/U */

#define ak  <input_value>  
#define bk  <input_value>  /* These are the constants for k/U^2 */
#define ck  <input_value>

#define avr <input_value>  
#define bvr <input_value> 
#define cvr <input_value> /* These are the constants for \beta */
#define dvr <input_value>

#define rho <input_value> /* Density */
#define mu  <input_value> /* Viscosity */
#define D   <input_value> /* Diameter */
#define Re  <input_value> /* Reynolds number */

DEFINE_PROFILE(vz,t,i) /* Function for the v_z/U profile */
{
	real x[ND_ND];                /* this will hold the position vector */
	real r;
	face_t f;
	
	real U=Re*mu/(D*rho); /* Calculates the velocity from the Reynolds */
	
	begin_f_loop(f,t)
	{
		F_CENTROID(x,f,t);
		r = x[1]/(D/2);
		F_PROFILE(f,t,i) = U*(az*pow(r,3)+bz*exp(-pow(cz*r,2)))*(-tanh(dz*(r-1.0)))/2.0;
	}
	end_f_loop(f,t)
}
DEFINE_PROFILE(vt,t,i) /* Function for the v_t/U profile */
{
	real x[ND_ND];                
	real r;
	face_t f;
	
	real U=Re*mu/(D*rho);
	
	begin_f_loop(f,t)
	{
		F_CENTROID(x,f,t);
		r = x[1]/(D/2);
		F_PROFILE(f,t,i) = U*at*pow(r,bt);
	}
	end_f_loop(f,t)
}
DEFINE_PROFILE(k,t,i) /* Function for the k/U^2 profile */
{
	real x[ND_ND];                
	real r;
	face_t f;
	
	real U=Re*mu/(D*rho);
	
	begin_f_loop(f,t)
	{
		F_CENTROID(x,f,t);
		r = x[1]/(D/2);
		F_PROFILE(f,t,i) = U*U*(0.05327/(ak+exp(pow(r,1.2)*ck)))*(tanh(bk*(r-1)));
	}
	end_f_loop(f,t)
}

DEFINE_PROFILE(omega,t,i) /* Function for the \omega profile from k and
* \beta profiles, as the turbulence model to be used is k-\omega */
{
	real x[ND_ND];     /* this will hold the position vector */
	real r,k,tvr;
	face_t f;
	
	real U=Re*mu/(D*rho);
	
	begin_f_loop(f,t)
	{
		F_CENTROID(x,f,t);
		r = x[1]/(D/2);
		k=U*U*(0.05327/(ak+exp(pow(r,1.2)*ck)))*(tanh(bk*(r-1)));
		tvr=avr*exp(-bvr*pow(r,dvr))*(exp(-cvr*(pow(r,13)-1))-1);
		F_PROFILE(f,t,i) = rho*k/(tvr*mu); 
		/* This is the expression to obtain \omega */
	}
	end_f_loop(f,t)
}
			\end{verbatim}
		\end{center}
		%\vspace{-12pt}
%		\caption{Example header text.\label{F1}}
%	\end{figure}

\end{appendices}

\end{document}